\documentclass[11pt,a4paper]{article}
\pdfoutput=1
\usepackage{ifpdf}
\usepackage{jheppub}
\usepackage{comment}

\usepackage{tikz-cd}
\usepackage{subcaption}
\usepackage{graphicx} 
\usepackage{mathrsfs}
\usepackage{amsmath,amsfonts,amssymb}
\usepackage{mathtools}
\usepackage{graphicx}
\usepackage{dcolumn}
\usepackage{bm}
\usepackage{hyperref}
\usepackage[mathlines]{lineno}
\usepackage{xcolor}
\usepackage{float}
\usepackage{blindtext}
\usepackage{titlesec}
\title{Sections and Chapters}
\usepackage[toc,page]{appendix}
\usepackage{xcolor}
\usepackage{graphicx}
\graphicspath{ {./images/} }
\usepackage[thinc]{esdiff}
\usepackage[euler]{textgreek}
\usepackage{braket}
\usepackage{physics}
\usepackage{xfrac}
\usepackage{soul}
\usepackage{stmaryrd}
\usepackage{pifont}
\usepackage[dvipsnames]{xcolor}
\usepackage{multirow}
\usepackage{slashed}
\usepackage{orcidlink}
\usepackage{amsmath}
\usepackage{relsize}
\usepackage[normalem]{ulem}
\usepackage{tikz-feynman}
\tikzfeynmanset{compat=1.1.0}
\usetikzlibrary{arrows,automata,positioning}
\tikzset{
  vertex/.style={circle,draw, minimum size=1.5em},
    edge/.style={->,> = latex'}
}
\tikzfeynmanset{
  fermion1/.style={
    /tikz/postaction={
      /tikz/decoration={
        markings,
        mark=at position 0.5 with {
          \node[
            transform shape,
            xshift=-0.5mm,
            fill,
            dart tail angle=100,
            inner sep=1.3pt,
            draw=none,
            dart
          ] { };
        },
      },
      /tikz/decorate=true,
    },
  },
  fermion2/.style={
    /tikz/postaction={
      /tikz/decoration={
        markings,
        mark=at position 0.5 with {
          \arrow{>[length=5pt, width=4pt]};
        },
      },
      /tikz/decorate=true,
    },
  },
}

\allowdisplaybreaks

\title{\textcolor{black}{A Non-Holomorphic Modular $A_4$ Framework for Resonant Leptogenesis with Gravitational Wave Signatures}}
\author[a]{Mitesh Kumar Behera,\,\orcidlink{0000-0002-2137-3100}\,}
\author[b]{Jaydeb Das,\,\orcidlink{0000-0001-6335-9377}\,}
\author[b]{Niloy Mondal,\,\orcidlink{0009-0006-5837-9772}\,}

\affiliation[a]{Department of Physics, School of Advanced Sciences, Vellore Institute of Technology, Tiruvalam Rd, Katpadi, Vellore, Tamil Nadu 632014, India.}
\affiliation[b]{Department of Physics, Indian Institute of Technology Guwahati,
North Guwahati, Assam-781039, India.}

\emailAdd{miteshbehera1304@gmail.com}
\emailAdd{jaydebphys@rnd.iitg.ac.in}
\emailAdd{niloy18@iitg.ac.in}

\abstract{We study a type-I seesaw framework based on non-holomorphic $A_4$ modular symmetry, where polyharmonic Maa\ss\ forms construct the Yukawa couplings and right-handed neutrino (RHN) Majorana mass matrix. The use of non-holomorphic modular forms yields highly constrained neutral lepton mass matrices with a more restrictive lepton-sector structure and naturally generates a quasi-degenerate RHN mass spectrum, enabling resonant leptogenesis at an intermediate scale with RHN masses of $\mathcal{O}(10^3)$ TeV without requiring an ad hoc mass degeneracy. We further extend the model by introducing a complex scalar field $(\Phi)$ charged under $\mathbb{Z}_3$ symmetry. The spontaneous breaking of the discrete symmetry after the phase transition associated with $\Phi$ leads to domain-wall (DW) formation. A radiatively induced bias term associated with the RHN sector triggers DW annihilation, resolving the cosmological DW problem, and producing a stochastic gravitational wave (GW) signal that indirectly probes the RHN mass scale. The accompanying first-order phase transition produces a second GW peak, yielding a characteristic double-peaked spectrum with frequencies separated by several orders of magnitude and potentially observable by complementary future GW detectors.
}

\begin{document}
\maketitle
\flushbottom

\preprint{}

 \section{Introduction}
The discovery of neutrino masses and flavor mixing~\cite{Fukuda:1998mi, Fukuda:2001nj, Ahmad:2001an, Ahmad:2002jz} stands as the primary laboratory evidence of physics beyond the Standard Model (SM). Over the past two decades, neutrino physics has entered a precision era, with the mass-squared differences, $\Delta m^2_{21}$ and $|\Delta m^2_{31}|$, and the three mixing angles, $\theta_{12}$, $\theta_{23}$, and $\theta_{13}$, measured with remarkable accuracy~\cite{Esteban:2020cvm}. Nevertheless, several fundamental questions remain unresolved, including neutrino mass ordering (normal or inverted), the value of the leptonic CP-violating phase $\delta$, and whether neutrinos are Dirac or Majorana particles.
In addition to neutrino physics, cosmological observations provide equally compelling evidence for the new physics (NP) beyond the SM (BSM). The observed baryon asymmetry of the Universe (BAU) is conventionally quantified by the baryon yield~\cite{Planck:2018vyg,ParticleDataGroup:2024cfk},
$$
Y_B \equiv \frac{n_B-n_{\bar B}}{s} \simeq 8.7\times10^{-11},
$$
where $n_B$, $n_{\bar B}$, and $s$ denote the baryon number density, antibaryon number density, and entropy density, respectively.
The dynamical generation of the observed baryon asymmetry from an initially baryon-symmetric Universe requires the three Sakharov conditions~\cite{Sakharov:1967dj}: baryon number violation, C and CP violation, and a departure from thermal equilibrium. Among the various proposed mechanisms of baryogenesis~\cite{Kuzmin:1985mm, Cohen:1990it, Turok:1990in, Dine:1990fj, Cohen:1991iu, Trodden:1998ym, Affleck:1984fy, Thomas:2022hyj, Fukugita:1986hr}, leptogenesis~\cite{Fukugita:1986hr, Pilaftsis:1997jf, Pilaftsis:2003gt, Dev:2017wwc, Alanne:2018brf, Hugle:2018qbw, Kusenko:2014uta, Hambye:2016sby, Datta:2022jic, Bhandari:2023wit, King:2024idj, Bhattacharya:2023kws, Bhattacharya:2024ohh, Bhattacharya:2025uaa, Choudhury:2026laq, Barman:2026yhc} provides a well-motivated framework, since it naturally connects the neutrino mass generation mechanism to the generation of the baryon asymmetry. In this scenario, primordial lepton asymmetry is partially converted into the observed baryon asymmetry through electroweak sphaleron processes~\cite{Rubakov:1996vz,Klinkhamer:1984di,Manton:1983nd}. Furthermore, astrophysical and cosmological evidence for dark matter (DM)~\cite{Planck:2018vyg}, together with its absence in the SM framework~\cite{Bertone:2004pz}, indicates the need for BSM physics.

The stark contrast between leptonic mixing and the quark sector has motivated the use of non-Abelian discrete symmetries (e.g., $A_4$, $S_4$, $A_5$) to explain the origin of flavor~\cite{Ma:2001dn}. In traditional frameworks, these symmetries are spontaneously broken by scalar fields (flavons), typically predicting a baseline Tri-Bimaximal (TBM) mixing pattern. However, the discovery of a non-zero reactor angle $\theta_{13}$~\cite{An:2012eh, Ahn:2012nd} necessitates structural corrections to this zeroth-order approximation. 

While conventional non-Abelian discrete groups successfully constrain neutrino mixing patterns via group product rules, they suffer from significant drawbacks. They typically demand a large pool of flavon fields whose specific vacuum expectation values (VEVs) compromise model predictability through high parameter sensitivity. Furthermore, eliminating unwanted Lagrangian operators requires the ad hoc introduction of auxiliary symmetries. To resolve these limitations, modular non-Abelian symmetries pioneered by Feruglio~\cite{Feruglio:2017spp} have emerged as an elegant alternative. By promoting Yukawa couplings to modular forms of a finite modular group $\Gamma_N$, flavor symmetry breaking is driven solely by the VEV of the complex modulus $\tau$. In supersymmetric (SUSY) frameworks, enforcing a holomorphic superpotential yields highly predictive architectures that accommodate lepton masses and mixings with minimal parameters.

Nevertheless, holomorphic modular forms are restricted to non-negative integral weights and span a finite function space, limiting their utility in non-SUSY or radiative neutrino mass models. Motivated by the absence of low-energy SUSY signals, non-SUSY realizations based on non-holomorphic modular symmetry have recently gained traction~\cite{Cremades:2004wa}. In particular, harmonic or polyharmonic Maaß forms satisfy Laplace-type equations rather than strict holomorphicity conditions. This expanded function space spans both positive and negative weights, enabling flexible, phenomenologically viable Yukawa structures while strictly preserving modular invariance and predictivity. Such non-holomorphic frameworks have been successfully deployed in Type II seesaw~\cite{Nomura:2024atp}, scotogenic~\cite{Nomura:2025ovm, Nomura:2025raf, Nomura:2024vzw}, and Zee-Babu models~\cite{Nomura:2024nwh}, offering a robust setup for exploring neutrino masses and flavor hierarchies.

In this work, we investigate the type-I seesaw framework embedded in a non-holomorphic $A_4$ modular symmetry. The model contains three right-handed neutrinos (RHNs), which are singlets under the SM gauge group and transform as an $A_4$ triplet. The charged-lepton mass matrix is diagonal by construction, while the Dirac and Majorana neutrino mass matrices are non-diagonal and are determined by non-holomorphic modular forms of different modular weights. These modular forms depend on the complex modulus $\tau$, whose allowed values are constrained by the low-energy neutrino oscillation data. A remarkable feature of this framework is that, after diagonalising the Majorana mass matrix, the RHN mass spectrum is naturally quasi-degenerate. Consequently, the tiny mass splittings required for resonant leptogenesis~\cite{Pilaftsis:1997jf, Pilaftsis:2003gt, Dev:2017wwc}, which can explain the BAU, emerge as a prediction of the underlying non-holomorphic modular symmetry, rather than being introduced by hand, as is commonly assumed in conventional resonant leptogenesis scenarios. This constitutes the central result of our analysis: the non-holomorphic modular $A_4$ symmetry provides a natural origin for the quasi-degenerate RHN spectrum, thereby offering a well-motivated framework for resonant leptogenesis. As a consequence, successful leptogenesis can be achieved at an intermediate scale of $\mathcal{O}(10^3)\,\mathrm{TeV}$, while remaining consistent with current neutrino oscillation data.

Previously, resonant leptogenesis in $A_4$ frameworks has been investigated both with holomorphic modular symmetry~\cite{Kang:2022psa} in SUSY framework and without modular symmetry~\cite{Borah:2017qdu}\footnote{For representative studies of leptogenesis in $A_4$ models with and without modular symmetry, see Refs.~\cite{Branco:2009by, Adhikary:2014qba, Karmakar:2015jza, Datta:2021zzf, Singh:2024imk, Parriciatu:2024dhb, Pathak:2025zdp, Nanda:2025lem, Tavartkiladze:2025oiq, Tapender:2026ets, Priya:2026ehe, Priya:2026vpo, Batra:2026pef}.}. In particular, Ref.~\cite{Kang:2022psa} considered a radiative neutrino mass model (with extra scalar doublets), where the leptogenesis scale can naturally be reduced to the TeV regime~\cite{Hugle:2018qbw} without relying on resonant enhancement. In contrast, our framework employs non-holomorphic modular symmetry within a minimal type-I seesaw setup, where the resonant enhancement of the CP asymmetry itself lowers the leptogenesis scale to the intermediate regime.

An important question is whether the RHN mass scale can be probed, either directly or indirectly. In the present framework, this becomes possible through the introduction of a complex scalar field, $\Phi$, carrying a non-trivial charge under the discrete $\mathbb{Z}_3$ symmetry. Following the phase transition (PT)\footnote{The formation and evaluation of DW after the PT and its associated GW is also discussed in Refs.~\cite{Wei:2022poh, Borboruah:2022eex, Fornal:2024avx, Roshan:2026xpf}} associated with the scalar sector, the scalar field $\Phi$ acquires a non-zero VEV, spontaneously breaking the $\mathbb{Z}_3$ symmetry and resulting in the formation of a network of domain walls (DWs)~\cite{Saikawa:2017hiv, Roshan:2024qnv}.  
These DWs can remain stable on cosmological timescales and eventually dominate the energy density of the Universe, leading to a cosmological evolution that is incompatible with the cosmic microwave background (CMB) observations reported by the \textit{Planck} Collaboration~\cite{Planck:2018vyg}. Consequently, the DW network must disappear before it comes to dominate the energy budget of the Universe.

This requirement establishes a direct connection with the RHN sector via the explicit $\mathbb{Z}_3$-breaking Yukawa interaction, $y_N \Phi \overline{N_R^c}N_R + \text{h.c.}$, which induces radiative corrections to the scalar potential, lifting the degeneracy among the $\mathbb{Z}_3$-related vacua and generating a finite bias energy. This bias renders the DWs metastable and drives their eventual annihilation. The annihilation process sources a stochastic gravitational wave (GW) background whose amplitude and peak frequency depend on the magnitude of the bias and, consequently, on the RHN mass scale. Therefore, the resulting GW signal provides an indirect probe of the RHN sector.

Furthermore, if the PT is strongly first order, it generates an additional stochastic GW background through bubble nucleation, expansion, and collisions, followed by sound waves and magnetohydrodynamic turbulence in the thermal plasma~\cite{Kamionkowski:1993fg,Ellis:2018mja,Croon:2018erz,Beniwal:2018hyi,Mazumdar:2018dfl,Kobakhidze:2016mch,Kang:2017mkl,Kannike:2019mzk,Chakrabarty:2022yzp,Ellis:2022lft,Choudhury:2026laq,Das:2026zuo,Srivastava:2025oer,Chaudhuri:2022sis,Chaudhuri:2025cjp,Borah:2023zsb,Borah:2024emz,Borah:2024lml,Bhattacharyya:2026esv}. Since the PT occurs at a high energy scale in our framework, the corresponding nucleation temperature is also high, resulting in a GW signal that peaks at relatively high frequencies. In contrast, the subsequent annihilation of the domain walls takes place at a later epoch, producing a GW signal that peaks at much lower frequencies. Consequently, the superposition of these two contributions yields a characteristic double-peaked GW spectrum with two widely separated peaks. This distinctive feature is a unique prediction of the present framework and provides a promising target for future space- and ground-based gravitational wave observatories.

The rest of the paper is organized as follows. In Sec.\,\ref{sec:model}, we present the theoretical framework of the model, outlining the non-holomorphic $A_4$ modular symmetry, the field content, and the lepton mass matrices. In Sec.\,\ref{sec: neu_phneo}, we study the neutrino phenomenology and identify the regions of parameter space consistent with the current neutrino oscillation data. The realization of resonant leptogenesis within the present framework is discussed in Sec.\,\ref{sec: reso_lepto}. In Sec.\,\ref{sec:gw}, we analyze the scalar potential, the associated PT and DW formation, and the resulting stochastic GW signals generated by both the strong first-order PT (FOPT) and DW annihilation. In subsec.\,\ref{sec:DM}, we discuss the possibility of realizing a viable dark matter candidate in the present framework. Finally, we summarize our main results and conclude in Sec.\,\ref{sec:summary}.

\section{Theoretical Framework}
\label{sec:model}
\subsection{Non-Holomorphic Modular Symmetry}
\label{subsec: non-holomorphic}

The modular approach to flavor provides an elegant framework in which the transformation properties of fields determine the observed flavor structure under the modular symmetry. In contrast to conventional flavor models, where the flavor symmetry is spontaneously broken by flavon fields, the Yukawa couplings are promoted to modular functions of the complex modulus $\tau$. In this work, we adopt the non-supersymmetric framework based on polyharmonic Maa\ss\ forms, following the formalism developed in Ref.~\cite{Qu:2024rns}.

The modular group $SL(2,\mathbb{Z})$ acts on the modulus $\tau$ through the fractional linear transformation
\begin{equation}
\tau \rightarrow \gamma\tau=
\frac{a\tau+b}{c\tau+d},
\qquad
\gamma=
\begin{pmatrix}
a & b\\
c & d
\end{pmatrix}
\in SL(2,\mathbb{Z}),
\end{equation}
where $a$, $b$, $c$, and $d$ are integers satisfying $ad-bc=1$. Restricting the modular group to level $N$ gives rise to the finite modular group $\Gamma_N$ ($N=2,3,4,\ldots$), which has been extensively employed in flavor model building
\cite{Kobayashi:2018wkl,Okada:2019xqk,Mishra:2020gxg,Kang:2026osw,Meloni:2023aru,Marciano:2024nwm,Belfkir:2024wdn,Nomura:2023usj,RickyDevi:2024ijc,Gogoi:2023jzl,Pathak:2024sei,Nomura:2024vus,Kashav:2021zir,Kashav:2022kpk,Kobayashi:2019gtp,Nomura:2023kwz,Kim:2023jto,Devi:2023vpe,Behera:2024ark,Behera:2025tpj,Dasgupta:2021ggp,CentellesChulia:2023osj,deMedeirosVarzielas:2023crv,Ding:2021zbg,King:2019vhv,deMedeirosVarzielas:2022ihu,Yao:2020zml}.

The matter fields $\psi$, $\psi^c$, the Higgs field $H$, and the polyharmonic Maa\ss\ forms $Y_r^{(k_Y)}(\tau)$ carry modular weights $k_\psi$, $k_{\psi^c}$, $k_H$, and $k_Y$, respectively, and transform in the irreducible representations $\rho_\psi$, $\rho_{\psi^c}$, $\rho_H$, and $\rho_Y$ of the finite modular group. Modular invariance of the Yukawa interactions requires that the total modular weight vanish and that the tensor product of the corresponding representations contains the trivial singlet \cite{Qu:2024rns},
\begin{equation}
k_Y = k_\psi + k_{\psi^c} + k_H,
\qquad
\rho_Y \otimes \rho_\psi \otimes \rho_{\psi^c} \otimes \rho_H
\supset \mathbf{1}.
\end{equation}
Once the modular weights and representation assignments are specified, these conditions uniquely determine the allowed Yukawa structures.

\subsection{Model Framework}

\begin{table}[h]
\label{mod_table}
\centering
\begin{tabular}{||c||c|c|c||c|c|c|c||c||c|}
\hline
Fields & $e_R$ & $\mu_R$ & $\tau_R$ & $L_1$ & $L_2$ & $L_3$ & $N_R$ & $H$  \\
\hline \hline
$SU(2)_L$ & $1$ & $1$ & $1$ & $2$ & $2$ & $2$& $1$ & $2$\\
\hline
$U(1)_Y$ & $-1$ & $-1$ & $-1$ & $-1/2$ & $-1/2$ & $-1/2$& $0$ & $1/2$\\
\hline
$A_4$ & $\mathbf{1}$ & $\mathbf{1'}$ & $\mathbf{1''}$ & $\mathbf{1}$ & $\mathbf{1'}$ & $\mathbf{1''}$ & $\mathbf{3}$ & $\mathbf{1}$  \\
\hline
$k_I$ & $1$ & $1$ & $1$ & $-1$ & $-1$ & $-1$ & $-1$ & $0$ \\
\hline \hline
\end{tabular}
\caption{Field content of the model and their charges under $SU(2)_L \times U(1)_Y\times A_4$ and their respective modular weights $(k_I)$. }\label{tab:modulus}
\end{table}
This framework corresponds to a minimal realization of the type-I seesaw mechanism embedded in a $A_4$ modular symmetric setup, with the particle content and charge assignments summarized in Tab.\,\ref{tab:modulus}. To investigate neutrino phenomenology in a predictive manner, the model is extended by incorporating modular symmetry, under which the relevant beyond-Standard-Model fields transform non-trivially. The flavor structure is controlled by modular forms that depend on the complex modulus $\tau$. Once $\tau$ acquires a vacuum expectation value in the fundamental domain, the modular symmetry is effectively broken, fixing the numerical values of the modular forms $Y^{(k)}_r(\langle\tau\rangle)$ and thereby determining the structure of the Yukawa couplings. All fields are assigned specific modular weights $k_I$, chosen appropriately to forbid unwanted operators and ensure modular invariance of the Lagrangian. A key advantage of the modular $A_4$ symmetry, compared to conventional discrete flavor models, is that Yukawa couplings themselves transform non-trivially under the flavor group, significantly reducing the need for multiple flavon fields. As a result, the model remains economical while yielding highly constrained and predictive neutrino mass matrices once the modulus $\tau$ is fixed.

The most general renormalizable Lagrangian which is invariant under the SM gauge group and $A_4$ symmetry\footnote{For a brief discussion of the kinetic terms, we refer to Ref.\cite{Zhang:2025dsa}}:
\begin{eqnarray}
    \mathcal{L}\,\supset\,  \mathcal{L}_\ell + \mathcal{L}_D + 
    \mathcal{L}_M 
\end{eqnarray} 
The modular-invariant Lagrangian for the charged lepton and neutrino sectors is expressed as
\begin{eqnarray}\label{eq:LagD} 
&&\mathcal{L}_{\ell} = - \left(\alpha_\ell (\overline{L}_1 H Y_1^{(0)}) e_R 
+ \beta_\ell (\overline{L}_2 H Y_1^{(0)}) \mu_R 
+ \gamma_\ell (\overline{L}_3 H Y_1^{(0)}) \tau_R + \text{h.c.}\right), \nonumber\\[2pt]
&&\mathcal{L}_D= -\left(\alpha_D \overline{L}_1 (Y_3^{(-2)} N_R)_1 \tilde{H} 
+ \beta_D \overline{L}_2 (Y_3^{(-2)} N_R)_{1'} \tilde{H} 
+ \gamma_D \overline{L}_3 (Y_3^{(-2)} N_R)_{1''} \tilde{H} + \text{h.c.}\right),\nonumber \\[2pt]
\tiny &&\mathcal{L}_{M}= -\frac{1}{2}\left(  M_0\beta_R \overline{N_R^c} N_R Y_3^{(-2)}+ M_0\gamma_R \overline{N_R^c} N_R Y_1^{(-2)}+ \text{h.c.}\right).
\end{eqnarray}
The corresponding mass matrix for charged leptons is diagonal 
\begin{equation}
  M_\ell = \frac{v_h}{\sqrt2} 
  \begin{pmatrix}
        \alpha_\ell & 0 & 0\\
        0 & \beta_\ell & 0\\
        0 & 0 & \gamma_\ell
  \end{pmatrix}
  \begin{pmatrix}
        Y_1^{(0)} & 0 & 0 \\
        0 & Y_1^{(0)} & 0 \\
        0 & 0 & Y_1^{(0)} \\
  \end{pmatrix} 
  =\begin{pmatrix}
      m_e & 0 & 0 \\
      0 & m_\mu & 0 \\
      0 & 0 & m_\tau
  \end{pmatrix}, 
\end{equation}
with $m_e, m_\mu, m_\tau$ the masses of observed charged leptons,  $Y_1^{(0)}=1$ \cite{Qu:2024rns} and $v_h(\approx 246 \text{GeV})$ being the Higgs VEV. In the neutral lepton sector, the induced mass matrix term in the basis of $\mathcal{N}= (\nu_L^c, N_R)^T$ can be written as
 \begin{eqnarray}
     \mathcal{L}_{\mathcal{M}} \supset -\frac{1}{2} (\overline{\nu_L}\quad \overline{N_R^c}) \begin{pmatrix}
        0 &M_D\\ M_D^T &M_R
    \end{pmatrix}\begin{pmatrix}
        \nu_L^c\\N_R
    \end{pmatrix} + \text{h.c.}.
 \end{eqnarray}
Here, the Dirac mass matrix is given as,
\begin{equation}
    M_D = \frac{v_h}{\sqrt2}
    \begin{pmatrix}
        \alpha_D & 0 & 0\\
        0 & {\beta}_D & 0\\
        0 & 0 & {\gamma}_D
    \end{pmatrix} 
    \begin{pmatrix}
        Y_{3,1}^{(-2)} & Y_{3,3}^{(-2)} & Y_{3,2}^{(-2)}  \\
        Y_{3,2}^{(-2)} & Y_{3,1}^{(-2)} & Y_{3,3}^{(-2)} \\
        Y_{3,3}^{(-2)} & Y_{3,2}^{(-2)} & Y_{3,1}^{(-2)}
    \end{pmatrix}=\frac{v_h}{\sqrt2}\,y,
\end{equation}
where $y$ is a complex $3\times3$ matrix  and the mass matrix of the Majorana neutrinos is represented as

\begin{equation}\label{eq:RHNmass}
   M_{R} = M_0 \left[ 
   \frac{\beta_R}{3}\begin{pmatrix}
       2 Y_{3,1}^{(-2)} & -Y_{3,3}^{(-2)} & -Y_{3,2}^{(-2)}\\
       -Y_{3,3}^{(-2)} & 2 Y_{3,2}^{(-2)} & -Y_{3,1}^{(-2)}\\
       -Y_{3,2}^{(-2)} & -Y_{3,1}^{(-2)} & 2 Y_{3,3}^{(-2)}
   \end{pmatrix}  +
       {\gamma}_R \begin{pmatrix}
           Y_1^{(-2)} & 0 & 0\\
           0 & 0 & Y_1^{(-2)}\\
           0 & Y_1^{(-2)} & 0
       \end{pmatrix} \right]\,.
   \end{equation}
In the limit where $|M_R|\gg |M_D|$, the effective light-neutrino mass matrix is generated through the type-I seesaw mechanism and is given by
\begin{equation}
m_\nu \simeq -M_D M_R^{-1} M_D^T
=
\left(\frac{v_h^2}{2M_0}\right)\tilde m_\nu
\equiv
\kappa\,\tilde m_\nu,
\qquad
\kappa=\frac{v_h^2}{2M_0},
\end{equation}
where $\tilde m_\nu$ is a dimensionless complex symmetric $3\times3$ matrix. It is diagonalized by the Takagi factorization \cite{Hahn:2006hr},
\begin{equation}
U_\nu^{T}\tilde m_\nu U_\nu
=
\mathrm{diag}
\left(
\tilde D_{\nu_1},
\tilde D_{\nu_2},
\tilde D_{\nu_3}
\right),
\qquad
U_\nu^\dagger U_\nu=\mathbf{1},
\end{equation}
with $\tilde D_{\nu_i}\ge 0$ $(i=1,2,3)$ denoting the singular values of $\tilde m_\nu$.
The physical light-neutrino masses are therefore given by
\begin{equation}
m_i=\kappa\,\tilde D_{\nu_i}.
\end{equation}
The overall scale $\kappa$ is fixed by the atmospheric mass-squared difference:
\begin{equation}
(\mathrm{NO}) :
\kappa^2=
\frac{|\Delta m^2_{\rm atm}|}
{\tilde D_{\nu_3}^{\,2}-\tilde D_{\nu_1}^{\,2}},
\qquad
(\mathrm{IO}) :
\kappa^2=
\frac{|\Delta m^2_{\rm atm}|}
{\tilde D_{\nu_2}^{\,2}-\tilde D_{\nu_3}^{\,2}},
\label{eq:kappa}
\end{equation}
where $\Delta m^2_{\rm atm}$ denotes the atmospheric neutrino mass-squared difference, and NO (IO) corresponds to the normal (inverted) mass ordering.

The solar mass-squared difference is subsequently obtained as
\begin{equation}
\Delta m^2_{\rm sol}
=
\kappa^2
\left(
\tilde D_{\nu_2}^{\,2}
-
\tilde D_{\nu_1}^{\,2}
\right),
\label{eq:solar}
\end{equation}
which can be directly compared with experimental observations.
In this framework, the overall mass scales of the charged leptons and neutrinos are controlled by $v_h/\sqrt{2}$ and $M_0$, respectively, while the flavor structure is dictated by the polyharmonic Maa\ss\ forms $Y_r^{(k)}$, ensuring modular $A_4$ invariance of the Lagrangian.

\section{Neutrino Phenomenology}
\label{sec: neu_phneo}
Neutrino oscillation experiments measure six independent parameters: the solar and atmospheric mass-squared differences, $\Delta m^2_{\rm sol}$ and $\Delta m^2_{\rm atm}$, the three leptonic mixing angles $\theta_{12}$, $\theta_{23}$ and $\theta_{13}$, and the Dirac CP-violating phase $\delta$. After diagonalizing the charged-lepton and light-neutrino mass matrices, these observables can be extracted from the Pontecorvo--Maki--Nakagawa--Sakata (PMNS) matrix \cite{Hochmuth:2007wq}.

The solar mass-squared splitting is given by
\begin{equation}
\Delta m^2_{\rm sol}=m_2^2-m_1^2,
\end{equation}
both normal ordering (NO) and inverted ordering (IO), while the atmospheric mass-squared splitting is defined as
\begin{equation}
\Delta m^2_{\rm atm}=
\begin{cases}
m_3^2-m_1^2, & \text{NO},\\
m_2^2-m_3^2, & \text{IO}.
\end{cases}
\end{equation}
The leptonic mixing angles are determined from the PMNS matrix elements according to,
\begin{equation}
\sin^2\theta_{12} = \frac{|U_{e2}|^2}{1-|U_{e3}|^2}, \qquad
\sin^2\theta_{23} = \frac{|U_{\mu3}|^2}{1-|U_{e3}|^2}, \qquad
\sin^2\theta_{13} = |U_{e3}|^2
\end{equation}
while the Dirac CP phase is extracted from the rephasing-invariant combination $U_{e2}U_{\mu3}U_{e3}^{\ast}U_{\mu2}^{\ast}$.

In the present framework, the light-neutrino mass matrix depends on a few independent parameters like the modulus $\tau$, the dimensionless free parameters $\alpha_D,~\beta_D,~\gamma_D,~\beta_R,~\gamma_R$, and the mass parameter $M_0$. The viable parameter space is determined by fitting these parameters to the first five measured neutrino oscillation observables. Throughout this work, we employ the latest global-fit results for neutrino oscillation parameters~\cite{deSalas:2020pgw,Capozzi:2021fjo,Esteban:2024eli}. The corresponding best-fit values and $3\sigma$ allowed ranges are summarized in Tab.~\ref{table:data_nufit_NO}.

\begin{table}[ht]
    \centering
    \begin{tabular}{||l||c||c||}
    \hline\hline
    Parameters  & Best-fit value & $3\sigma$ range \\
    \hline \hline 
$\sin^2\theta_{12}$ & $0.3088^{+0.0067}_{-0.0066}$ & $0.2893 - 0.3295$ \\
$\sin^2\theta_{23}$ & $0.470^{+0.017}_{-0.014}$ & $0.435 - 0.584$ \\
$\sin^2\theta_{13}$ & $0.02248^{+0.00055}_{-0.00059}$ & $0.02064 - 0.02418$ \\
\hline
$\Delta m^2_{21}/10^{-5}\,\mathrm{eV}^2$ & $7.537^{+0.094}_{-0.10}$ & $7.236 - 7.823$ \\
$\Delta m^2_{31}/10^{-3}\,\mathrm{eV}^2$ & $+2.511^{+0.021}_{-0.020}$ & $+2.450 - +2.576$ \\
\hline
$\delta_{\rm CP}/^\circ$ & $212^{+26}_{-36}$ & $125 - 365$ \\
    \hline\hline
    \end{tabular}
    \caption{Best-fit values and their $3\sigma$ ranges for neutrino oscillation parameters in normal ordering obtained from NuFIT 6.1~\cite{Esteban:2024eli}.}
    \label{table:data_nufit_NO}
\end{table}

\subsection{Numerical analysis and model predictions}

In this section, we quantitatively evaluate the phenomenological viability of the proposed modular symmetry framework. We identify the parameter space regions consistent with the NuFIT under the normal ordering (NO) at the $3\sigma$ level, referencing the standard global benchmarks summarized in Tab.\,\ref{table:data_nufit_NO}. 

To evaluate the statistical compatibility of our model against empirical data, we perform a numerical scan over the involved parameters of the framework. The test statistic is governed by a rigorous $\chi^2$ minimization program \cite{iminuit} constructed as follows:
\begin{equation}
    \chi^2 = \sum_{i} \left( \frac{\mathcal{O}_i^{\text{th}} - \mathcal{O}_i^{\text{exp}}}{\sigma_i} \right)^2 \,,
\end{equation}
where $\mathcal{O}_i^{\text{th}}$ denotes the model-predicted value for a given neutrino observable, $\mathcal{O}_i^{\text{exp}}$ represents the central experimental value from global fits, and $\sigma_i$ is the associated $1\sigma$ experimental uncertainty.

We identify the region of our model parameter space consistent with neutrino oscillation data at the 3$\sigma$ level; see Tab.~\ref{table:data_nufit_NO} and facilitate resonant leptogenesis by creating a small mass splitting between the three mass eigenstates of the right-handed neutrinos. Recall that in our setup, we have parameters: the modulus $\tau$ and the free parameters, as mentioned above. In our scan, the real and imaginary part of $\tau$ is varied within as defined below,
\begin{equation}\label{eq:RItau}
    \Big|\text{Re}(\tau)\Big|\in [0,0.5],\quad \text{Im}(\tau) \in [1, 3.5].
\end{equation} 
while the dimensionless free parameters and the mass parameter are taken to be in the range, respectively, as shown below:
\begin{eqnarray}\label{eq:mpv}
    \mathscr{x} \alpha_D, \beta_D, \gamma_D = [10^{-8}-10^{-3}], &&\quad \beta_R=[10^{-13}-10^{-9}], \quad \gamma_R = [10^{-10}-10^{-2}] \nonumber \\
    && M_0 = [10^{10}-10^{13}]~ \rm GeV.
\end{eqnarray} 

\begin{figure}[htbp]
    \centering
    \begin{subfigure}{0.48\linewidth}
        \centering
        \includegraphics[width=.97\linewidth]{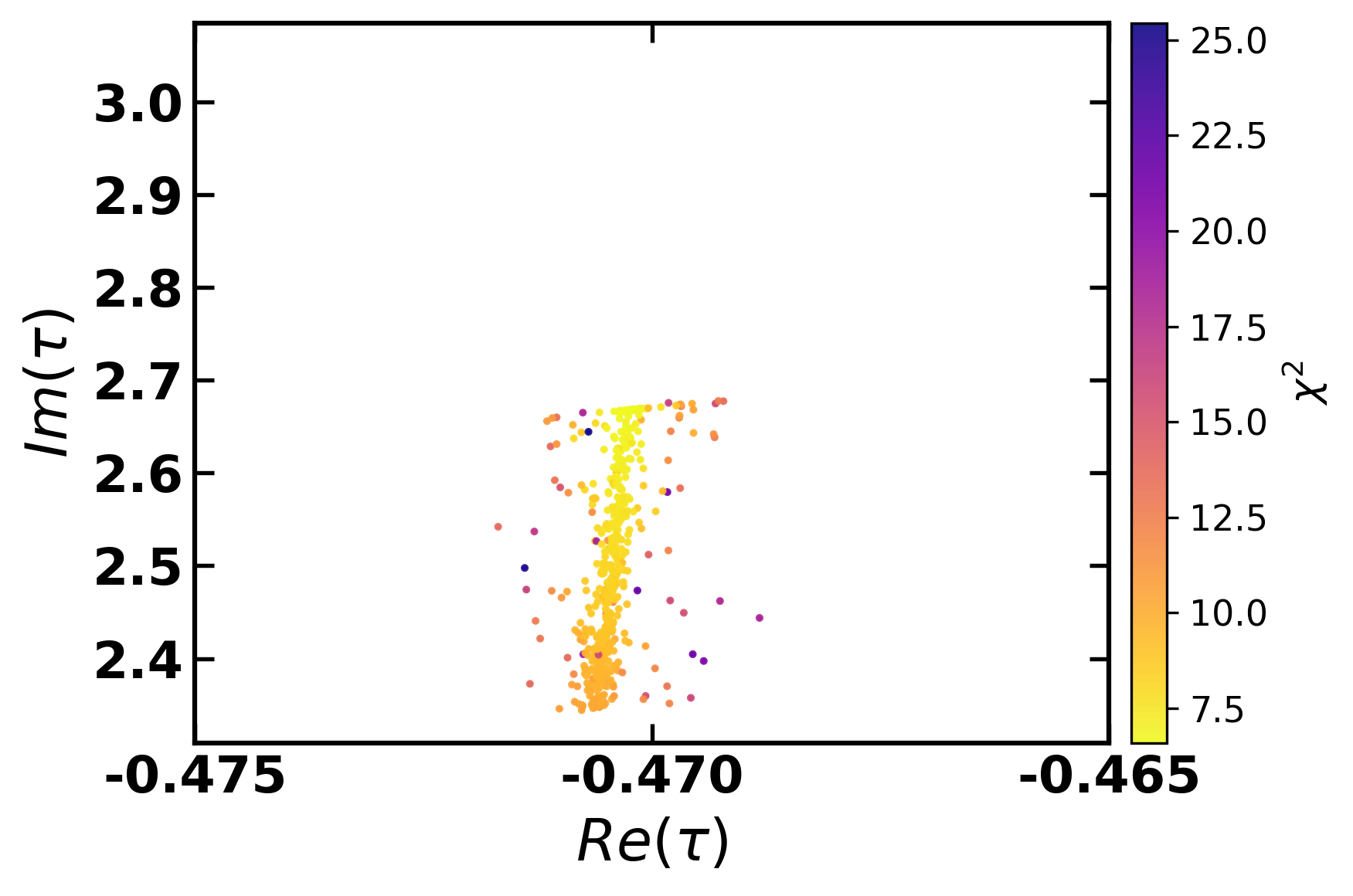}
        \caption{}
    \end{subfigure}
    \hfill
    \begin{subfigure}{0.48\linewidth}
        \centering
        \includegraphics[width=\linewidth]{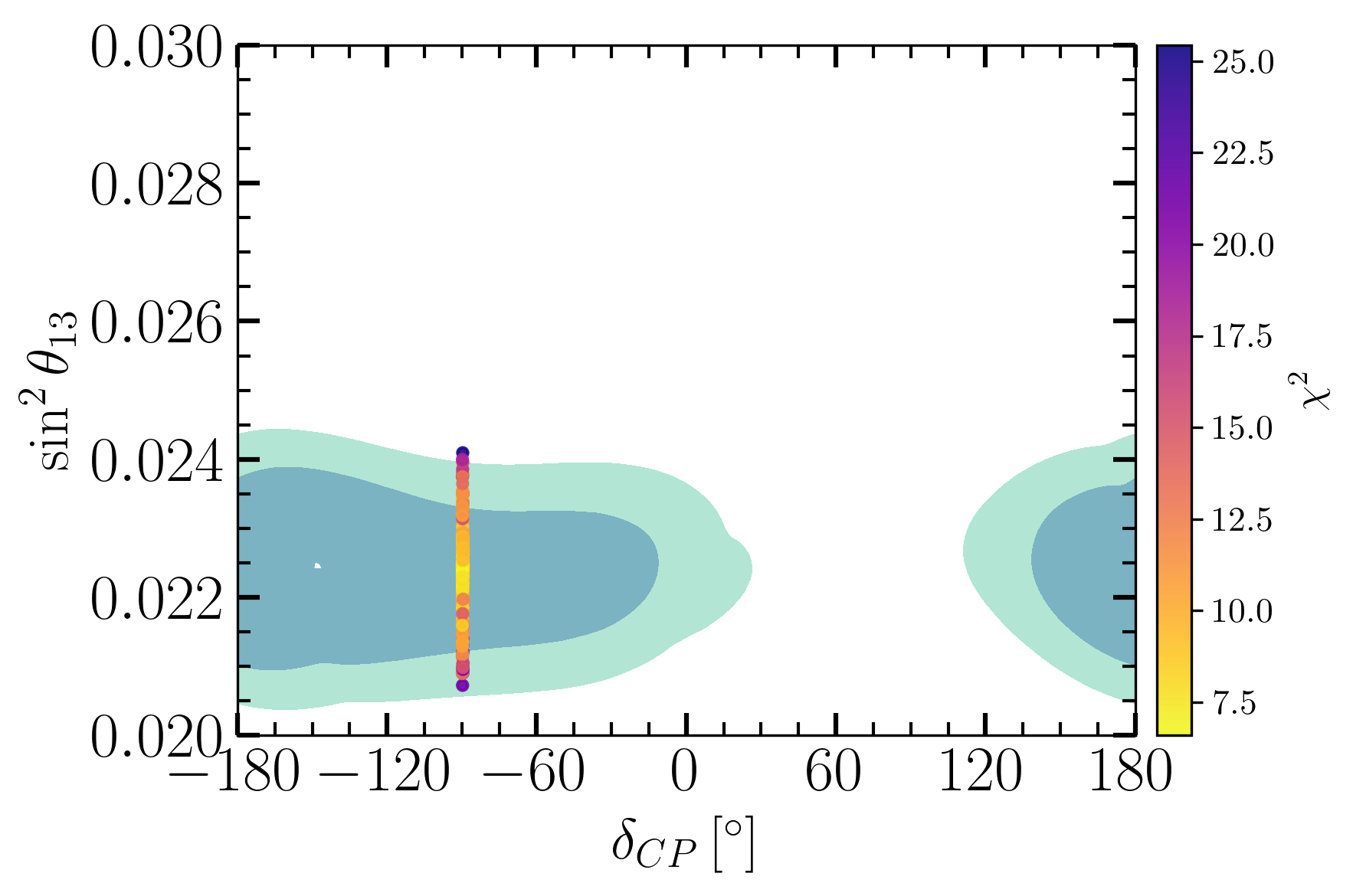}
        \caption{}
    \end{subfigure}
    \caption{Panel (a) illustrates the allowed region in the $\text{Re}(\tau)$--$\text{Im}(\tau)$ plane. Panel (b) shows the model predictions of $\delta_{cp}$ against $\sin^2 \theta_{13}$ overlaid with the $2\sigma$ (dark blue) and $3\sigma$ (light green) confidence level (C.L.) contours from the NuFIT.}
    \label{fig:re_im_ss13_dcp}
\end{figure}

Fig.\,\ref{fig:re_im_ss13_dcp} summarizes the regions of parameter space that successfully reproduce the observed
neutrino oscillation data. Panel~(a) displays the allowed values of the complex modulus $\tau$ in the $\mathrm{Re}(\tau)$--$\mathrm{Im}(\tau)$ plane. The viable solutions are confined to a narrow region around $\mathrm{Re}(\tau)\simeq -0.470$, while the imaginary part is restricted to $2.3\lesssim\mathrm{Im}(\tau)\lesssim2.7$. The strong localization of the modulus reflects the highly constrained nature of the modular flavor structure, with the preferred region lying close to the boundary of the fundamental domain, where realistic fermion mass textures
naturally emerge. Panel~(b) shows the correlation between the Dirac CP-violating phase $\delta_{\rm CP}$ and the reactor mixing angle $\sin^2\theta_{13}$. The model predicts
$\delta_{\rm CP}$ close to $-90^\circ$, while the allowed values of $\sin^2\theta_{13}$ lie in the interval $0.0207\lesssim\sin^2\theta_{13}\lesssim0.0241$, fully consistent with the NuFIT allowed region. Most of the benchmark points populate the $2\sigma$ region, with the remaining points lying inside the corresponding $3\sigma$ contour.

\begin{figure}[htbp]
    \centering
    \begin{subfigure}{0.48\linewidth}
        \centering
        \includegraphics[width=\linewidth]{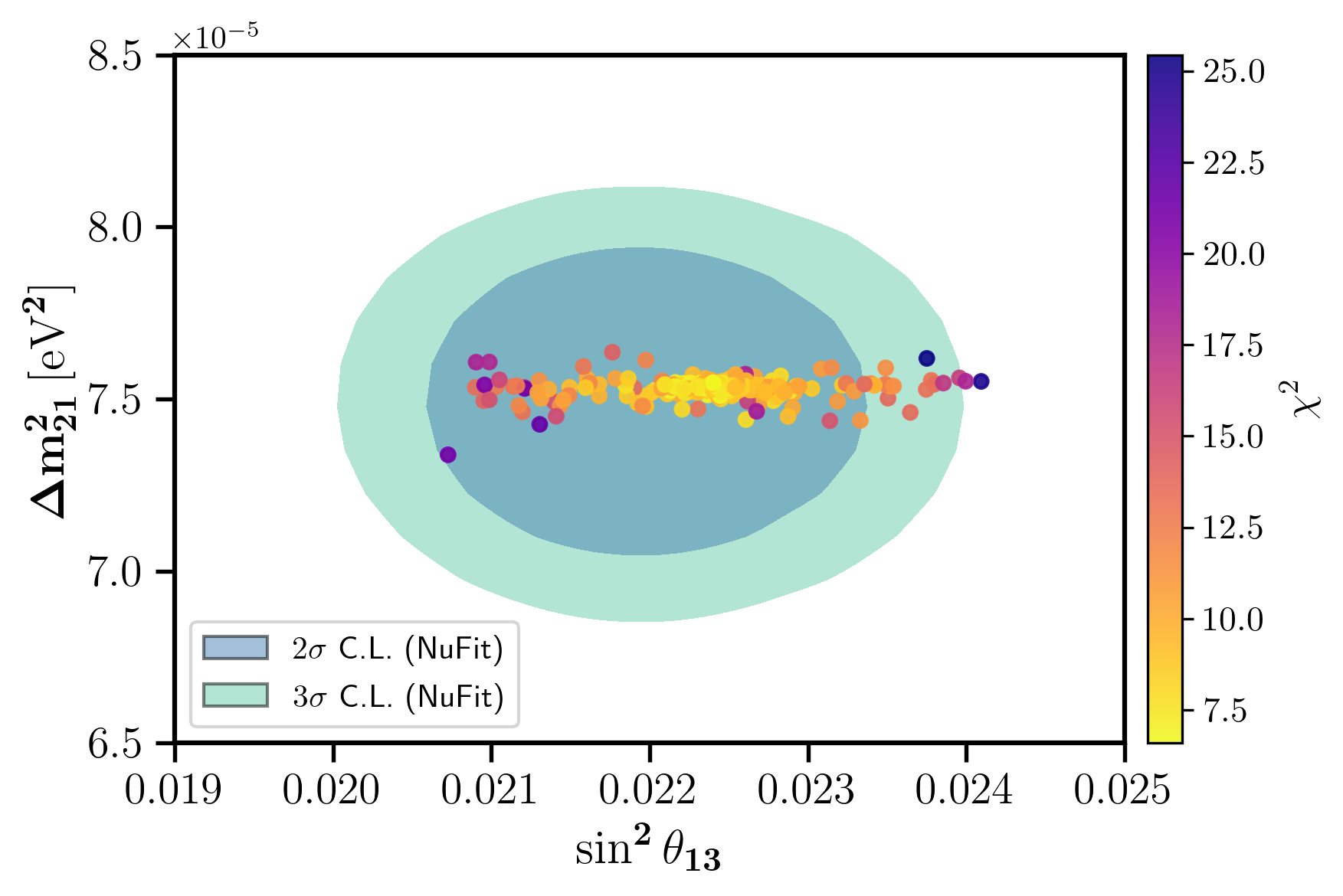}
        \caption{}
    \end{subfigure}
    \hfill
    \begin{subfigure}{0.48\linewidth}
        \centering
        \includegraphics[width=\linewidth]{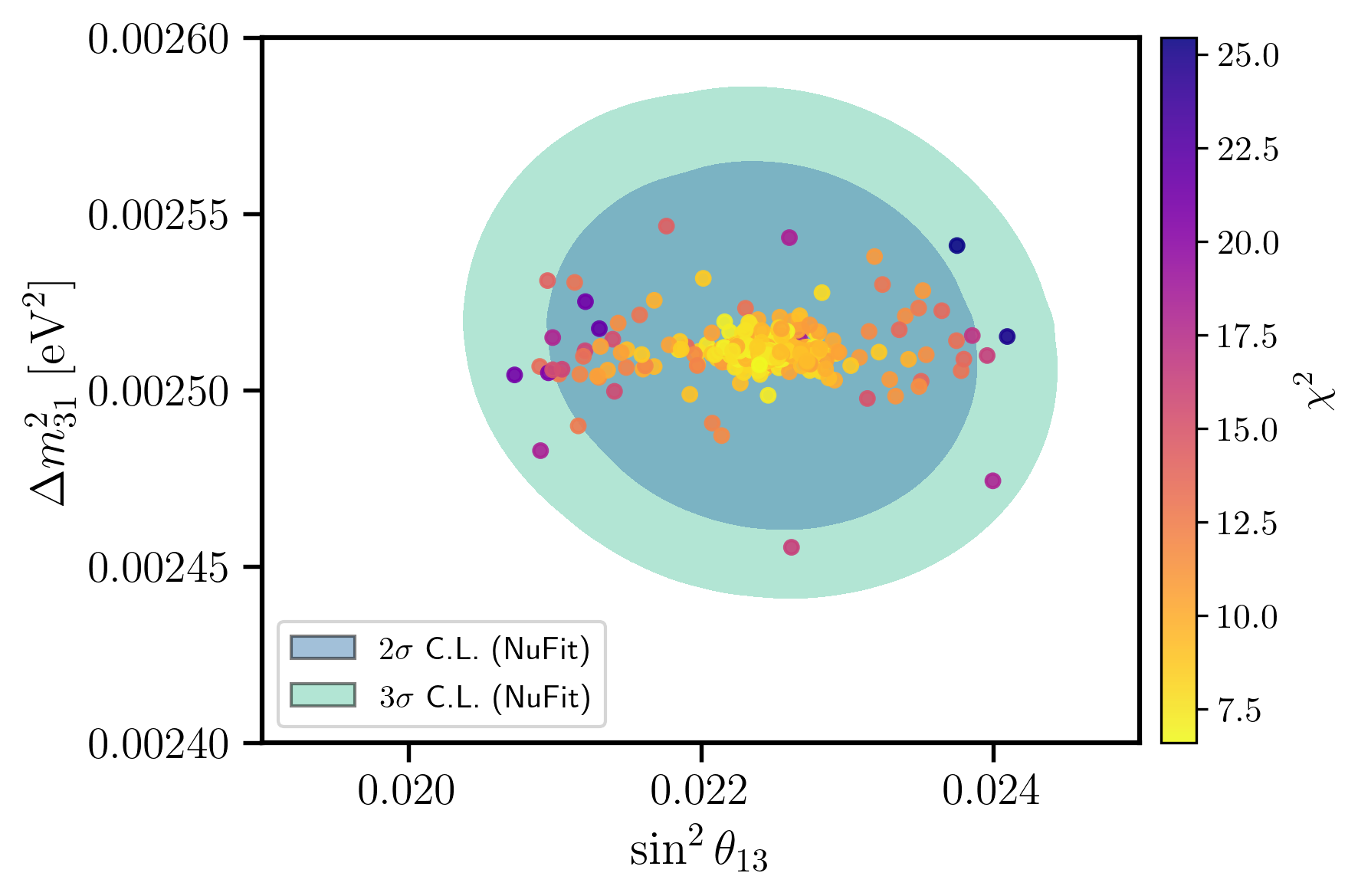}
        \caption{}
    \end{subfigure}
   \caption{Panel (a) displays the solar mass splitting $\Delta m^2_{21}$ ($10^{-5}$~eV$^2$), while panel (b) shows the atmospheric mass splitting $\Delta m^2_{31}$ (eV$^2$) alongside $\sin^2 \theta_{13}$. Scattered points represent the model predictions, shown relative to the $2\sigma$ and $3\sigma$ NuFIT data.}
    \label{fig:ss13_dm12_dm13}
\end{figure}

Fig.\,\ref{fig:ss13_dm12_dm13} demonstrates that the model simultaneously reproduces the observed neutrino
mass-squared splittings. In panel~(a), the predicted values of the solar mass-squared difference occupy the range $\Delta m^2_{21}\simeq(7.34-7.62)\times10^{-5}\,\mathrm{eV}^2$,
remaining entirely within the experimentally allowed region. The narrow spread in $\Delta m^2_{21}$ indicates that the solar sector is strongly constrained by the underlying modular symmetry. Likewise, panel~(b) shows that the atmospheric mass-squared difference falls in the interval $\Delta m^2_{31}\simeq(2.456-2.543)\times10^{-3}\,\mathrm{eV}^2$, again consistent with the NuFIT $2\sigma$ preferred region, with a few points extending into the $3\sigma$ contour. The simultaneous agreement with both mass-squared splittings is a non-trivial consequence of the model, since these observables emerge from the same
set of underlying parameters without requiring independent adjustments. In addition to the oscillation observables, the absolute neutrino mass scale is subject to constraints from cosmological observations. The corresponding upper bound on the sum of the light-neutrino masses depends on the adopted cosmological framework and the datasets included in the analysis. The predictions of the present model are compatible with cosmological constraints obtained in scenarios beyond the minimal $\Lambda$CDM framework~\cite{Planck:2018vyg,DiValentino:2019dzu}.

\begin{figure}[htbp]
    \centering
    \begin{subfigure}{0.5\linewidth}
        \centering
        \includegraphics[width=\linewidth]{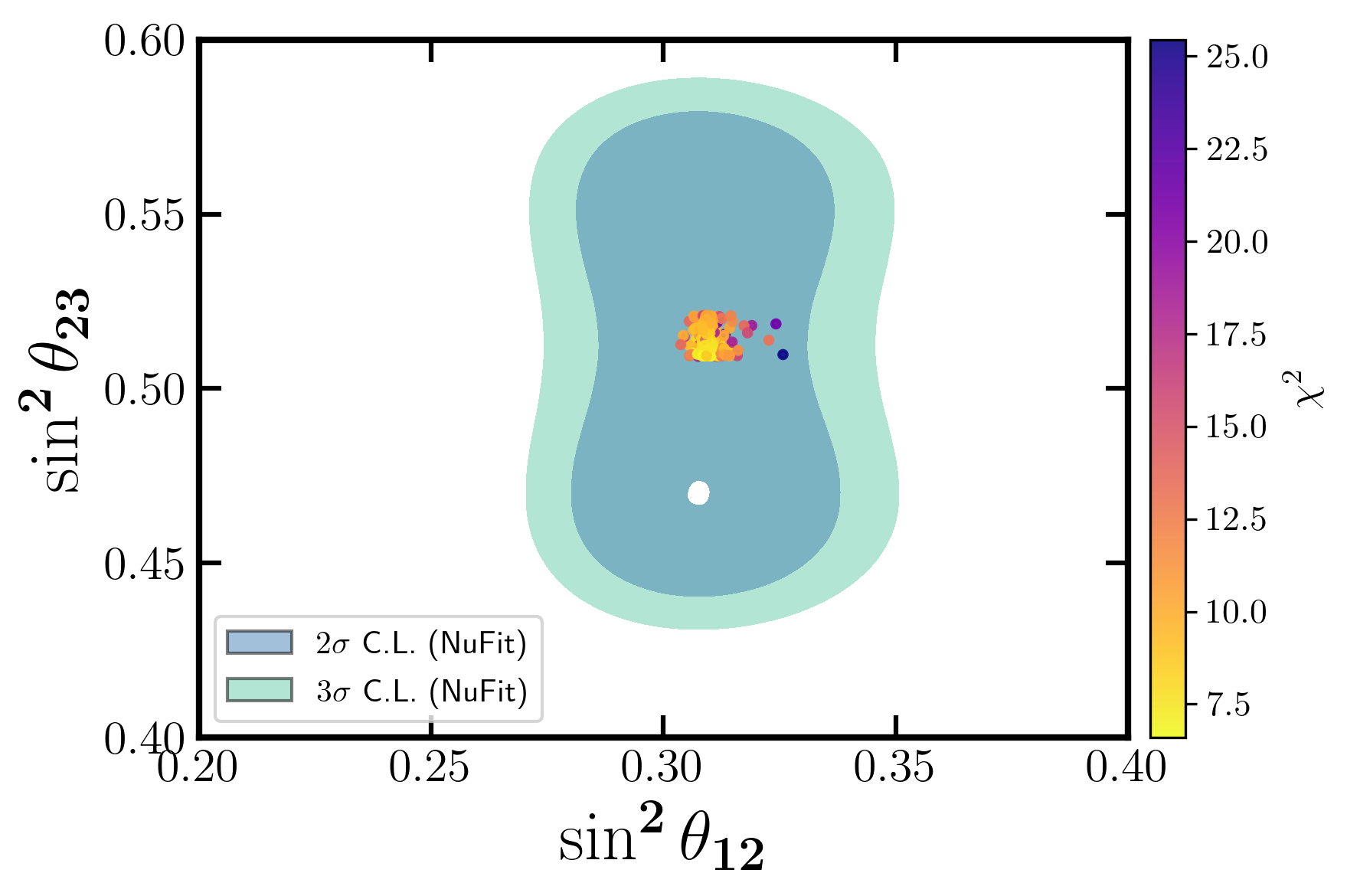}
        \caption{}
    \end{subfigure}
    \hfill
    \begin{subfigure}{0.48\linewidth}
        \centering
        \includegraphics[width=\linewidth]{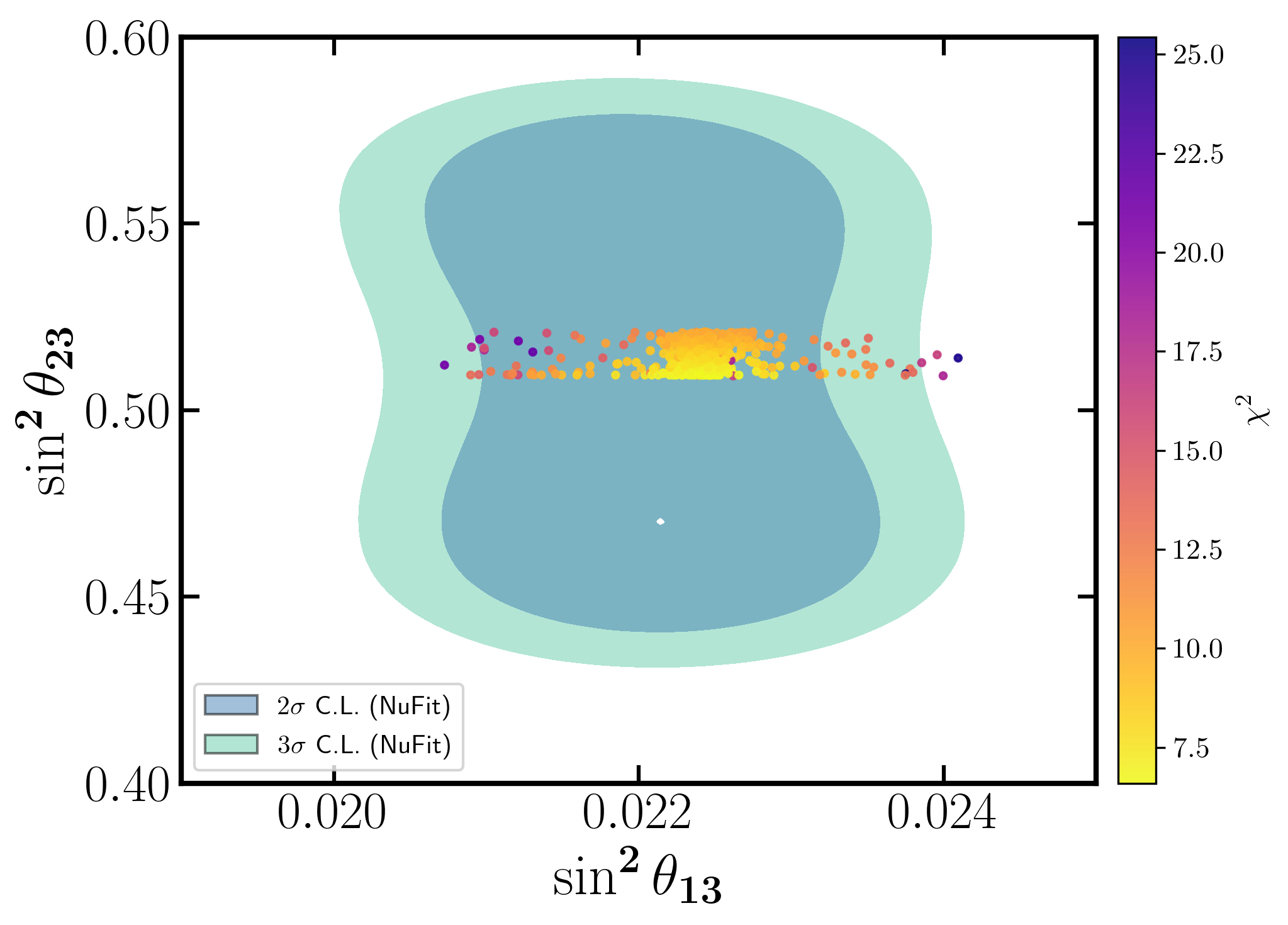}
        \caption{}
    \end{subfigure}
    \caption{Model predictions for the atmospheric neutrino mixing parameter $\sin^2\theta_{23}$ projected against (a) the solar mixing angle $\sin^2\theta_{12}$ and (b) the reactor mixing angle $\sin^2\theta_{13}$. Shaded regions delineate the $2\sigma$ and $3\sigma$ preferred parameter boundaries from NuFIT data.}
    \label{fig:ss12_ss13_ss23}
\end{figure}

The correlations among the leptonic mixing angles are presented in Fig.\,\ref{fig:ss12_ss13_ss23}. Panel~(a)
shows the prediction for $\sin^2\theta_{23}$ as a function of $\sin^2\theta_{12}$, where the
allowed solutions are concentrated within
$0.304\lesssim\sin^2\theta_{12}\lesssim0.326$ and
$0.509\lesssim\sin^2\theta_{23}\lesssim0.521$. Panel~(b) displays the corresponding
correlation between $\sin^2\theta_{23}$ and $\sin^2\theta_{13}$. Since $\theta_{13}$ is
currently the most precisely measured leptonic mixing angle, agreement with its allowed
region provides a stringent test of the model. All viable solutions lie within the
experimentally preferred region, with the majority clustering around the global best-fit
values, illustrating the predictive nature of the proposed framework.

\section{Resonant Leptogenesis }
\label{sec: reso_lepto}
Within the type-I seesaw scenario, the Majorana nature of the RHNs provides the necessary source of lepton-number violation. Their CP-violating out-of-equilibrium decays can generate an overall lepton asymmetry, which is subsequently transformed into a baryon asymmetry via the electroweak sphaleron process. For the study of leptogenesis, it is convenient to express all interactions in the mass basis of the heavy Majorana states. The $3\times 3$ complex symmetric Majorana mass matrix $M_R$ can be diagonalized by a unitary matrix $V$ through takaji autotune diagonalization as
\begin{eqnarray}
    V^T M_R V = D_R \equiv \mathrm{diag}(M_{N_1}, M_{N_2}, M_{N_3}), \qquad V^\dagger V = V V^\dagger = I.
\end{eqnarray}
Accordingly, we define the mass eigenstate ($N_R^\prime$) of the heavy fields as
\begin{eqnarray}
    N_R = V\, N_R', \qquad \text{or equivalently} \qquad N_R' = V^\dagger N_R.
\end{eqnarray}
In the right-handed neutrino mass basis, the Majorana mass term and the Yukawa term of Eq.~\eqref{eq:LagD} can be written as
\begin{eqnarray}
    \mathcal{L}_M +\mathcal{L}_{D}= - \frac{1}{2} \sum_i M_{N_i}\, \overline{N_{R_i}^{\prime c}}\, N_{R_i}' - h_{\alpha k}\,\bar{L}_\alpha\,  N_{R_k}'\, \tilde{H} + \text{h.c.}\,,
\end{eqnarray}
where the Yukawa couplings in the mass basis are defined as
\begin{eqnarray}
    h_{\alpha k} = y_{\alpha j} V_{j k},\quad \alpha =e,\mu,\tau,\quad j,k =1,2,3\,.
\end{eqnarray}
Note that $\alpha$ is the generation index for the lepton doublets, whereas $j,k$ are the generation indices for the heavy RHNs.
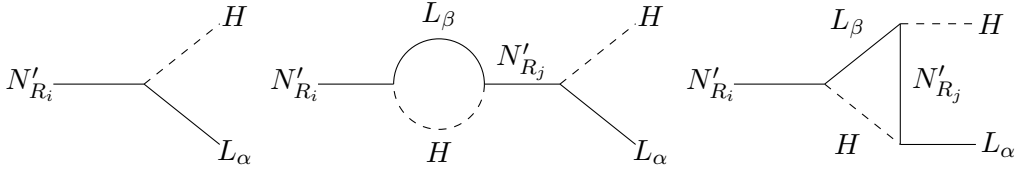
\begin{figure}[!htb]
\centering
\begin{tikzpicture}[scale=1.0]

\begin{scope}
\node at (-0.5,0) {$N_{R_i}^\prime$};
\draw (-0.2,0) -- (1,0);
\draw[dashed] (1,0) -- (2,0.8);
\draw (1,0) -- (2,-0.8);        
\node at (2.2,0.9) {$H$};
\node at (2.2,-0.9) {$L_\alpha$};
\end{scope}
\begin{scope}[xshift=3.5cm]
\node at (-0.5,0) {$N_{R_i}^\prime$};
\draw (-0.2,0) -- (0.8,0);
\draw (0.8,0) arc (180:0:0.6);        
\draw[dashed] (2.0,0) arc (0:-180:0.6); 
\node at (1.4,0.9) {$L_\beta$};
\node at (1.4,-0.9) {$H$};
\node at (2.5,0.3) {$N_{R_j}^\prime$};
\draw (2.0,0) -- (3.0,0);
\draw[dashed] (3.0,0) -- (4.0,0.8); 
\draw (3.0,0) -- (4.0,-0.8);       
\node at (4.2,0.9) {$H$};
\node at (4.2,-0.9) {$L_\alpha$};
\end{scope}

\begin{scope}[xshift=9cm]
\node at (-0.5,0) {$N_{R_i}^\prime$};
\draw (-0.2,0) -- (1,0);
\draw (1,0) -- (2,0.8);          
\draw[dashed] (1,0) -- (2,-0.8); 
\draw (2,0.8) -- (2,-0.8);       
\node at (1.3,0.8) {$L_\beta$};
\node at (1.3,-0.8) {$H$};
\node at (2.5,0) {$N_{R_j}^\prime$};
\draw[dashed] (2,0.8) -- (3,0.8);
\draw (2,-0.8) -- (3,-0.8);       
\node at (3.2,0.8) {$H$};
\node at (3.3,-0.8) {$L_\alpha$};
\end{scope}
\end{tikzpicture}
\caption{Diagrams contributing to asymmetric decay of $N_{R_i}^\prime$.\label{fig:diags}}
\end{figure}
The CP asymmetry generated in the decay of the $i$-th RHN is defined as
\begin{equation}
\varepsilon_i = \sum_{\alpha =e, \mu, \tau}  \frac{\left[{\Gamma(N_i \rightarrow L_\alpha H) - \Gamma(N_i \rightarrow \bar{L}_\alpha H^\dagger)}\right]}{\left[ \Gamma(N_i \rightarrow L_\alpha H) + \Gamma(N_i \rightarrow \bar{L}_\alpha H^\dagger) \right]}
\end{equation}
In general, the CP asymmetry generated in RHN decays receives contributions from both the vertex and self-energy corrections~\cite {Flanz:1996fb, Pilaftsis:1997jf, Pilaftsis:2003gt, Iso:2010mv, Qi:2022fzs, Das:2024gua, King:2024idj}. The relative importance of these two contributions, however, depends on the mass spectrum of the RHNs.

As mentioned in the previous section, we vary the model parameters according to Eq.~\eqref {eq:mpv}, together with the real and imaginary components of the modulus $\tau$ (as Eq.~\eqref{eq:RItau}). We identify viable parameter points, characterized by $\mathrm{Re}(\tau)\approx -0.470$ and $\mathrm{Im}(\tau)\in[2.3,2.7]$, that are consistent with current neutrino oscillation data. 
A notable feature of these parameter points is the emergence of a naturally small mass splitting between the second and third RHN mass eigenstates after diagonalizing the mass matrix in Eq.~\ref{eq:RHNmass}. The resulting RHN mass spectrum satisfies $M_{N_1}>M_{N_2}\approx M_{N_3}$, placing $N_2$ and $N_3$ naturally in the resonant leptogenesis regime, where $\Delta M_{23}\equiv |M_{N_2}-M_{N_3}|\sim \Gamma_{N_3}/2$. In this quasi-degenerate limit, the self-energy contribution to the CP asymmetry is resonantly enhanced and dominates over the vertex contribution, which can therefore be safely neglected in the present analysis. The CP asymmetry is then given by
\begin{equation}
\varepsilon_i = \frac{1}{8\pi} \sum_{j \neq i} \frac{\text{Im} \left[ ( h^\dagger h)_{ij} \right]^2 f_{ij}}{(h^\dagger h)_{ii}}, \quad i,j=2,3,
\label{eq:cpasym}
\end{equation}
where the self-energy loop function $f_{ij}$ is given by:
\begin{equation}
f_{ij} = \frac{(M_{N_i}^2 - M_{N_j}^2) M_{N_i} M_{N_j}}{(M_{N_i}^2 - M_{N_j}^2)^2 + (M_{N_i}\Gamma_{N_j})^2}\,,\quad \text{with} \quad \Gamma_{N_i} = \frac{( h^\dagger h)_{ii} M_{N_i}}{8\pi}
\label{eq:fij}
\end{equation}
are the decay widths of the $N^\prime_i$'s. 
The naturally generated near-degeneracy between the $M_{N_2}$ and $M_{N_3}$ substantially enhances the CP asymmetry through the self-energy loop function in Eq.~\eqref{eq:fij}. As a result, successful leptogenesis can be achieved at an intermediate RHN mass scale of $M_{N_i}\sim10^6,\mathrm{GeV}$, significantly below the conventional Davidson-Ibarra bound~\cite{Davidson:2002qv}, $M_{N_i}\gtrsim10^9,\mathrm{GeV}$, associated with hierarchical thermal leptogenesis.

The dynamical evolution of the lepton asymmetry of the Universe is governed by the following sets of Boltzmann equations (BEQs)
\begin{align}
     \frac{dY_{N_{i}}}{dz}&=-\frac{1}{sHz}\left[\gamma_{i}^{}\left(\frac{Y_{N_i^{}}}{Y_{N_{i}^{}}^{eq}}-1\right)\right]\,,\nonumber\\
    \frac{dY_{\Delta L}}{dz}&= \frac{1}{sHz} \mathlarger{\sum}_{i=2,3}\gamma_i^{}\left[ \varepsilon_i\left(\frac{Y_{N_i^{}}}{Y_{N_{i}^{}}^{eq}}-1\right)-\frac{Y_{\Delta L}}{2Y_L^{eq}}\right]\,,\label{eq:beql}
\end{align}
where $s=0.44 g_s^\ast T^3$ is the entropy density, while $H=1.66 (g_\rho^\ast)^{1/2} T^2/M_{Pl}$ denotes the Hubble expansion rate. The quantities $g^\ast_{\rho}$ and $g^\ast_{s}$ represent the effective relativistic numbers of degrees of freedom associated with the energy and entropy density, respectively. Furthermore, $z\equiv M_{N_3}/T$, and $Y^{eq}=n^{eq}/s$ denote the equilibrium comoving number density of a particle, where $n^{eq}$ is the equilibrium number density of the corresponding particle. The reaction density of the ith RHN is defined as 
\begin{equation}
\gamma_{i}=n^{eq}_{N_{i}^{}}\,\langle\Gamma_{D_i}\rangle\quad\text{with}\quad\langle\Gamma_{D_i}\rangle\equiv\langle\Gamma_{N_i}\rangle=\Gamma(N_{i}^{}\rightarrow lH)\,{\mathcal{K}_1(z)}/{\mathcal{K}_2(z)}\,,
\end{equation}
where $\mathcal{K}_{j}$ is the modified Bessel function of the second kind of jth order. The departure of RHNs from thermal equilibrium is quantified by the wash-out factor $K$ as 
\begin{equation}
    K_i=\frac{\langle\Gamma_{D_i}\rangle_{z=\infty}}{H(z=1)}=\frac{\langle\Gamma_{ID_i}\rangle_{z=\infty}n^{eq}_{l}}{H(z=1)\,n^{eq}_{N_i}}\quad\text{with}\quad\langle\Gamma_{ID_i}\rangle=\langle\Gamma_{D_i}\rangle\frac{n^{eq}_{N_i}}{n^{eq}_{l}}\,,
\end{equation}
where $\langle\Gamma_{ID}\rangle$ denotes the thermally averaged inverse decay rate. We find that the viable parameter space that simultaneously reproduces the observed neutrino oscillation data and naturally predicts a quasi-degenerate RHN spectrum consistently lies in the strong washout regime. Following Refs.\,\cite{Buchmuller:2003gz, Buchmuller:2002rq, Davidson:2008bu, Marciano:2024nwm, Priya:2025wdm}, we include only the inverse decay ($lH\rightarrow N$) process in the washout term, neglecting the $\Delta L=1$ and off-shell $\Delta L=2$ scattering contributions. To avoid double counting, however, the on-shell contribution from the $\Delta L=2$ process (RHN-mediated $lH\leftrightarrow\bar{l}\bar{H}$ scattering) is subtracted, ensuring the correct evolution equation for $Y_{\Delta L}$. After the lepton asymmetry ($Y_{\Delta L}$) freezes at high temperature ($z\equiv M_{N_3}/T\sim10$), it then partially converts to baryon asymmetry ($Y_{\Delta B}$) via the $B+L$ violating (and $B-L$  conserving) sphaleron processes, and the conversion rate can be written as
\begin{equation}
    Y_{\Delta B}=-\frac{4N_{H}+8N_{f}}{13N_{H}+22N_{f}}\,Y_{\Delta L}=-\frac{28}{79}\,Y_{\Delta L}\,,
\end{equation}
where $N_f=3$ is the number of RHNs and $N_{H}=1$ is the number of Higgs doublet.

A distinctive feature of our model is that the required mass degeneracy is not imposed by hand, as is commonly done in conventional resonant leptogenesis scenarios. Instead, \textit{it emerges dynamically from the non-holomorphic structure of the modular $A_{4}$ symmetry through the modular forms $Y_{1}^{(-2)}$ and $Y_{3,i}^{(-2)}$ $(i=1,2,3)$ for suitable values of the modulus $\tau$.} The small mass splitting is therefore an intrinsic prediction of the model rather than an external assumption, providing a well-motivated realization of resonant leptogenesis \footnote{Beyond leptogenesis, this near-degenerate mass spectrum as a natural consequence of the model may have important phenomenological implications. In particular, it can play a significant role in dark matter co-annihilation dynamics~\cite{Griest:1990kh, Baker:2015qna} and offers an interesting avenue for collider searches involving long-lived particles, thereby considerably extending the phenomenological reach of the model.}. 

\begin{figure}[htb!]
    \centering
    \includegraphics[width=0.475\linewidth]{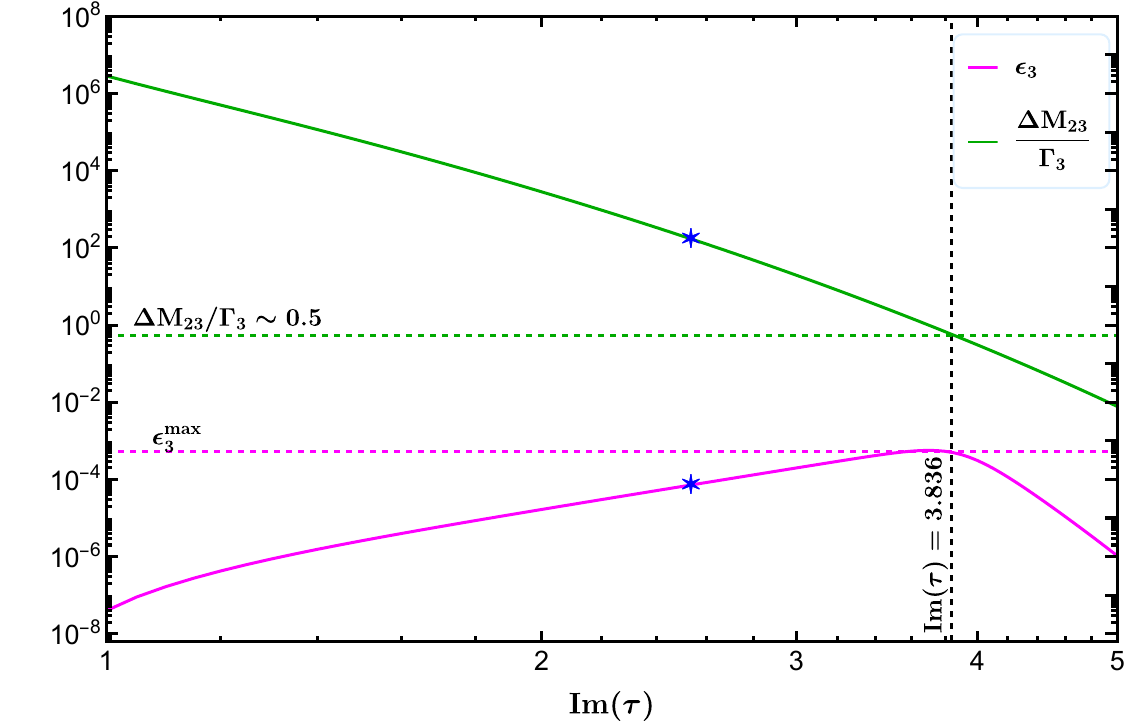}
    \includegraphics[width=0.49\linewidth]{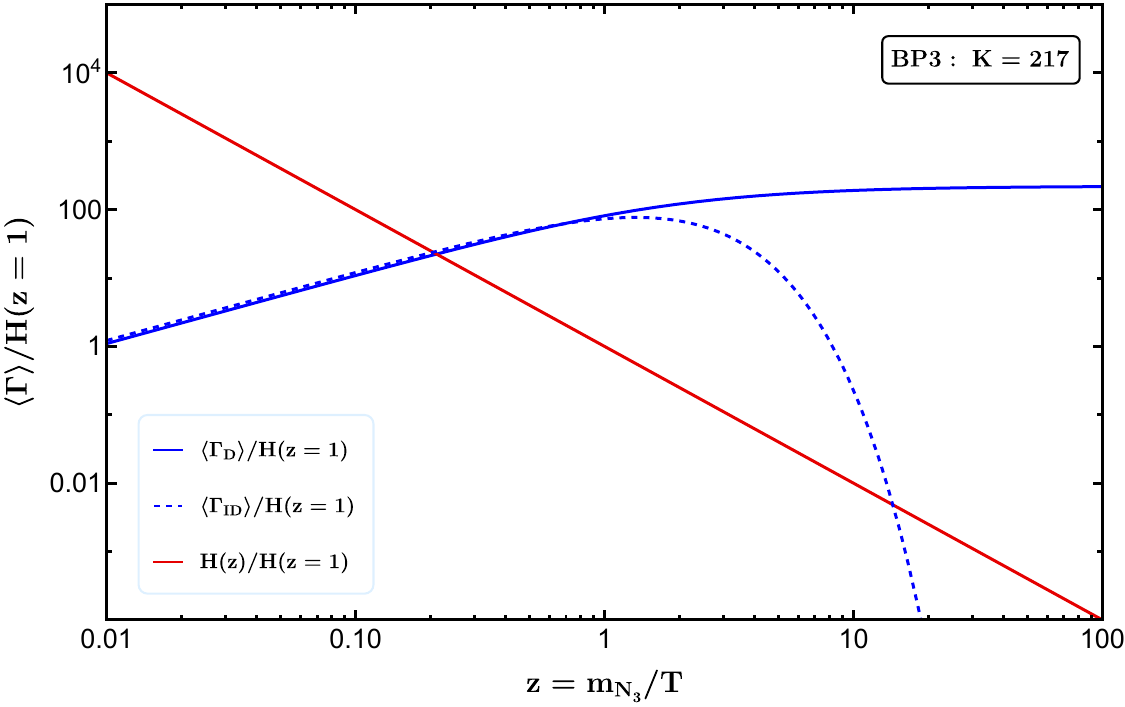}
    \caption{Left panel: Variation of $\Delta M_{23}/\Gamma_3$ (green curve) and $\epsilon_3$ (magenta curve) with $\mathrm{Im}[\tau]$. The star marks the benchmark point BP3, which successfully reproduces the observed BAU. Right panel: Evolution of the $N_{3}$ decay and inverse decay rates as a function of $z\equiv M_{N_3}/T$. In both panels, the model parameters ($\alpha_D,\beta_D,\gamma_D,\alpha_R,\gamma_R,$ and $
    M_0$) are fixed to the BP3 values listed in Tab.\,\ref{tab:cpasy}.}
    \label{fig:imtau}
\end{figure}

It is worth emphasizing that the resonant enhancement of the CP asymmetry is governed not simply by the mass splitting $\Delta M_{23}$, but more precisely by the ratio $\Delta M_{23}/\Gamma_{N_{2,3}}$. The resonance enhancement of CP asymmetry reaches its maximum when $\Delta M_{23}/\Gamma_{N_3}\simeq0.5$, as illustrated in the left panel of Fig.\,\ref{fig:imtau}. For fixed model parameters, increasing $\mathrm{Im}(\tau)$ leads to a progressively finer mass splitting, $\Delta M_{23}$, between the quasi-degenerate RHNs, while the decay width $\Gamma_{N_3}$ increases due to the enhancement of the Yukawa couplings, $h_{ii}$. Consequently, the combined effect of the finer mass splitting and the larger decay width drives the ratio $\Delta M_{23}/\Gamma_{N_3}$ to smaller values as $\mathrm{Im}(\tau)$ increases.

Although the CP asymmetry attains its maximum near $\Delta M_{23}/\Gamma_{N_i}\simeq0.5$, the benchmark points (see Tab.\,\ref{tab:cpasy}) selected in this work do not lie precisely at this resonance. This is because the same parameter region that maximizes the CP asymmetry also leads to significantly larger decay widths ($\Gamma_{N_i}$) and, consequently, much larger inverse-decay and hence stronger washout. Therefore, the observed baryon asymmetry is determined by the interplay between CP-asymmetry generation and washout rate, rather than by maximizing the CP asymmetry alone. Across the viable parameter space consistent with neutrino oscillation data and the naturally generated quasi-degenerate RHN mass spectrum, the washout parameter satisfies $K\gg100$, placing our parameter-space firmly in the strong washout regime. This, in turn, favors successful leptogenesis by reproducing observed BAU only for RHN masses in the intermediate scale $M_{N_3}\simeq10^{6},\mathrm{GeV}$. The strong washout behavior (with $K=217$) for the benchmark point BP3 is illustrated in the right panel of Fig.\,\ref{fig:imtau}.
 
\begin{table}[htb!]
\centering
\small
\begin{tabular}{|c|c|c|c|}
\hline
BPs &$\tau$ &
$(\alpha_D,\beta_D,\gamma_D)$ &
$(\beta_R,\gamma_R)$ \\
\hline
BP1& $-0.470+2.476i$ & $(9.89\times10^{-4},1.67\times10^{-3},6.05\times10^{-4})$ & $(1.29\times10^{-11},6.13\times10^{-3})$ \\
\hline
BP2& $-0.470+2.483i$ & $(1.46\times10^{-4},2.46\times10^{-4},8.95\times10^{-5})$ & $(1.18\times10^{-11},9.70\times10^{-5})$ \\
\hline
BP3& $-0.470+2.538i$ & $(1.38\times10^{-5},2.33\times10^{-5},8.41\times10^{-6})$ & $(1.55\times10^{-12},1.70\times10^{-7})$ \\
\hline
BP4& $-0.470+2.377i$ & $(7.99\times10^{-6},1.35\times10^{-5},4.93\times10^{-6})$ & $(3.91\times10^{-11},1.54\times10^{-6})$ \\
\hline
\end{tabular}
\end{table}
\begin{table}[htb!]
\centering
\begin{tabular}{|c|c|c|c|c|}
\hline
BPs & $M_0$~(\text{GeV}) & $M_{N_3}$~(\text{GeV}) & $\Delta M_{23}/\Gamma_{N_3}$ & $\epsilon_{3}$ \\
\hline
BP1 & $2.02\times10^{11}$ & $5.89\times10^{9}$ & $1.03\times10^{-5}$ & $9.07\times10^{-7}$ \\
\hline
BP2 & $2.76\times10^{11}$ & $1.29\times10^{8}$ & $2.66\times10^{-2}$ & $1.52\times10^{-3}$ \\
\hline
BP3 & $1.40\times10^{12}$ & $1.23\times10^{6}$ & $1.69\times10^{2}$ & $7.19\times10^{-5}$ \\
\hline
BP4 & $5.36\times10^{10}$ & $3.46\times10^{5}$ & $7.28\times10^{2}$ & $2.48\times10^{-5}$ \\
\hline
\end{tabular}
\caption{Values of the model parameters relevant for leptogenesis are shown for four representative choices of $M_{N_3}$, corresponding to the scenarios illustrated in Fig.\,\ref{fig:leptoline}. The corresponding Yukawa matrix for the BP3 that reproduces the observed $Y_{\Delta L}$, is provided in App.\,\ref{app:modulerF}.}
\label{tab:cpasy}
\end{table}

\begin{figure}[htb!]
    \centering
    \includegraphics[width=0.65\linewidth]{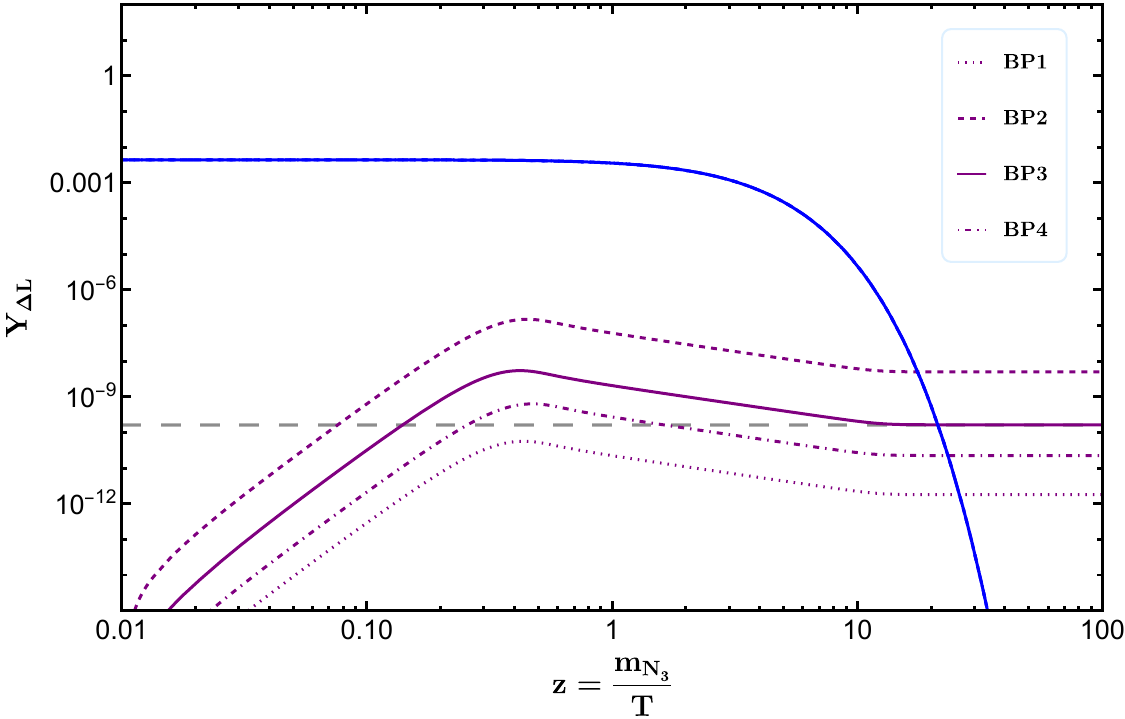}
    \caption{Evolution of lepton asymmetry with temperature is illustrated for different smallest RHN masses, $M_{N_{3}^{}}$. The gray dashed line corresponds to the observed BAU.}
    \label{fig:leptoline}
\end{figure}

Fig.\,\ref{fig:leptoline} shows the evolution of the lepton asymmetry as a function of $z\equiv M_{N_3}/T$ for the four benchmark points listed in Tab.\,\ref{tab:cpasy}. Since the washout parameters for all benchmark points are of the same order, $K_{2,3}\sim\mathcal{O}(10^{2})$, the washout efficiencies remain broadly comparable. Consequently, the differences in the final lepton asymmetry are predominantly driven by the corresponding CP asymmetries, $\epsilon_{2,3}$, with larger values of $\epsilon_{2,3}$ yielding larger final values of $Y_{\Delta L}$.

\begin{figure}[htbp]
    \centering
    \begin{subfigure}{0.49\linewidth}
        \centering
        \includegraphics[width=\linewidth]{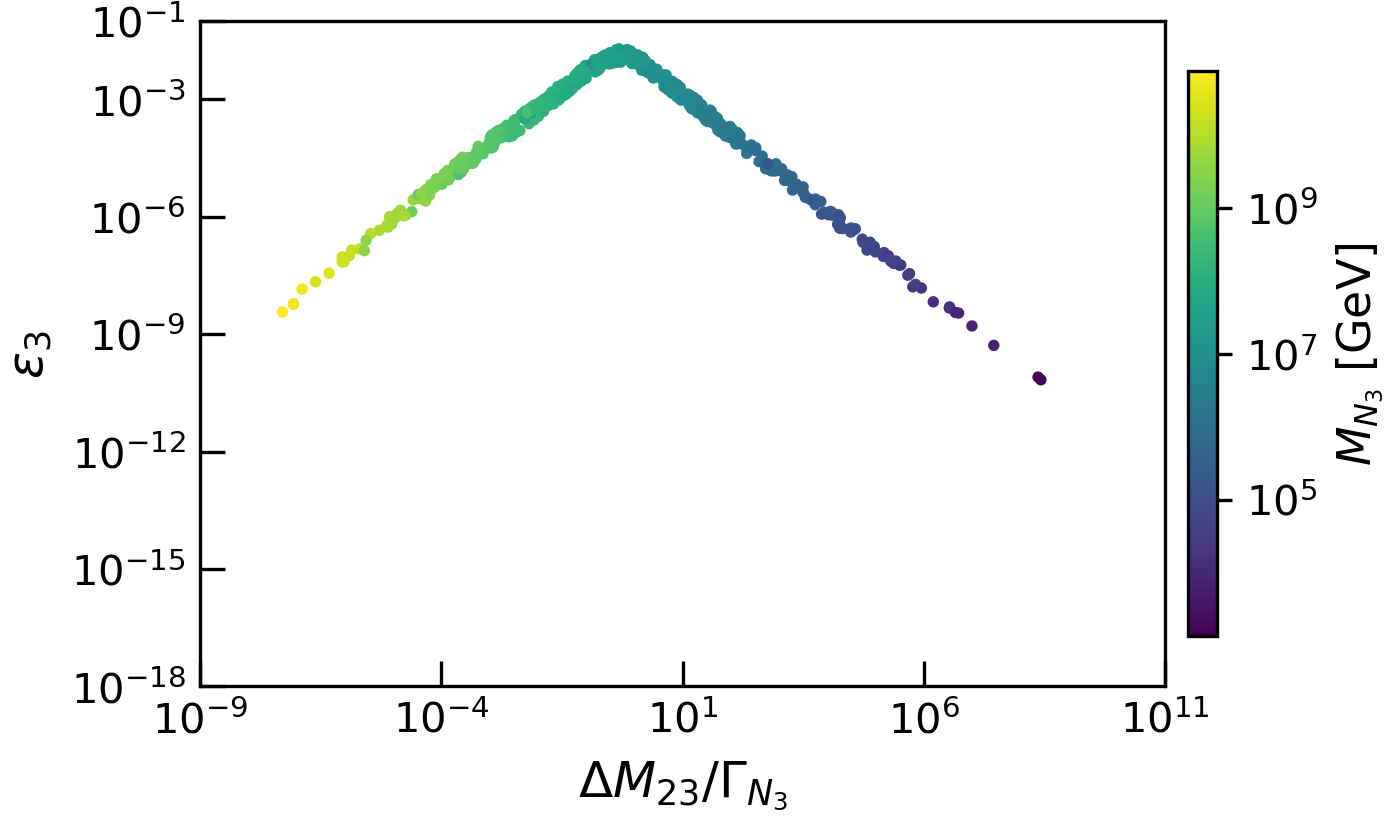}
        \caption{}
    \end{subfigure}
    \hfill
    \begin{subfigure}{0.49\linewidth}
        \centering
        \includegraphics[width=\linewidth]{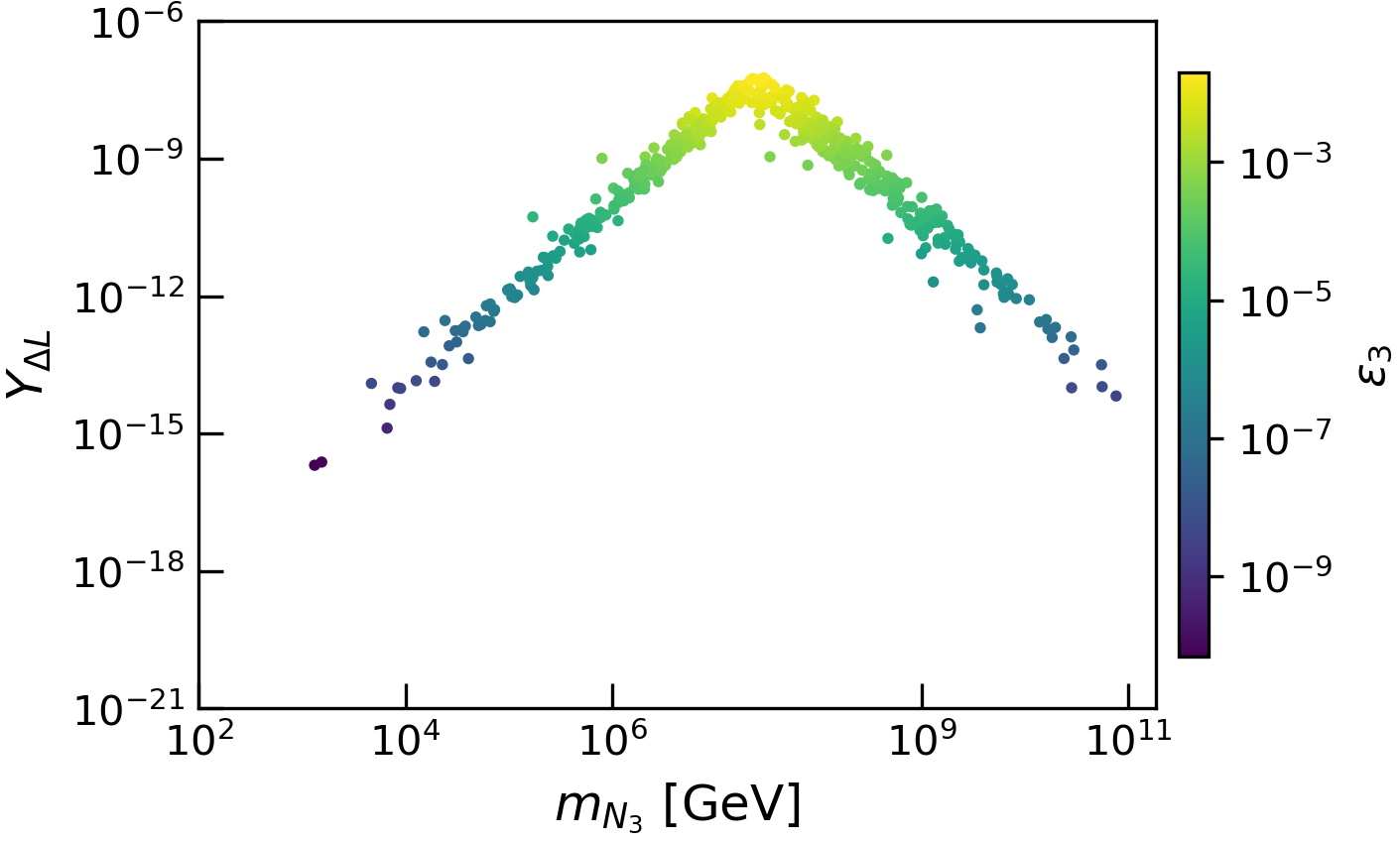}
        \caption{}
    \end{subfigure}
    \caption{Panel (a) shows the variation of CP-asymmetry $\epsilon_3$ with $\Delta M_{23}/\Gamma_{N_3}$, the corresponding $M_{N_3}$ is presented in colour bar. Panel (b) illustrates the change in $Y_{\Delta L}$ in terms of $M_{N_{2}^{}}$, where the color bar represents the CP asymmetry parameter $\epsilon_{3}$.}
    \label{fig:YBL}
\end{figure}

To generate Fig.\,\ref{fig:YBL}, we scan the model parameters over the ranges specified in Eqs.~\eqref{eq:RItau} and \eqref{eq:mpv}, and retain only the parameter points consistent with the observed neutrino oscillation data. Panel (a) shows that the CP asymmetry reaches its maximum when $\Delta M_{23}/\Gamma_{N_3}\simeq 0.5$, corresponding to the resonant condition for leptogenesis. The color gradient indicates that this condition is realized for $M_{N_3}\sim 10^{6}$--$10^{7}\,\mathrm{GeV}$, where the resonant enhancement of the CP asymmetry is most effective.

For values of $M_{N_3}$ below this range, the mass splitting $\Delta M_{23}$ becomes increasingly fine. However, the decay width $\Gamma_{N_3}$ decreases even more rapidly, causing the ratio $\Delta M_{23}/\Gamma_{N_3}$ to increase above its resonant value. Consequently, the CP asymmetry is suppressed despite having a smaller mass splitting. On the other hand, for $M_{N_3}$ above this mass range, the mass splitting $\Delta M_{23}$ increases, but the decay width $\Gamma_{N_3}$ grows more rapidly than the splitting. As a result, the ratio $\Delta M_{23}/\Gamma_{N_3}$ decreases below the resonant value of $1/2$, moving the system away from the resonance condition and thereby reducing the CP asymmetry.

Panel (b) of Fig.\,\ref{fig:YBL} shows the variation of the generated lepton asymmetry, $Y_{\Delta L}$, with $M_{N_3}$, while the color gradient represents the corresponding CP asymmetry. As expected, the lepton asymmetry closely follows the behavior of the CP asymmetry, attaining its maximum for $M_{N_3}\sim 10^{6}$--$10^{7},\mathrm{GeV}$, where the resonance condition is best satisfied. For both lower and higher values of $M_{N_3}$, the departure from the resonance condition suppresses the CP asymmetry, leading to a corresponding reduction in the generated lepton asymmetry.

\section{Possible Gravitational Wave Signatures in the presence of RHNs}\label{sec:gw}
Having identified the parameter space consistent with neutrino oscillation data and demonstrated that the modular structure of the model naturally induces the small mass splittings in the RHN sector, required for resonant leptogenesis, we now turn to its phenomenological implications. Specifically, we examine the GW signatures associated with the right-handed neutrino sector. The details of this analysis and corresponding results are presented in this section. 

After an elaborate discussion on the seesaw sector of the Lagrangian in previous sections, we now focus on the scalar potential structure of our setup. In addition to the Standard Model Higgs doublet ($H$), we introduce a complex scalar field, $\Phi\equiv(\phi+i\,\eta)/\sqrt{2}$ which carries a non-trivial charge under the discrete symmetry $\mathbb{Z}_3$ (with $A_4$ charge $\mathrm{1}$) and transforms as $\Phi\rightarrow e^{i\,2\pi/3}\Phi$. Subject to the symmetry, the most general renormalizable scalar potential is given by \footnote{In addition to the $A_4\otimes\mathbb{Z}_3$ symmetry of the potential, we impose a CP symmetry on $\Phi$ under which $\Phi\rightarrow\Phi^\ast$. As will be discussed in Sec.~\ref{sec:DM}, this symmetry stabilizes the CP-odd scalar, making it a viable dark matter candidate. The choice of a $\mathbb{Z}_{3}$ symmetry over $\mathbb{Z}_2$ is motivated by the strong first-order phase transition dynamics, which will be discussed in the following subsections.}
\begin{eqnarray}\label{eq:pot}
V(H,\Phi) &=& -\mu_H^{2}\,(H^\dagger H) 
+ \lambda_H\,(H^\dagger H)^{2} 
- \mu_\Phi^{2}\,(\Phi^\dagger \Phi) 
+ \lambda_\Phi\,(\Phi^{\dagger}\Phi)^2 \nonumber\\
&+& \lambda_{H\Phi}\,(H^\dagger H)(\Phi^\dagger \Phi)- \frac{\sqrt{2}\mu_3}{3}\left( \Phi^3 +\text{h.c}\right).
\end{eqnarray}
At high temperatures, thermal corrections restore the $\mathbb{Z}_3$ symmetry, and the scalar field $\Phi$ remains in the symmetric phase with $\langle \Phi \rangle = 0$. As the Universe cools below the critical temperature, the interplay between the mass parameter\footnote{A detailed discussion is provided in the next subsection.} $\mu_\Phi^2(T)=\mu_\Phi^2-c_\Phi T^2$, the quartic coupling $\lambda_\Phi$, and the cubic coupling $\mu_3$ drives $\Phi$ to acquire a nonzero VEV, thereby spontaneously breaking the $\mathbb{Z}_3$ symmetry.

\subsection{Gravitational waves from domain wall annihilation} The spontaneous breaking of the $\mathbb{Z}_3$ symmetry gives rise to three degenerate vacuum states $\langle\Phi\rangle=\tilde{v}_{\phi}e^{i\theta}$ (where $\tilde{v}_{\phi}=v_{\phi}/\sqrt{2}$ and $\theta=2\pi k/3$, with $k=0,1,2$). Since these vacua are energetically equivalent, different regions of the Universe can independently possess different vacuum configurations during the phase transition. The interfaces separating neighbouring domains then form domain walls (DWs), which correspond to smooth field configurations interpolating between distinct vacuum states. The energy density stored in the DW evolved as $\rho_{\rm DW}\propto t^{-1}$~\cite{Saikawa:2017hiv}, while the radiation energy density decreases more rapidly, $\rho_R\propto t^{-2}$. Consequently, the DWs become the dominant component of the Universe at late times. Such a scenario is cosmologically unacceptable, as it is incompatible with the CMB observation. This situation is known as the domain wall problem. 

\section*{Bias term}

This problem can be avoided by introducing a soft, explicit $\mathbb{Z}_3$-breaking term, known as the bias term, that removes the vacuum degeneracy by introducing a tiny energy splitting among the minima. The resulting pressure difference renders the DWs unstable, causing them to annihilate before they dominate the energy density of the Universe. In our model, this can be achieved via the interactions \footnote{One may be concerned that the explicit breaking term could induce additional contributions to the RHN mass matrix. However, for the parameter region relevant to observable GW signals, $\alpha_{R}=\alpha_{R}^{\prime}\lesssim\mathcal{O}(10^{-21})$ and $v_{\phi}\lesssim\mathcal{O}(10^{10})$ GeV, the resulting correction is exceedingly small and can be safely neglected.}
\begin{equation}
   \mathcal{L}_{\slashed{\mathbb{Z}}_3} = -\frac{1}{2}\left(  \alpha_R \Phi\overline{N_R^c} N_R Y_3^{(-2)}+ \alpha_R^\prime \Phi\overline{N_R^c} N_R Y_1^{(-2)}+ \text{h.c.}\right)\,,
\end{equation}
which not only provides the required explicit symmetry breaking but also naturally couples the scalar sector to the right-handed neutrino sector, yielding an intriguing interplay between domain wall dynamics and neutrino phenomenology.  
The explicit symmetry-breaking term lifts the vacuum degeneracy through one-loop radiative corrections to the scalar potential induced by the heavy RHNs at both zero and finite temperatures. The corresponding zero-temperature Coleman-Weinberg (CW) potential in the $\overline{\mathrm{MS}}$ regularization scheme is given by~\cite{Manohar:2020nzp, Gelmini:2020bqg,Quiros:1999jp}
\begin{eqnarray}
    V_{\rm CW}(v_\phi,\theta) = -\frac{1}{64\pi^2} \text{Tr}\left[\left(M_N^\dagger M_N\right)^2\log\left(\frac{M_N^\dagger M_N}{\mu^2}\right)-\frac{3}{2}\right],
\end{eqnarray}
where $\mu$ is the renormalization scale of the theory. Here, the heavy Majorana mass matrix is given by
\begin{eqnarray}
   M_N\equiv M_N(v_{\phi},\theta) = M_R + M_R^\prime(v_{\phi},\theta),
\end{eqnarray}
with~\footnote{For simplicity, we consider $\alpha_{R}=\alpha_{R}^\prime$ throughout the rest of our analysis.}
\begin{eqnarray}
\small
  M_R^\prime \equiv \frac{v_{\phi}e^{i\theta}}{\sqrt{2}}Y=\frac{v_{\phi}e^{i\theta}}{\sqrt2}\frac{\alpha_R}{3} \begin{pmatrix}
       \left(2 Y_{3,1}^{(-2)} +3 Y_1^{(-2)}\right)& -Y_{3,3}^{(-2)} & -Y_{3,2}^{(-2)}\\
       -Y_{3,3}^{(-2)} & 2 Y_{3,2}^{(-2)} & \left(3 Y_1^{(-2)}-Y_{3,1}^{(-2)}\right)\\
       -Y_{3,2}^{(-2)} & \left(3 Y_1^{(-2)}-Y_{3,1}^{(-2)}\right) & 2 Y_{3,3}^{(-2)}
   \end{pmatrix}.
\end{eqnarray}
 Expanding $V_{\rm CW}$ around $M_R^\dagger M_R$, and retaining the leading contribution yields the simplified expression 
\begin{equation}\label{eq:Vcw}
   V_{\rm CW}(v_{\phi},\theta)
   =\tilde{V}_{\rm CW}-\frac{v_\phi|K|}{16\sqrt2\pi^2}\cos(\theta+\delta) \,, 
\end{equation}
where 
\begin{eqnarray}
K={\rm Tr}\left[M_R^\dagger M_R\left(\ln\frac{M_R^\dagger M_R}{\mu^2}-1\right)M_R^\dagger Y\right],  \quad \text{and}\quad\delta = \text{Arg}(K).
\end{eqnarray}
For the three $\mathbb{Z}_3$ vacua, $\theta=0,2\pi/3,$ and $4\pi/3$. The total vacuum energy difference from the zero-temperature one-loop correction is denoted as
\begin{eqnarray}
\Delta V_{\rm bias}^{\rm CW}&=&\sum_{ j>i=1}^{3} V_{\rm CW}\left(v_{\phi},\theta_{j}\right)-V_{\rm CW}(v_{\phi},\theta_{i})\,.
\end{eqnarray}
Here, $\theta_i$ denotes the $\theta$ values corresponding to different vacuua, and also note that the $v_{\phi}$ independent term $\tilde{V}_{\rm CW}$ will cancel from each term in $\Delta V_{\rm bias}^{\rm CW}$. On the other hand, the one-loop finite temperature contribution is~\cite{Quiros:1999jp}
\begin{eqnarray}
V_T(v_{\phi},\theta)
=
-\frac{T^4}{2\pi^2}
\,{\rm Tr}
\left[
J_F\!\left(\frac{M_N^\dagger M_N}{T^2}\right)
\right],
\end{eqnarray}
where $J_F\left(\frac{m^2}{T^2}\right)$ is the thermal fermionic function (see App.\,\ref{app:thT}). Now, following the similar prescription of Taylor expansion around $M_R^\dagger M_R$, one can get 
\begin{equation}
V_{T}(v_{\phi},\theta)
=\tilde{V}_T
-\frac{ T^2v_\phi|K_T|}
{\sqrt2\pi^2}
\cos(\theta+\delta_T)\,,
\end{equation}
where
\begin{eqnarray}
K_T
=
{\rm Tr}
\left[
J_F'\!\left(\frac{M_R^\dagger M_R}{T^2}\right)
M_R^\dagger Y
\right],\quad \quad \delta_T = \text{Arg}\left(K_T\right).
\end{eqnarray}
Therefore, the total vacuum energy difference due to the temperature correction is denoted as
\begin{equation}\label{eq:vbiasT}
    \Delta V^T_{\rm bias} = \sum_{ j>i=1}^{3}V_{T}\left(v_{\phi},\theta_{j}\right) - V_{T}(v_{\phi},\theta_{i}).
\end{equation}
For $M_{N_i}^2 \ll T^2$, the thermal function admits a high-temperature expansion. In this limit, the bias term is provided in App.\,\ref{app:thT}.
Finally, one can write the total bias term as 
\begin{equation}
    \Delta V_{\rm bias}=\Delta V_{\rm bias}^{\rm CW}+\Delta V_{\rm bias}^{T}\,.
\end{equation}
For our numerical analysis, we fix the RG scale at $\mu=10^{13}$ GeV.
\section*{Domain wall annihilation}
The resulting energy difference from the bias term induces an effective pressure, known as volume pressure, $p_{V}\sim\Delta V_{\rm bias}$, across the walls, which continuously shrinks the regions occupying the vacuum with higher energy. As the pressure imbalance becomes larger than the force associated with the wall tension, $p_T\sim\mathcal{A}\sigma/t$~\cite{Everett:1982nm, Press:1989yh, Garagounis:2002kt} (here $\sigma$ and $\mathcal{A}$ are the surface tension and area parameter of the DW, respectively), the domain walls become unstable and eventually annihilate, and ultimately the Universe settles to the true vacuum with no DWs. The temperature of the Universe at the time of DW annihilation is given by~\cite{Borah:2025bfa}
\begin{equation}\label{eq:Tann}
    T_{\rm ann}=\frac{3.39\times 10^{-2}\,\mathcal{C}_{\rm an}^{-\frac{1}{2}}\mathcal{A}_{}^{-\frac{1}{2}}\left(\frac{10}{g_{\rho}^{\ast}(T_{\rm ann})}\right)^{\frac{1}{4}}\,\Big(\frac{\sigma}{{\rm TeV^3}}\Big)^{-\frac{1}{2}}\,\left(\frac{\Delta V^{\rm CW}_{\rm bias}}{10^{-12}\,{\rm GeV^4}}\right)^{\frac{1}{2}}}{\sqrt{1-1.14\times10^{-3}\,\mathcal{C}_{\rm an}^{-1}\mathcal{A}^{-1}\,\left(\frac{10}{g_{\rho}^{\ast}(T_{\rm ann})}\right)^{\frac{1}{2}}\,\Big(\frac{\sigma}{{\rm TeV^3}}\Big)^{-1}\,\left(\frac{C_1}{10^{-12}\,{\rm GeV^2}}\right)}}\, {\rm GeV}\,.
\end{equation}
where $\mathcal{A}=1.10\pm0.20$ and $\mathcal{C}_{\rm an}=5.01\pm0.44$ are the dimension-less constants taken from Ref.~\cite{Kawasaki:2014sqa} based on the study of axionic DW network from $\mathbb{Z}_N$ symmetry, here we take the case for $N=3$. In addition, $C_1$ represents the coefficient of the $T^2$ term enclosed within the parentheses in the third term of Eq.~\ref{eq:vbiasT}. When $\Delta V_{\rm bias}$ is too small, the DWs survive long enough to dominate the energy density of the Universe. The corresponding domination temperature $T_{\rm dom}$ is~\cite{Saikawa:2017hiv}
\begin{equation}
    T_{\rm dom}=1.625\times 10^{-5}\,\mathcal{A}^{\frac{1}{2}}\,\left(\frac{10}{g_{\rho}^{\ast}(T_{\rm ann})}\right)^{\frac{1}{4}}\,\Bigg(\frac{\sigma}{{\rm TeV^3}}\Bigg)^{\frac{1}{2}}\,.
\end{equation}\label{eq:Tdom}
The annihilation of DWs must take place before it starts to dominate the energy budget of the Universe. This puts a bound on $T_{\rm ann}$ as $T_{\rm ann}>T_{\rm dom}$. In addition, consistency with standard cosmology demands that DW annihilation must occur before the onset of Big Bang nucleosynthesis (BBN), leading to the further constraint on annihilation temperature as $T_{\rm ann}>T_{\rm BBN}$, where $T_{\rm BBN}$ is the BBN temperature. Percolation theory predicts that an extensive network of false-vacuum domains can form provided the bias term satisfies the condition $\Delta V_{\rm bias}<0.79\,V_{b}$~\cite{Stauffer:1978kr}, where $V_{b}$ denotes the height of the potential barrier between two adjacent minima.

The annihilation and collapse of DWs releases a substantial amount of energy in the form of gravitational waves (GWs), which can survive until today as a stochastic GW background. The present-day peak frequency ($f_{\rm peak}$) and peak amplitude of the GWs ($\Omega_{\rm peak}h^2$) can be expressed as~\cite{Saikawa:2017hiv, Wu:2022stu}
\begin{align}\label{eq:fop}
   & f_{\rm peak}\simeq1.1\times 10^{-9}\,\text{Hz}\,\left(\frac{g_\rho^{\ast}(T_{\rm ann})}{10}\right)^{\frac{1}{2}}\left(\frac{g_{s}^{\ast}(T_{\rm ann})}{10}\right)^{-\frac{1}{3}}\left(\frac{T_{\rm ann}}{0.01 \,\text{GeV}}\right)\nonumber\,,\\
  &  \Omega_{\rm peak}h^2\simeq7.2\times10^{-18}\,\tilde{\epsilon}\mathcal{A}\,\left(\frac{g_\rho^{\ast}(T_{\rm ann})}{10}\right)^{-\frac{4}{3}}\left(\frac{\sigma}{\text{TeV}^3}\right)^2\,\left(\frac{T_{\rm ann}}{0.01 \,\text{GeV}}\right)^{-4}\,,
\end{align}
where $\tilde{\epsilon}=0.7\pm0.4$~\cite{Hiramatsu:2013qaa}. The amplitude of the GW follows a broken power-law spectrum which can be parametrize as 
\begin{eqnarray}
 \Omega_{\rm GW}h^2_{} = \Omega_{\rm peak} h^2 \frac{(a+b)^c}{\left(a x^{b / c}+b x^{-a / c}\right)^c} \ ,
\label{eq:spec-par}
\end{eqnarray}
where $x\equiv f/f{\rm peak}$. The positive real parameters $a,b,$ and $c$ determine the spectral shape, with $a$ and $b$ corresponding to the low and high-frequency spectral indices, respectively, while $c$ controls the smoothness of the transition around the peak. Causality requires the low-frequency slope to satisfy $a=3$, while numerical simulations favor $b \simeq c \simeq 1$~\cite{Hiramatsu:2013qaa}. 

The peak frequency, $f_{\rm peak}$, characterizes the annihilation time of the domain wall. The resulting GW background may act as an additional source of radiation and is therefore constrained by CMB and BBN measurements of the effective number of relativistic species, $\Delta N_{\rm eff}$. These observations translate into an upper limit on the GW abundance, $\Omega_{\rm GW}h^2\lesssim10^{-6}$~\cite{Planck:2018vyg, Cyburt:2015mya, Abazajian:2019eic}.
\begin{figure}[!htb]
\centering
\includegraphics[width=0.48\linewidth]{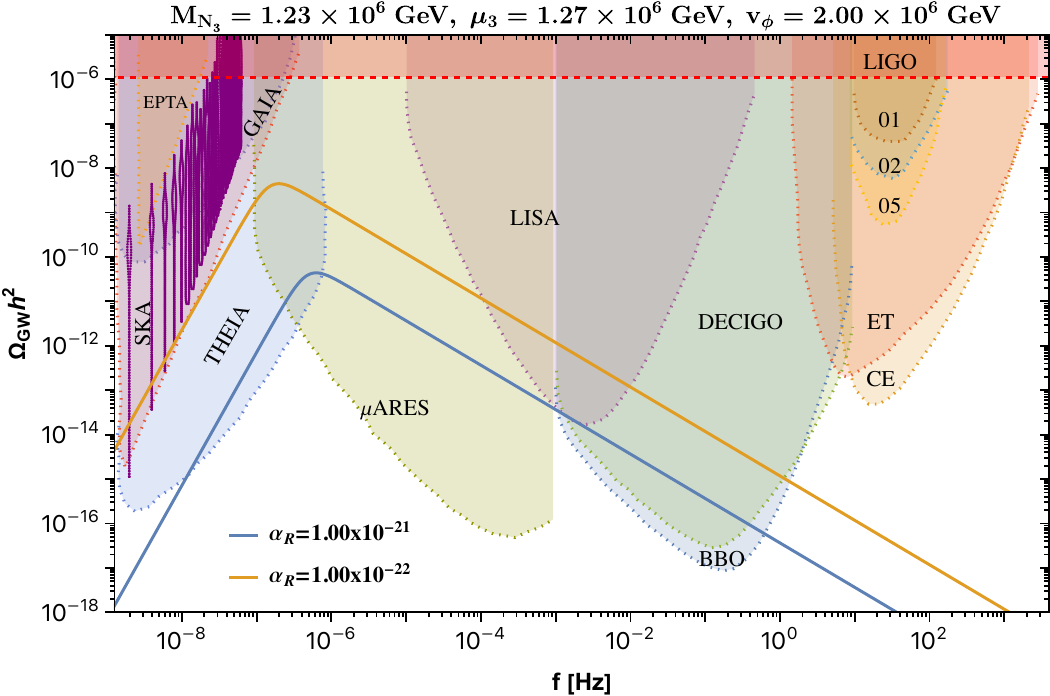}
\includegraphics[width=0.48\linewidth]{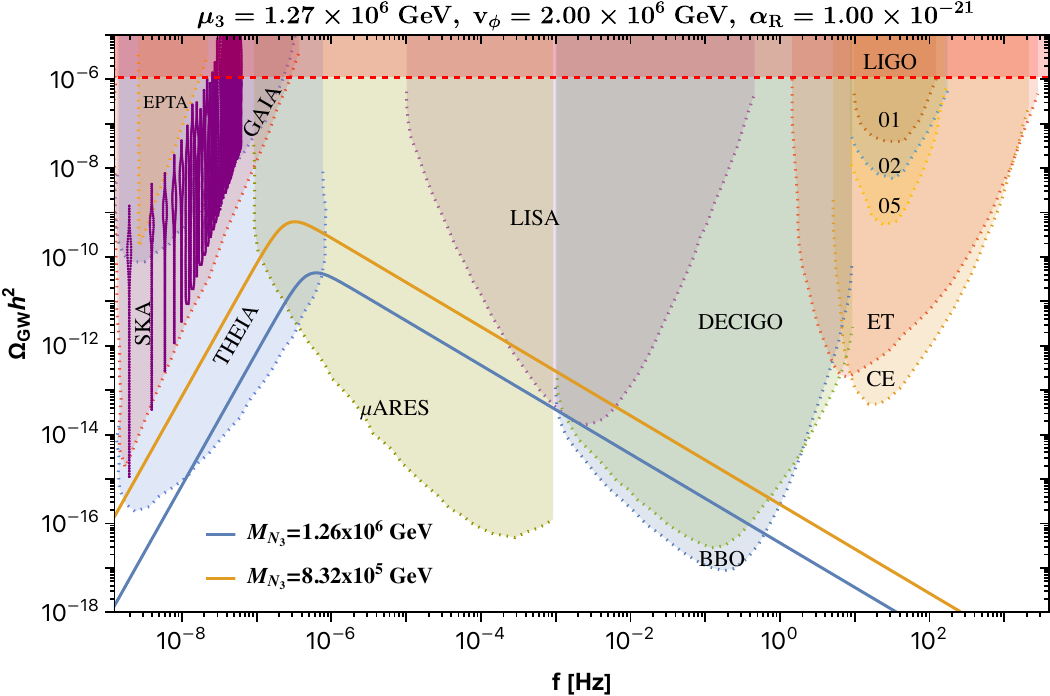}
\includegraphics[width=0.48\linewidth]{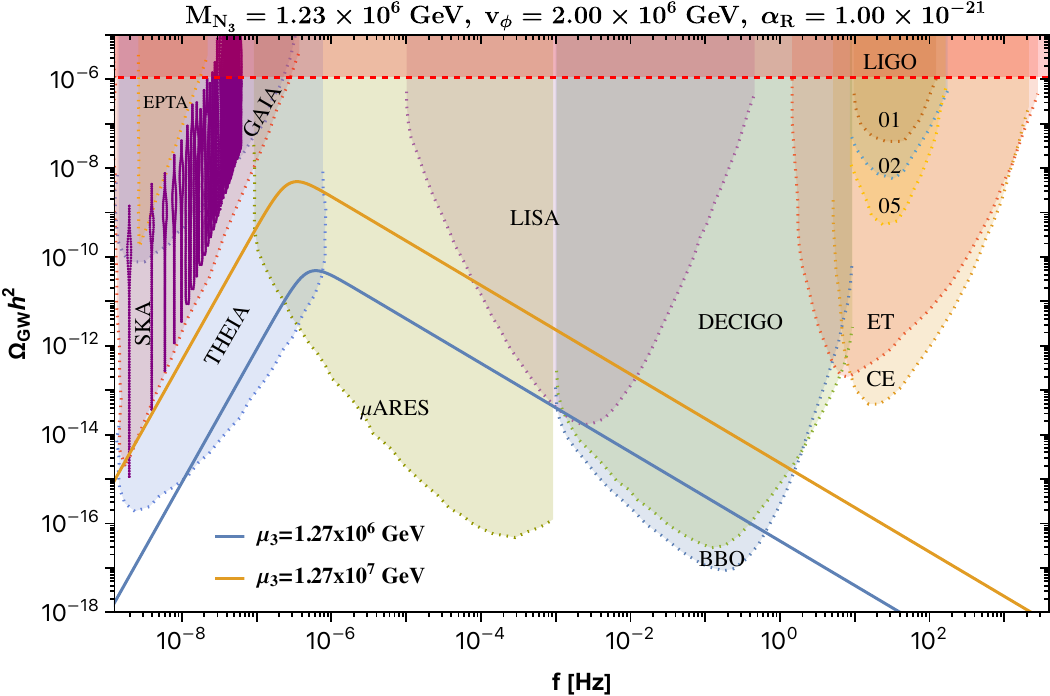}
\includegraphics[width=0.48\linewidth]{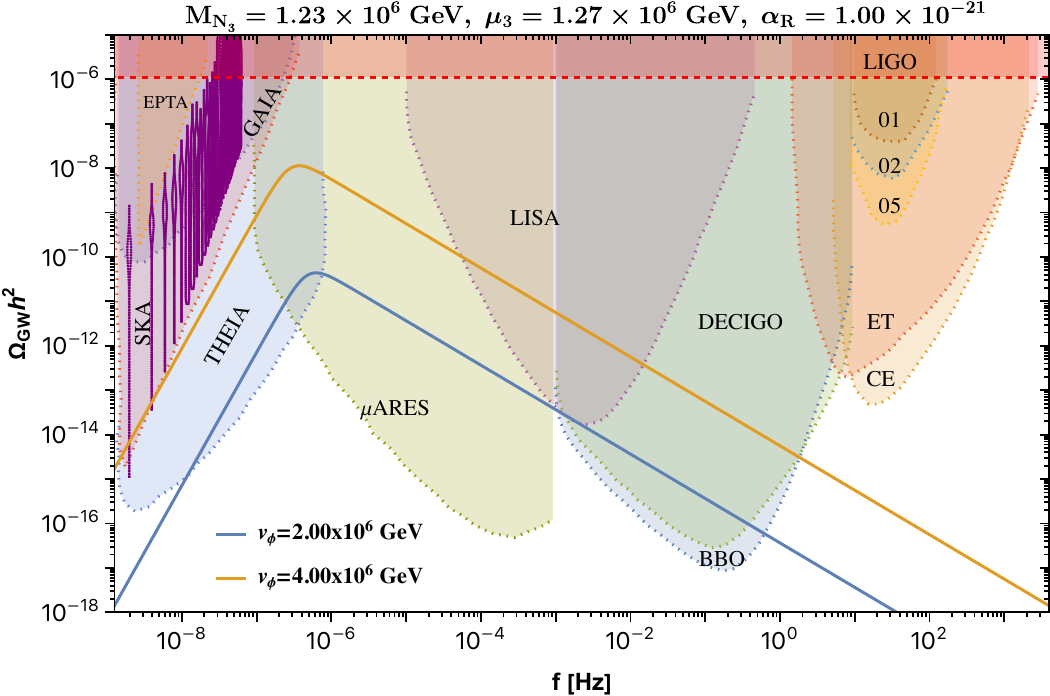}
\caption{Different GW spectra from DW annihilation are shown for representative choices of the model parameters. The fixed parameter values used in each panel are listed in the corresponding plot labels, while the two line curves (shown in blue and yellow) demonstrate the dependence of the GW spectrum on the model parameter identified in the legends.}
\label{fig:GW_spectrum}
\end{figure}
The GW spectra for various model parameters are presented in Fig.\,\ref{fig:GW_spectrum}. The projected sensitivities of SKA~\cite{Weltman:2018zrl}, GAIA~\cite{Garcia-Bellido:2021zgu}, EPTA~\cite{Moore:2014eua}, THEIA~\cite{Garcia-Bellido:2021zgu}, $\mu$ARES~\cite{Sesana:2019vho}, LISA~\cite{amaroseoane2017laserinterferometerspaceantenna}, DECIGO~\cite{Seto:2001qf,Kawamura:2006up,Yagi:2011wg}, BBO~\cite{Crowder:2005nr,Corbin:2005ny,Harry:2006fi}, ET~\cite{Punturo:2010zz,Hild:2010id,Sathyaprakash:2012jk,Maggiore:2019uih}, CE~\cite{LIGOScientific:2016wof,Reitze:2019iox}, and LIGO~\cite{LIGOScientific:2016wof,LIGOScientific:2014qfs,LIGOScientific:2016jlg} are indicated by the shaded regions in different colours while the purple spike-shaped band represents the recent NANOGrav result~\cite{NANOGrav:2023gor}. The red shaded region is excluded by the current constraints on the effective number of relativistic species, $N_{\rm ef}$~\cite{Planck:2018vyg,Cyburt:2015mya,Abazajian:2019eic}. 

As discussed above, DW annihilation is initiated once the volume pressure generated by the bias energy overcomes the restoring force associated with the wall tension, i.e., $p_V>p_T$. Since $p_V\sim\Delta V_{\rm bias}$ and $p_T\sim\sigma/t$, Eq.~\ref{eq:Tann} establishes that $T_{\rm ann}$ increases with the bias energy while decreasing with the DW surface tension, $\sigma$. Employing the relation $\sigma\simeq\mu_3v_\phi^2$~\cite{Wu:2022stu, Deng:2020dnf,Hattori:2015xla}, together with $\Delta V_{\rm bias}\equiv f(\alpha_R,v_\phi,M_{N_i})$ (cf. Eq.~\ref{eq:Vcw}), one finds that $T_{\rm ann}$ is governed by the four model parameters $\{\mu_3,v_\phi,M_{N_i},\alpha_R\}$. Consequently, the characteristic observables of the resulting GW signal, namely the peak frequency $f_{\rm peak}$ and the peak amplitude $\Omega_{\rm peak}h^2$, are likewise controlled by the same parameter set. In particular, increasing $\alpha_R$ or $M_{N_i}$ enhances the bias energy, thereby raising the annihilation temperature, whereas larger values of $v_\phi$ or $\mu_3$ increase $\sigma$ and consequently delay the annihilation. Since $f_{\rm peak}\propto T_{\rm ann}$ (see Eq.~\ref{eq:fop}), the peak frequency shifts to higher (lower) values with increasing $\alpha_R$ and $M_{N_i}$ (increasing $v_\phi$ and $\mu_3$), Fig.\,\ref{fig:GW_spectrum} clearly depicts these behaviour of $f_{\rm peak}$ on the model parameters. 

On the other hand, an earlier DW annihilation time corresponds to a larger $T_{\rm ann}$, implying that the generated GWs undergo larger redshift. As a result, we have a smaller peak amplitude $\Omega_{\rm peak}h^2$. Conversely, a later annihilation at lower temperatures leaves less time for redshift, yielding a larger $\Omega_{\rm peak}h^2$. Therefore, $\Omega_{\rm peak}h^2$ exhibits the opposite dependence on the underlying model parameters, decreasing with increasing $\alpha_R$ and $M_{N_i}$, while increasing for larger values of $v_\phi$ and $\mu_3$. 

\begin{figure}[!htb]
\centering
\includegraphics[scale=0.36]{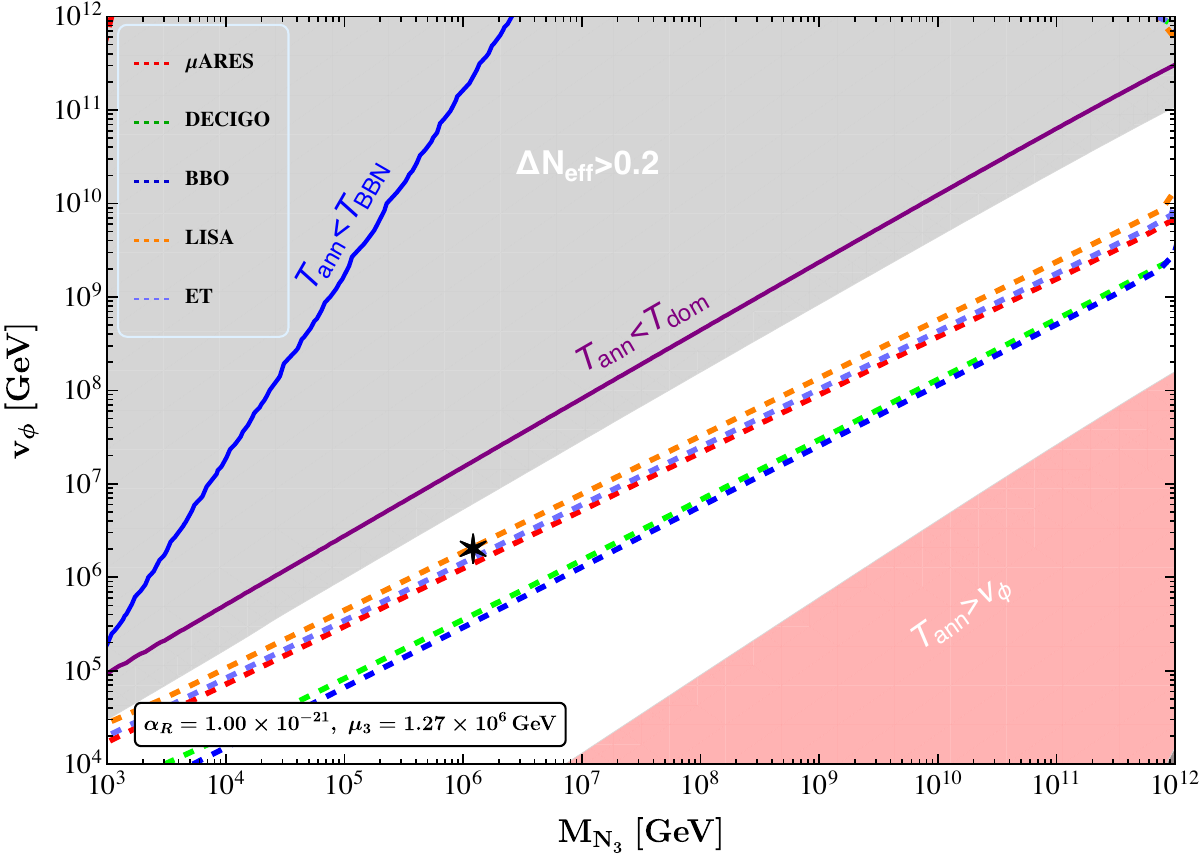}
\includegraphics[scale=0.36]{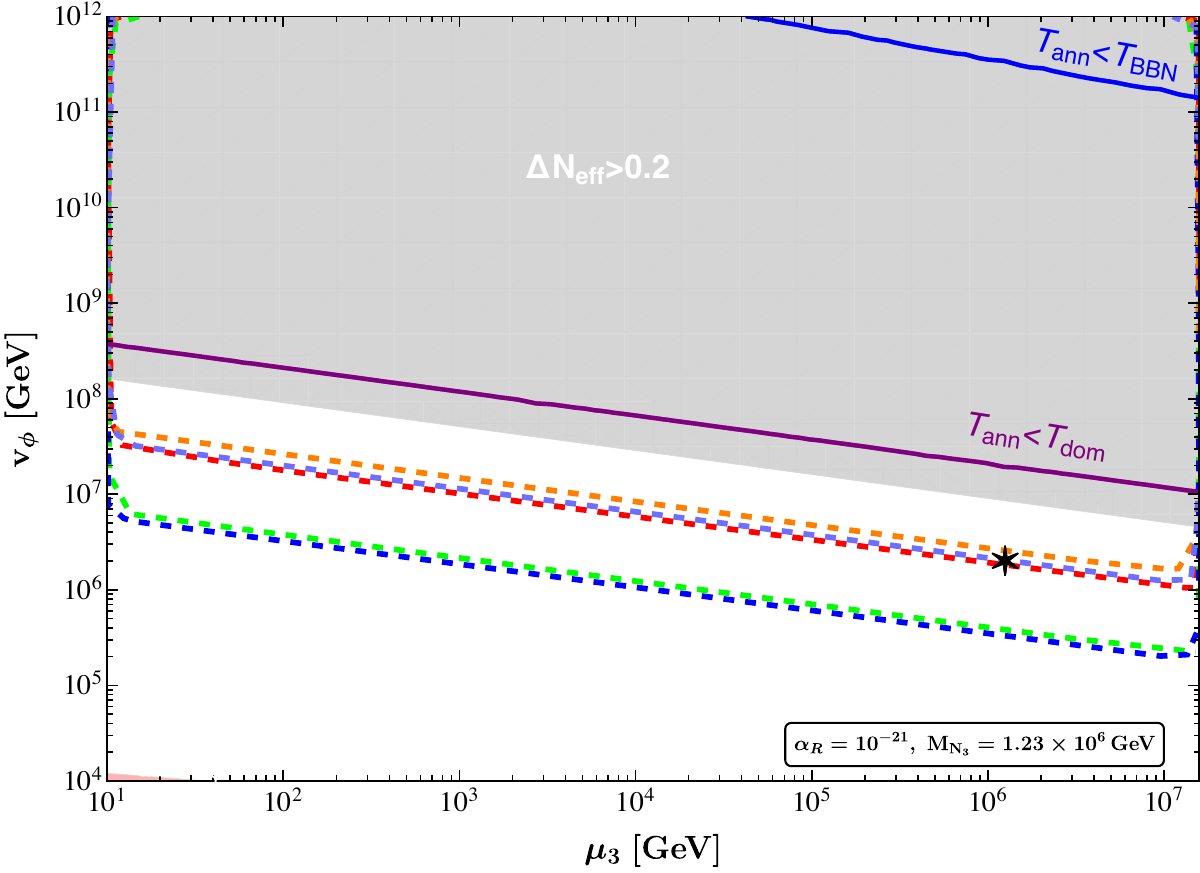}
\caption{Allowed region of parameter space is depicted in $M_{N_3^{}}-v_{\phi}$ (left panel) and $\mu_{3}-v_{\phi}$ (right panel) planes. The gray shaded region is disallowed from the PLANCK bound on $\Delta N_{\rm eff}$. The point depicted in the star shape simultaneously satisfies two benchmark points: BP3 of Tab.~\ref{tab:cpasy} (responsible for GW from DW) and $\rm BP4~(3c)$ of Tab.~\ref{tab:FOPT} (responsible for GW from FOPT). The same point also satisfies the neutrino oscillation data and reproduces the observed BAU.}
\label{fig:Summary}
\end{figure}

Fig.\,\ref{fig:Summary} summarizes the viable parameter space in the $M_{N_3}-v_{\phi}$ (left panel) and $\mu_3-v_{\phi}$ (right panel) planes from the perspective of the GW signatures associated with DW annihilation. The gray shaded regions are excluded by the Planck bound~\cite{Planck:2018vyg} on $\Delta N_{\rm eff}$, while the red shaded regions are ruled out by the requirement $T_{\rm ann}<v_{\phi}$, assuming that the spontaneous $\mathbb{Z}_3$-breaking temperature satisfies $T_{\slashed{\mathbb{Z}}_3}\simeq v_{\phi}$.

The blue and purple solid curves correspond to the conditions $T_{\rm ann}=T_{\rm BBN}$ and $T_{\rm ann}=T_{\rm dom}$, respectively. In the $M_{N_3}-v_{\phi}$ plane, the regions to the left of these curves are excluded, whereas in the $\mu_3-v_{\phi}$ plane, the excluded regions lie above them. The behaviour of these boundaries can be readily understood from Eq.~\eqref{eq:Tann} and Eq.~\eqref{eq:Tdom}. For fixed $\mu_3$, the annihilation temperature increases with increasing $M_{N_3}$ ($T_{\rm ann}\propto M_{N_3}^{3/2}/\sqrt{v_{\phi}}$ from Eq.~\eqref{eq:Tann} using $\sigma\sim\mu_3v_{\phi}^2$). Consequently, preserving either $T_{\rm ann}=T_{\rm BBN}$ or $T_{\rm ann}=T_{\rm dom}$ requires a corresponding increase in $v_{\phi}$, giving rise to the positively sloped boundaries in the left panel. On the other hand, for fixed $M_{N_3}$, $T_{\rm ann}$ decreases with increasing $\mu_3$ ($T_{\rm ann}\propto \hat{m}^{2}/\sqrt{\mu_3v_{\phi}}$, from Eq.~\eqref{eq:Tann}, where $\hat{m}$ is a constant with mass-dimension). Consequently, satisfying the condition $T_{\rm ann}=T_{\rm BBN}$ requires progressively smaller values of $v_\phi$. Likewise, the boundary defined by $T_{\rm ann}=T_{\rm dom}$ also shifts toward lower $v_\phi$ with increasing $\mu_3$. Because increasing $\mu_3$, $T_{\rm dom}$ increases (see Eq.~\eqref{eq:Tdom}) and to maintain $T_{\rm ann}=T_{\rm dom}$ we have to decrease $v_{\phi}$.

As a quantitative measure of the observability of the predicted GW signal, we calculate the signal-to-noise ratio (SNR)~\cite{Maggiore:1999vm, Allen:1997ad, Roshan:2026yon},
\begin{equation}
\varrho=\left[n_{\mathrm{det}} t_{\mathrm{obs}} \int_{f_{\min }}^{f_{\max }} d f\left(\frac{\Omega_{\text {signal }}(f)}{\Omega_{\text {noise }}(f)}\right)^2\right]^{1 / 2}\,,
\label{eq:SNR}
\end{equation}
 where $n_{\rm det}$ distinguishes between experiments that probe the stochastic GW background through auto-correlation ($n_{\rm det}=1$) and those employing cross-correlation measurements ($n_{\rm det}=2$). Here, $t_{\rm obs}$ denotes the total observing time of the experiment, while $\Omega_{\rm noise}(f)$ is the detector noise spectrum expressed in terms of the GW energy density~\cite{Schmitz:2020syl}. For each experiment ET~\cite{Punturo:2010zz}, LISA~\cite{LISA:2017pwj}, DECIGO~\cite{Kawamura:2020pcg}, $\mu$Ares~\cite{Sesana:2019vho}, and THEIA~\cite{Garcia-Bellido:2021zgu}, the SNR is obtained by integrating the squared ratio $\Omega_{\rm signal}(f)/\Omega_{\rm noise}(f)$ over its corresponding sensitive frequency band. The projected sensitivity curves for $\rm SNR=5$ are depicted in Fig.\,\ref{fig:Summary} as dashed curves in different colours. 
 
 It is worth mentioning that the star-shaped benchmark point in Fig.\,\ref{fig:Summary}, corresponding simultaneously to BP3 in Tab.\,\ref{tab:cpasy}, which successfully reproduces the observed BAU and predicts the GW signal from DW annihilation for $\alpha_R=1.00\times10^{-21}$ and $v_\phi=2.00\times10^6$ GeV, and BP4 (3c) in Tab.\,\ref{tab:FOPT}, which gives rise to the GW signal from the first-order phase transition, as will be discussed in detail in the next subsection. This benchmark lies well within the projected sensitivities of several future GW observatories, highlighting the promising opportunity to probe the proposed framework through its predicted GW signatures indirectly.

\subsection{High temperature potential}
After expanding the potential in Eq.\,\eqref{eq:pot} about the background fields, the tree-level potential takes the form
\begin{eqnarray}\label{eq:treeU1}
V_0(h,\phi) &=& -\frac{\mu_H^2}{2}  h^2 - \frac{\mu^2_\Phi}{2} \phi^2 +\frac{\lambda_H}{4} h^4  + \frac{\lambda_\Phi }{4}\phi^4 
+ \frac{\lambda_{H\Phi}}{4} h^2 \phi^2 - \frac{\mu_3}{3}\phi^3 .
\end{eqnarray}
Tree-level minimization conditions give the relations:
\begin{eqnarray}
   \mu_H^2 = \frac{1}{2} \left(2\lambda_H v_h^2 +\lambda_{HS} v_\phi^2\right),\quad
 \mu_\Phi^2 = \frac{1}{2} \left(2\lambda_S v_\phi^2 +\lambda_{HS} v_h^2 - 2\mu_3 v_\phi \right).
\end{eqnarray}
After SSB, the two CP-even scalar fields mix at zero temperature, and the corresponding mixing matrix is given by
\begin{equation}\label{eq:massMatrix}
   M^2(v,v_\phi) = 
    \begin{pmatrix}
        2 \lambda_H  v_h^2&  \lambda_{HS}\,v_hv_\phi\\
     \lambda_{HS}\, v_hv_\phi& 2\lambda_S v_\phi^2+\mu_3v_\phi
    \end{pmatrix}.
\end{equation}
The physical scalar mass eigenstates, $\varphi_{1}$ and $\varphi_{2}$, are obtained by diagonalizing the scalar mass matrix through the orthogonal rotation matrix $\mathcal{R}$, such that
\begin{equation}
    \begin{pmatrix}
    \varphi_1\\
    \varphi_2\\
    \end{pmatrix}
    =
    \mathcal{R}
    \begin{pmatrix}
    h\\
    \phi\\
    \end{pmatrix} \, ,\qquad \mathcal{R} = \begin{pmatrix}
    \cos\theta & -\sin\theta\\
    \sin\theta & \cos\theta\\
    \end{pmatrix},
\end{equation}
which satisfy 
\begin{equation}
     M^2(v_h,v_\phi) = \mathcal{R}^T M_{\rm diag}^2 \mathcal{R}, \quad \text{with} \quad   M_{\rm diag}^2 = \begin{pmatrix}
    m^2_{\varphi_1} & 0\\
    0 & m^2_{\varphi_2}\\
    \end{pmatrix}.
\end{equation}
Using the above equations and Eq.~\eqref{eq:massMatrix}, the model parameters in terms of inputs are
\begin{eqnarray}
   &&\lambda_H = \frac{m_{\varphi_1}^2 \cos ^2\theta}{2 v_h^2}+\frac{m_{\varphi_2}^2 \sin ^2\theta}{2 v_h^2},\quad
   \lambda_\Phi = \frac{m_{\varphi_1}^2 \sin ^2\theta}{2 v_\phi^2}+\frac{m_{\varphi_2}^2 \cos ^2\theta}{2 v_\phi^2}+\frac{\mu_3 }{2v_\phi}\nonumber\, ,\\
 &&  \lambda_{H\Phi} = \frac{m_{\varphi_2}^2-m_{\varphi_1}^2 }{v_h v_\phi}\cos\theta\sin\theta.
\end{eqnarray}
From the global $U(1)$ symmetry-breaking term, the pseudo-Goldstone boson $\eta$ acquires a non-zero mass given by
\begin{equation}
    m_\eta^2(v,v_\phi)=3\mu_3v_\phi,
    \qquad \Rightarrow \qquad
    \mu_3=\frac{m_\eta^2}{3v_\phi}.
\end{equation}
Since the physical mass squared must satisfy $m_\eta^2>0$, it follows that $\mu_3$ and $v_\phi$ must have the same sign. Therefore, for the convention $v_\phi>0$, the cubic coupling is required to be positive, \textit{i.e.}, $\mu_3>0$.

We first consider the decoupling limit of the model, motivated by domain wall considerations and leptogenesis, in which the singlet VEV is assumed to be of the order of a few hundred TeV. In this limit, the Higgs-singlet mixing is highly suppressed, $\sin\theta \simeq 0$, resulting in an effectively vanishing portal coupling, $\lambda_{H\Phi} \simeq 0$.
\begin{figure}[htbp]
    \centering
    \begin{subfigure}{0.49\linewidth}
        \centering
        \includegraphics[width=\linewidth]{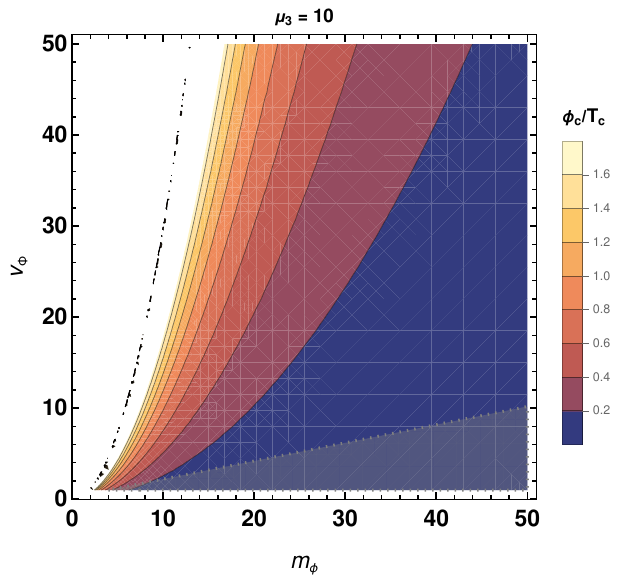}
    \end{subfigure}
    \hfill
    \begin{subfigure}{0.49\linewidth}
        \centering
        \includegraphics[width=\linewidth]{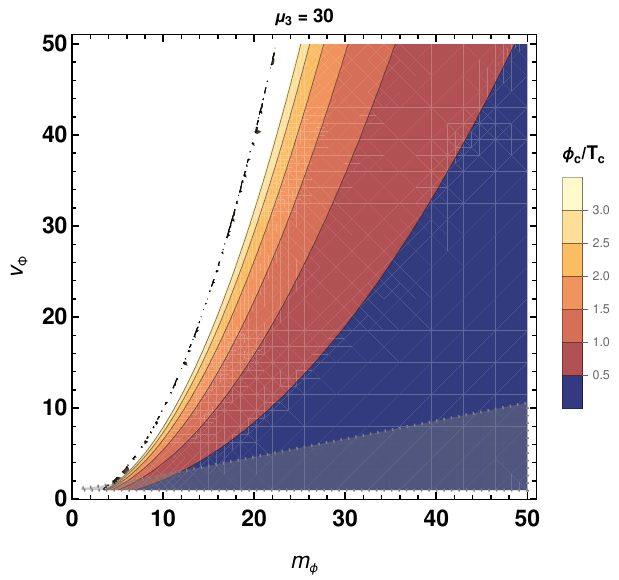}
    \end{subfigure}\caption{Variation of the singlet VEV, $v_\phi[10^5 \text{GeV}]$, with the CP-even scalar mass, $m_\phi[10^5 \text{GeV}]$. The color scale represents the strength of the phase transition, $\phi_c/T_c$. The left and right panels correspond to $\mu_3=10[10^5 \text{GeV}]$ and $\mu_3=30[10^5 \text{GeV}]$, respectively.}
    \label{fig:PT}
\end{figure}
To ensure the stability of the electroweak vacuum at zero temperature, the desired vacuum $(v_h,v_\phi)$ must be the global minimum of the tree-level scalar potential. To ensure that $(v_h,v_\phi)$ is the global minimum, its vacuum energy must be lower than that of any competing local minimum. In particular, if $(v_h,0)$ is a local minimum, one requires $V_0(v_h,v_\phi)<V_0(v_h,0)$.

At zero temperature, the large hierarchy between the singlet VEV ($\sim 10^3~\mathrm{TeV}$) and the electroweak VEV, $v_h = 246~\mathrm{GeV}$, causes the phase transition to proceed predominantly along the singlet scalar direction. Since the transition occurs at temperatures well above the electroweak scale, the Higgs field remains in the symmetric phase ($\langle H \rangle_T = 0$), and the finite-temperature evolution is governed almost entirely by the singlet field. As the temperature decreases to the electroweak scale, the Higgs field acquires a nonzero VEV through the standard smooth crossover, with only negligible effects from the singlet sector owing to the small Higgs-portal coupling. The one-loop finite-temperature effective potential is given by
\begin{equation}
    V_{\rm eff}(\phi,T)=V_{0}(\phi)+V_{\rm CW}(\phi)+V_T(\phi,T).
\end{equation}
We neglect the Coleman-Weinberg correction, $V_{\rm CW}(\phi)$, since it has a negligible effect on the phase transition in the parameter region considered. We therefore do not discuss it further.
On the other hand, the finite-temperature correction is
\begin{equation}
    V_T(\phi,T)=\frac{T^4}{2\pi^2}\sum_{i=\phi,\eta} n_i
    J_{B}\!\left(\frac{m_i^2(\phi)}{T^2}\right),
\end{equation}
with
\begin{equation}
    J_{B}(m^2/T^2)=
    \int_0^\infty dx\,x^2
    \ln\!\left[1 - e^{-\sqrt{x^2+m^2/T^2}}\right],
\end{equation}
where the degrees of freedom, $n_\phi, n_\eta =1$. In the high-temperature limit, $y\equiv m^2/T^2\ll1$, the thermal bosonic function admits the expansions
\begin{align}
J_B(y)&=
-\frac{\pi^4}{45}
+\frac{\pi^2}{12}y
-\frac{\pi}{6}y^{3/2}
-\frac{y^2}{32}\ln\!\left(\frac{y}{a_B}\right)
+\mathcal{O}(y^4),
\end{align}
where $a_B=16\pi^2e^{3/2-2\gamma_E}$, with the Euler–Mascheroni constant $\gamma_E$.
Retaining only the leading temperature-dependent terms, the finite temperature effective potential can be approximated by the high-temperature expansion
\begin{equation}
V_{\rm eff}(\phi,T)
\simeq
-\frac{1}{2}\left(\mu_\Phi^2-c_\Phi T^2\right)\phi^2
+\frac{\lambda_\Phi}{4}\phi^4
-\frac{\mu_3}{3}\phi^3,
\end{equation}
Here, $c_\Phi \simeq\lambda_\Phi/3$ denotes the thermal mass coefficient arising from the leading bosonic $T^2$ correction due to the $\phi$ and $\eta$ fields. For the large values of $\mu_3$ considered in this work, the thermally generated cubic term is much smaller than the tree-level cubic term proportional to $\mu_3$, which already provides the barrier between the minima. Therefore, the thermal cubic contribution can be safely neglected. The contribution of the heavy Majorana neutrinos to the effective potential is suppressed, since it is controlled by the Yukawa coupling $\alpha_R$, which is assumed to be sufficiently small.

\subsection{Phase transition}

As the Universe cools, the finite-temperature effective potential develops a second minimum at a nonzero value of the singlet field. The critical temperature, $T_c$, is defined as the temperature at which the symmetric and broken phases become degenerate. At $T=T_c$, the effective potential satisfies
\begin{equation}\label{eq:crtc}
V_{\rm eff}(0,T_c)=V_{\rm eff}(\phi_c,T_c),
\qquad
\left.\frac{\partial V_{\rm eff}(\phi,T)}{\partial\phi}\right|_{\phi=\phi_c,T=T_c}=0,
\end{equation}
where $\phi_c$ is the singlet vacuum expectation value at the critical temperature. Solving these equations determines $T_c$ and $\phi_c$, and the strength of the phase transition is characterized by the order parameter $\phi_c/T_c$, with a strongly first-order phase transition typically requiring $\phi_c/T_c \gtrsim 1$.

Solving the conditions in Eq.\eqref{eq:crtc} simultaneously yields the critical VEV and the corresponding critical temperature,
\begin{eqnarray}\label{eq:tcvc}
\phi_c = \frac{2\mu_3}{3\lambda_\Phi},\qquad \quad
T_c = \frac{\sqrt{2\mu_3^2+9\lambda_\Phi\mu_\Phi^2}}
{3\sqrt{c_\Phi\lambda_\Phi}}.    
\end{eqnarray}
For the critical temperature to be real, the physical parameter space is constrained by the requirement $2\mu_3^2+9\lambda_\Phi\mu_\Phi^2>0$.

In Fig.\,\ref{fig:PT}, we present the parameter space in the $(m_\phi,\,v_\phi)$ plane for two representative values of the trilinear coupling $\mu_3$, as indicated in the figure caption. Comparing the two panels, it is evident that the strength of the phase transition increases with increasing $\mu_3$. This behavior is expected, as the trilinear interaction enhances the tree-level potential barrier separating the symmetric and broken phases, thereby strengthening the first-order phase transition. The gray shaded region is excluded by the tree-level perturbativity constraint $(\lambda_\Phi \le 4\pi)$ on the singlet quartic coupling, while the white region corresponds to an unphysical parameter space where no real critical temperature exists.

We now turn to the stochastic GW background generated by the FOPT along the singlet direction. Such a transition proceeds through the nucleation, expansion, and collision of true-vacuum bubbles, sourcing a stochastic GW signal that can be probed by present and future space-based GW observatories. For the GW analysis, we select a set of benchmark points (BPs), listed in Tab.\,\ref{tab:FOPT}. These benchmark points are chosen to yield a FOPT while satisfying all the theoretical constraints discussed in the previous sections.

\begin{table}[htbp]
\centering
\renewcommand{\arraystretch}{1.3}
\begin{tabular}{|c|c|c|c|c|c|c|c|}
\hline
\hline
\multirow{2}{*}{BPs} &
\multicolumn{3}{c|}{Input Parameters} &
\multicolumn{4}{c|}{PT \& GW Quantities} \\
\cline{2-8}
& $m_\phi$   & $v_\phi$  & $\mu_3$ 
& $\alpha_n$ & $\beta/H_n$ & $T_n$ & $\phi_c/T_c$ \\
\hline
 BP1
& $4.0 $ & $5.9 $ &  $4.5 $ & $0.018$ & $468$ & $2.43$ & $1.53$ \\
\hline
BP2  &
$4.0 $ & $5.9 $ & $4.0 $ & $0.008$ & $1144$ & $ 3.09 $ & $1.30$ \\
\hline
BP3 &
$4.0 $ & $5.9 $ & $3.0$ & $0.003$ & $3907$ & $ 4.27$ & $0.92$ \\
\hline
BP4 
& 12.0  & 20.0  & 12.7  & 0.080 &  32 &  5.37 & 1.68\\
\hline
BP5
& 14.9 & 22.0 & 18.18 & 0.134 & 10 & 5.46 & 1.73 \\
\hline
BP6 
& 30.0  & 48.8  & 32.0  & 0.035 & 179 & 16.20 & 1.64 \\
\hline
BP7 
& 24.0 & 40.04  &  25.1  & 0.043 & 125 & 12.49 & 1.65 \\
\hline
\hline
\end{tabular}
\caption{Benchmark points and the corresponding input parameters and output quantities relevant for the stochastic GW spectra. All dimensionful quantities are given in units of $10^5\,\mathrm{GeV}$.}
\label{tab:FOPT}
\end{table}

The relevant parameters governing a FOPT and the resulting stochastic GW signal are the nucleation temperature $T_n$, the strength parameter $\alpha_n$, and the inverse duration parameter $\beta/H_n$. 
The nucleation temperature, $T_n$, is defined as the temperature at which one critical bubble is nucleated per Hubble volume per Hubble time, and is determined by the conditions~\cite{Linde:1981zj,Mazumdar:2018dfl}
\begin{equation}
\Gamma(T_n)\simeq H^4(T_n),
\qquad\qquad \text{where}\quad
\Gamma(T)=T^4\left(\frac{S_3(T)}{2\pi T}\right)^{3/2}
\exp\!\left[-\frac{S_3(T)}{T}\right],
\end{equation}
where $\Gamma(T)$ is the bubble nucleation rate per unit volume, $H(T)$ is the Hubble expansion rate, and $S_3(T)$ is the three-dimensional Euclidean bounce action. 
The strength of the phase transition is characterized by~\cite{Kamionkowski:1993fg}
\begin{equation}
\alpha_n=\frac{\Delta\rho}{\rho_{\rm rad}(T_n)},
\end{equation}
where $\Delta\rho$ is the released vacuum (latent) energy density, defined in Ref.\,\cite{Kehayias:2009tn}, and the radiation energy density $\rho_{\rm rad}(T_n)=\frac{\pi^2}{30}\,g_*(T_n)\,T_n^4$, with $g_*(T_n)$ is the relavistic degrees of freedom at nucleation temperature. The inverse duration of the phase transition is quantified by~\cite{Nicolis:2003tg}
\begin{equation}
\frac{\beta}{H_n}
=
\left.
T\frac{d}{dT}\left(\frac{S_3}{T}\right)
\right|_{T=T_n},
\end{equation}
where $H_n$ is the Hubble parameter at $T_n$. Larger values of $\alpha_n$ correspond to stronger phase transitions, while smaller values of $\beta/H_n$ indicate longer-lasting transitions that generally produce stronger GW signals. For each benchmark point (BP), we list the corresponding values of $T_n$, $\alpha_n$, and $\beta/H_n$, which determine the amplitude and spectral shape of the resulting stochastic GW background. 
To evaluate them, we use the publicly available package \texttt{CosmoTransitions}~\cite{Wainwright:2011kj}. The values of $\alpha$ and $\beta/H_n$ obtained from \texttt{CosmoTransitions}, as listed in Tab.\,\ref{tab:FOPT}, are found to be in excellent agreement with the corresponding analytical estimates presented in Ref.~\cite{Ellis:2020awk}. Furthermore, for the parameter space considered in this work, the value of the phase transition order parameter $\phi_c/T_c$ computed by \texttt{CosmoTransitions} exactly reproduces the analytical formula given in Eq.~\eqref{eq:tcvc}. 

\begin{figure}
    \centering
    \includegraphics[width=0.49\linewidth]{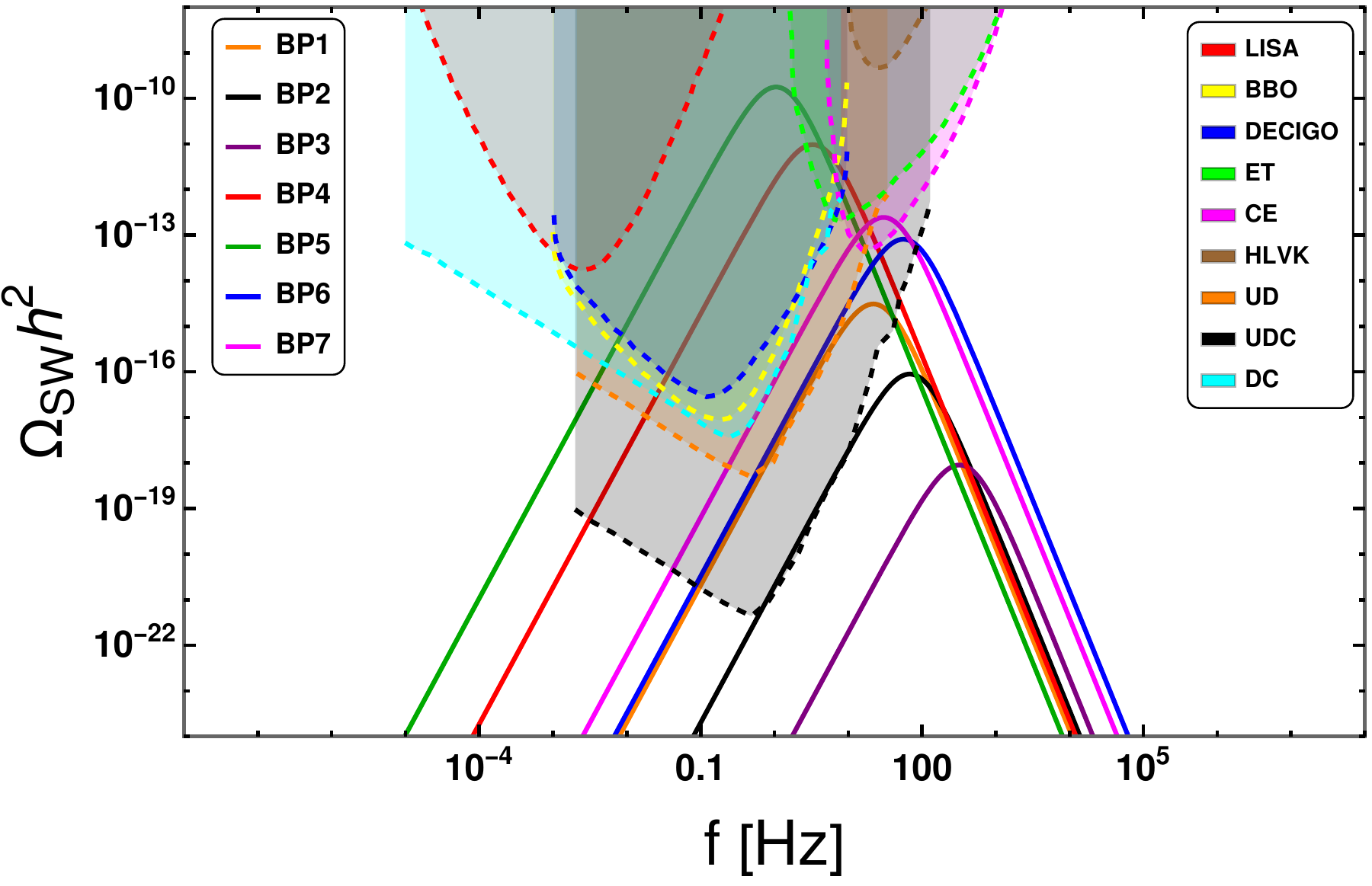}
    \includegraphics[width=0.49\linewidth]{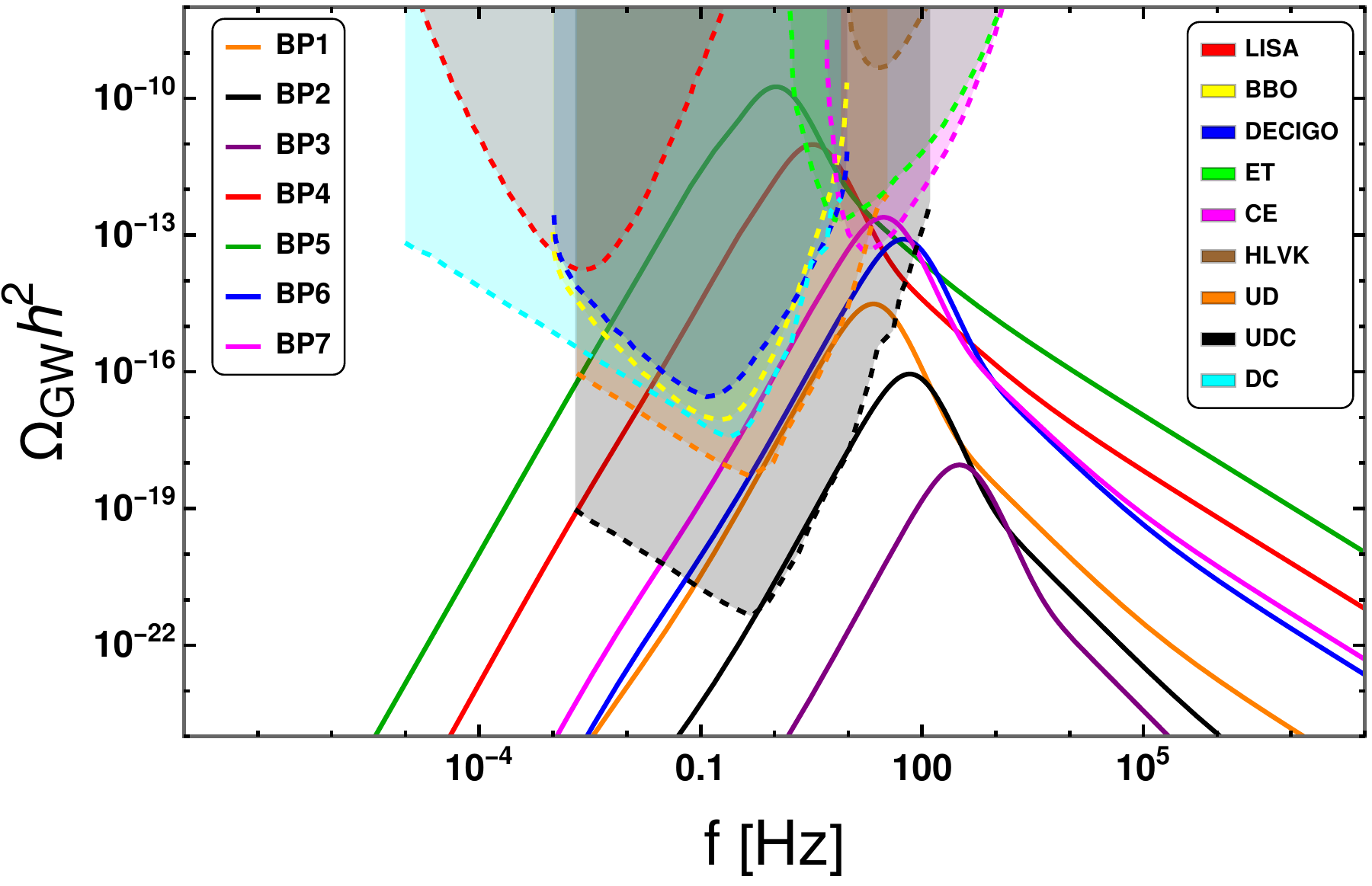}
    \caption{Variation of the gravitational wave amplitude with frequency. The left panel shows the contribution from sound waves only, while the right panel presents the total gravitational wave spectrum. The shaded regions indicate the projected sensitivities of future gravitational wave detectors, including LISA~\cite{LISA:2017pwj}, BBO~\cite{Yagi:2011wg}, the Einstein Telescope (ET)~\cite{Punturo:2010zz}, Cosmic Explorer (CE)~\cite{Reitze:2019iox}, the HLVK network (LIGO Hanford--Livingston, Virgo, and KAGRA)~\cite{LIGOScientific:2014pky,VIRGO:2014yos,KAGRA:2018plz}, DECIGO, Ultimate DECIGO (UD), DECIGO Correlation (DC), and Ultimate DECIGO Correlation (UDC)~\cite{Nakayama:2009ce}, as indicated in the legend.}
    \label{fig:ptgw}
\end{figure}

The stochastic GW energy density spectrum receives contributions from bubble wall collisions, sound waves, and magnetohydrodynamic (MHD) turbulence. The total GW spectrum can therefore be approximated as~\cite{Ellis:2020awk,Caprini:2015zlo}
\begin{equation}\label{eq:GWTotal}
\Omega_{\rm GW}h^2 \simeq \Omega_{\rm col}h^2+\Omega_{\rm sw}h^2+\Omega_{\rm tur}h^2,
\end{equation}
where $\Omega_{\rm col}$, $\Omega_{\rm sw}$, and $\Omega_{\rm tur}$ denote the contributions from bubble wall collisions, sound waves, and MHD turbulence, respectively, and $h\equiv H_0/(100\,\mathrm{km\,s^{-1}\,Mpc^{-1}})$ is the dimensionless Hubble parameter. The explicit expressions for the individual contributions are given in Refs.~\cite{Caprini:2015zlo,Das:2026zuo}. 

Fig.\,\ref{fig:ptgw} shows the stochastic GW spectra as a function of frequency for the selected benchmark points. The total GW spectrum and the dominant sound wave contribution are displayed, as indicated in the figure caption. Unlike the electroweak phase transition, the high-scale phase transition generates a stochastic GW background with peak frequencies ranging from a few Hz to a few hundred Hz. The predicted GW signals for BP4 and BP5 are within the projected sensitivity of DECIGO, BBO, ET, and CE. The GW signals for BP6 and BP7 fall within the sensitivity reach of Ultimate-DECIGO and CE, while those for BP1 are detectable only by Ultimate-DECIGO. In contrast, the GW signals predicted for BP2 and BP3 lie below the projected sensitivity of all future planned GW detectors.

It should be emphasized that BP1, BP2, and BP3 of Tab.~\ref{tab:FOPT} differ only in the value of trilinear coupling, $\mu_3$, allowing us to isolate its impact on the GW spectrum. As shown in Fig.\,\ref{fig:PT}, decreasing the value of $\mu_3$ weakens the FOPT. Consequently, $T_n$ increases, leading to a lower value of $\alpha$ and a higher value of $\beta/H_n$. Since the GW amplitude increases with $\alpha$ and decreases with $\beta$, a smaller value of $\mu_3$ results in a progressively weaker GW signal, explaining the suppression of the GW spectrum from BP1 to BP3.

\begin{figure}[htb!]
    \centering
    \includegraphics[width=0.658\linewidth]{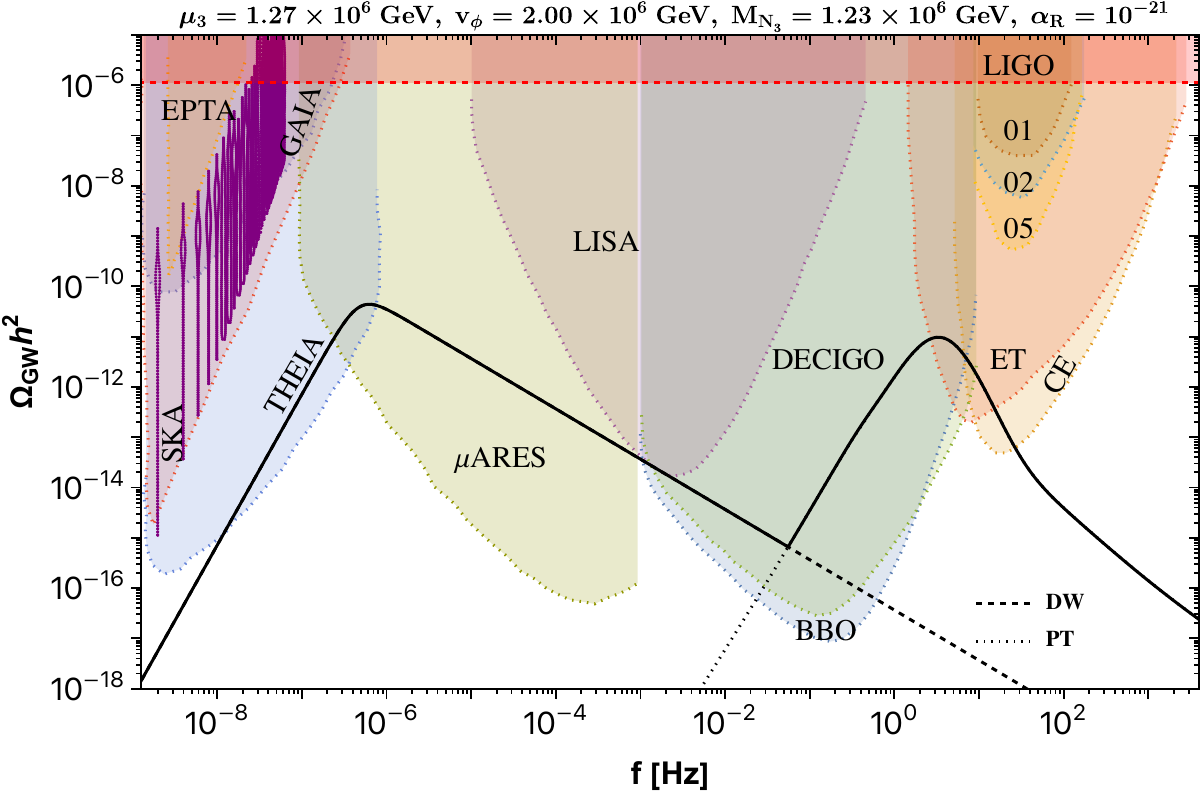}
    \caption{Two-peak GW spectrum for the benchmark point corresponding to BP3 in Tab.~\ref{tab:cpasy} and BP4 in Tab.~\ref{tab:FOPT}, marked by a star in Fig.\,\ref{fig:Summary}, consistent with neutrino oscillation data and the observed BAU.}
    \label{fig:2peak}
\end{figure}

Fig.\,\ref{fig:2peak} displays the predicted GW spectrum for the representative benchmark BP3 in Tab.\,\ref{tab:cpasy} and BP4 in Tab.\,\ref{tab:FOPT}. The low-frequency peak is generated by DW annihilation at the relatively low $T_{\rm ann}$, while the high-frequency peak originates from the FOPT occurring at a much higher $T_n$. The resulting large separation between the two peak frequencies is a characteristic prediction of our framework.
The low-frequency signal indirectly probes the leptogenesis scale through the correlation between $T_{\rm ann}$ and $M_{N_3}$. In contrast, the high-frequency peak is sensitive to the symmetry-breaking scale via $T_n$, and is further correlated with the DM mass\footnote{The pseudo Goldstone boson $\chi$ can serve as a viable DM candidate, as discussed in the next subsection.}, $m_\chi$, through the parameters $\mu_3$, $v_\phi$, and the scalar mass $m_\phi$. Therefore, the observation of both GW peaks would provide complementary insights into the leptogenesis and symmetry-breaking scales while constraining the underlying particle spectrum of the model.

\subsection{Comments on Dark Matter candidate}\label{sec:DM}

For a trilinear coupling with real value $\mu_3$, the scalar potential is CP invariant and consequently remains invariant under the discrete transformation $\chi \rightarrow -\chi$. Therefore, CP symmetry forbids the decay of $\chi$, rendering it stable.

In our model, we have a soft $\mathbb{Z}_3$ breaking Yukawa interaction term involving DM and the RHNs, whose coupling strength is governed by the dimensionless parameter $\alpha_R$. In general, this interaction induces a complex effective Yukawa coupling via modular functions, allowing the dark matter candidate $\chi$ to decay into a pair of heavy neutrinos\footnote{Here, $N^\prime = N^\prime_R + N^{\prime c}_R$.} ($N^\prime N^\prime $), a heavy and a light neutrino ($N^\prime\nu$), or a pair of light neutrinos ($\nu\nu$), depending on the mass hierarchy between $\chi$ and the heavy neutrinos.

Throughout this work, we consider the mass spectrum $m_\chi < 2M_N$, for which the decay channel $\chi \rightarrow N^\prime N^\prime$ is kinematically forbidden. Consequently, if $m_\chi > M_N$, the dominant two-body decay mode is $\chi \rightarrow N\nu$, whose decay width scales as
\begin{equation}
\Gamma(\chi \rightarrow N^\prime\nu)
\sim
\alpha_R^2
\left(\frac{m_D}{M_N}\right)^2
(m_\chi- M_N),
\end{equation}
where $m_D$ denotes the Dirac neutrino mass. The exact expression contains an additional phase space suppression factor, $\lambda^{1/2}$, where $\lambda$ is the Källén function.  On the other hand, the decay into two light neutrinos is additionally suppressed by two powers of the active--sterile mixing, with the corresponding decay width scaling as
\begin{equation}
\Gamma(\chi \rightarrow \nu\nu)
\sim
\alpha_R^2
\left(\frac{m_D}{M_N}\right)^4
m_\chi.
\end{equation}
Since $\alpha_R$ parametrizes a tiny explicit breaking of the $\mathbb{Z}_3$ symmetry, we choose a benchmark value of $\alpha_R \lesssim\mathcal{O}(10^{-20})$. Such a small value suppresses the decay widths of $\chi$ sufficiently, leading to a lifetime much longer than the age of the Universe. Therefore, despite the explicit breaking of the $\mathbb{Z}_3$ symmetry, the dark matter candidate $\chi$ remains effectively stable on cosmological timescales.

The pseudo Goldstone dark matter within the $\mathbb{Z}_3$ symmetric framework has been extensively studied in the context of collider phenomenology and the electroweak phase transition with direct detection signature in Refs.~\cite{Kannike:2019mzk,Kang:2017mkl,Ghosh:2024ing}.
In this case, however, no observable GW signal is produced from domain wall annihilation. Instead, the GW spectrum is generated solely by the electroweak phase transition, with its peak frequency shifted to lower values than that shown in Fig.\,\ref{fig:ptgw} owing to the lower nucleation temperature. Since the primary focus of the present work is the neutrino sector, particularly the realization of resonant leptogenesis, we do not pursue a detailed discussion of the DM phenomenology.
\section{Summary and Conclusion}
\label{sec:summary}
In this work, we have investigated the phenomenological implications of non-holomorphic $A_4$ modular symmetry within a type-I seesaw framework, with a particular focus on neutrino mass generation, resonant leptogenesis, and the production of a stochastic gravitational wave background from domain wall annihilation and its connection to the leptogenesis scale. We constructed a model in which the lepton flavor structure is governed by polyharmonic Maa\ss\ forms associated with the $A_4$ modular group, going 
beyond the conventional holomorphic modular symmetry framework and thereby avoiding the necessity of low-energy supersymmetry. The field content and 
charge assignments under $SU(2)_L \times U(1)_Y \times A_4$, together with the modular weights, were specified, and the resulting charged lepton and neutrino mass matrices were derived. Light neutrino masses are generated through the type-I seesaw mechanism, with the flavor structure entirely controlled by the weight $(-2)$ polyharmonic Maa\ss\ forms and the vacuum 
expectation value of the complex modulus $\tau$.

A comprehensive numerical scan over the model parameter space was performed to identify regions consistent with current neutrino oscillation data at the $3\sigma$ level, using the latest NuFIT~6.1 global fit results. The model was shown to successfully reproduce all six neutrino oscillation observables, namely the three leptonic mixing angles $\theta_{12}$, $\theta_{23}$, and $\theta_{13}$, the two mass-squared splittings $\Delta m^2_{21}$ and $\Delta m^2_{31}$, and the Dirac CP-violating phase $\delta_{CP}$, within the current experimental constraints for normal mass ordering. The allowed region of the modular parameter $\tau$ is strongly localized near the 
boundary of the fundamental domain, reflecting the constraining power of the modular symmetry.

We have also studied resonant leptogenesis to address the origin of the observed baryon asymmetry of the Universe. A distinctive feature of our framework is that, after diagonalising the Majorana mass matrix, the right-handed neutrinos exhibit a naturally quasi-degenerate mass spectrum with tiny mass splittings. These splittings resonantly enhance the self-energy contribution to the CP asymmetry when they become comparable to the corresponding decay widths, leading to the generation of a lepton asymmetry. The latter is subsequently converted into the observed baryon asymmetry through electroweak sphaleron processes. Thus, the tiny heavy-neutrino mass splittings required for successful resonant leptogenesis emerge naturally as a consequence of the underlying non-holomorphic modular symmetry, enhancing the predictive power of the framework.

Although the RHN sector plays a central role in neutrino mass generation and leptogenesis, its characteristic mass scale typically lies far beyond the reach of terrestrial experiments, making a direct experimental probe extremely challenging. This motivates the exploration of cosmological observables that can provide indirect access to the underlying RHN mass scale. In this context, we extend the framework by introducing a complex scalar field $\Phi$ charged under the discrete $\mathbb{Z}_3$ symmetry, whose spontaneous symmetry breaking establishes an indirect connection between the RHN mass scale and potentially observable gravitational wave (GW) signals. The spontaneous breaking of the discrete symmetry inevitably leads to the formation of DW networks, poses a severe cosmological problem, as such walls would come to dominate the energy density of the Universe. We addressed this challenge without invoking inflation by introducing a small explicit soft breaking of the $\mathbb{Z}_3$ symmetry via the $\Phi \overline{N^c}N$ term. This explicit breaking induces a bias energy between the otherwise degenerate vacua, which drives the eventual annihilation of the domain wall network. 
A key consequence of the relatively low domain-wall annihilation temperature is the generation of a stochastic GW background with a low-frequency peak, whose spectral properties are governed by the symmetry-breaking scale $v_\phi$ and the bias term generated from the heavy neutrino sector. We find that these GW signals can be probed over a broad range of symmetry-breaking scales by present and future GW observatories, including LISA, BBO, DECIGO, ET, CE, $\mu$ARES, and SKA.

We further investigated the GW signatures associated with the cosmological evolution of the model. The annihilation of the domain walls produces a stochastic GW background, while the high-scale FOPT, whose strength is governed by the tree-level cubic coupling $\mu_3$, generates an additional GW signal at much higher frequencies. Owing to the high nucleation temperature, the FOPT signal is shifted to frequencies ranging from a few hertz to a few hundred hertz, placing it within the projected sensitivity of future GW observatories such as DECIGO, ET, and CE.
A distinctive prediction of this framework is therefore a double-peak GW spectrum, with the low-frequency and high-frequency peaks originating from domain wall annihilation and the high-scale FOPT, respectively. The low-frequency peak probes the heavy Majorana neutrino sector and the leptogenesis scale, whereas the high-frequency peak is sensitive to the symmetry-breaking scale and the scalar sector. The simultaneous observation of both GW signals would provide complementary information on the underlying particle spectrum and cosmological history, offering a unique test of the framework.

In summary, we have presented a predictive framework based on non-holomorphic modular symmetry that unifies neutrino flavor, resonant leptogenesis, and gravitational wave phenomenology. A distinctive feature of the model is the natural emergence of tiny mass splittings among the heavy Majorana neutrinos after diagonalization of the Majorana mass matrix, providing the necessary ingredient for successful resonant leptogenesis. Furthermore, the cosmological evolution of the model predicts a characteristic double-peak gravitational wave spectrum, arising from domain-wall annihilation and the high-scale first-order phase transition.
%
\acknowledgments
JD acknowledges the ANRF (formerly Science and
Engineering Research Board), Government of
India, for the national postdoctoral fellowship (NPDF) grant PDF/2023/0015. NM acknowledges the Council for Scientific \& Industrial Research (CSIR), Government of India, for granting the Senior Research Fellowship.
\appendix

\section{Modular Yukawa couplings}
\label{app:modulerF}
We consider the modular forms $Y_{1}^{(-2)}$ \& $Y_{3,i}^{(-2)}$ $(i=1,2,3)$, which constitute an $A_4$ singlet and a triplet, respectively, both carrying the modular weight $-2$. Written in terms of the modulus $\tau=x+iy$ and the parameter $q=e^{2\pi i\tau}$, their $q$-series representations take the form

\begin{align}
Y^{(-2)}_{1}(\tau)
=&\;
\frac{y^3}{3}
-\frac{15\,\Gamma(3,4\pi y)}{4\pi^3\,q}
-\frac{135\,\Gamma(3,8\pi y)}{32\pi^3\,q^2}
-\frac{35\,\Gamma(3,12\pi y)}{9\pi^3\,q^3}
+\cdots
\nonumber\\[2mm]
&\;
-\frac{\pi\,\zeta(3)}{12\,\zeta(4)}
-\frac{15\,q}{2\pi^3}
-\frac{135\,q^2}{16\pi^3}
-\frac{70\,q^3}{9\pi^3}
-\frac{1095\,q^4}{128\pi^3}
-\frac{189\,q^5}{25\pi^3}
-\frac{35\,q^6}{4\pi^3}
+\cdots \, .
\end{align}

\begin{align}
Y^{(-2)}_{3,1}(\tau)
=&\;
\frac{y^3}{3}
+\frac{21\,\Gamma(3,4\pi y)}{16\pi^3\,q}
+\frac{189\,\Gamma(3,8\pi y)}{128\pi^3\,q^2}
+\frac{169\,\Gamma(3,12\pi y)}{144\pi^3\,q^3}
+\frac{1533\,\Gamma(3,16\pi y)}{1024\pi^3\,q^4}
+\cdots
\nonumber\\[2mm]
&\;
+\frac{\pi\,\zeta(3)}{40\,\zeta(4)}
+\frac{21\,q}{8\pi^3}
+\frac{189\,q^2}{64\pi^3}
+\frac{169\,q^3}{72\pi^3}
+\frac{1533\,q^4}{512\pi^3}
+\frac{1323\,q^5}{500\pi^3}
+\frac{169\,q^6}{64\pi^3}
+\cdots ,
\end{align}
\begin{align}
Y^{(-2)}_{3,2}(\tau)
=&\;
-\frac{729\,q^{1/3}}{16\pi^3}
\left(
\frac{\Gamma(3,8\pi y/3)}{16\,q}
+\frac{7\,\Gamma(3,20\pi y/3)}{125\,q^2}
+\frac{65\,\Gamma(3,32\pi y/3)}{1024\,q^3}
+\frac{74\,\Gamma(3,44\pi y/3)}{1331\,q^4}
+\cdots
\right)
\nonumber\\[2mm]
&\;
-\frac{81\,q^{1/3}}{16\pi^3}
\left(
1
+\frac{73\,q}{64}
+\frac{344\,q^2}{343}
+\frac{567\,q^3}{500}
+\frac{20198\,q^4}{2197}
+\frac{4681\,q^5}{4096}
+\cdots
\right) ,
\end{align}
\begin{align}
Y^{(-2)}_{3,3}(\tau)
=&\;
-\frac{81\,q^{2/3}}{32\pi^3}
\left(
\frac{\Gamma(3,4\pi y/3)}{q}
+\frac{73\,\Gamma(3,16\pi y/3)}{64\,q^2}
+\frac{344\,\Gamma(3,28\pi y/3)}{343\,q^3}
+\frac{567\,\Gamma(3,40\pi y/3)}{500\,q^4}
+\cdots
\right)
\nonumber\\[2mm]
&\;
-\frac{729\,q^{2/3}}{8\pi^3}
\left(
\frac{1}{16}
+\frac{7\,q}{125}
+\frac{65\,q^2}{1024}
+\frac{74\,q^3}{1331}
+\cdots
\right) .
\end{align}

\subsection*{Yukawa Matrices for BP3}
For BP3 of Tab.~\ref{tab:cpasy} which reproduces the correct BAU, the effective Yukawa matrix is

\begin{equation} h=
    \begin{pmatrix}
        -7.65\times10^{-5}+1.11\times10^{-9} i & -7.34\times10^{-8} + 4.71\times10^{-7}i& 4.85\times10^{-7} + 2.56\times10^{-7}i\\
        -1.15\times10^{-7} - 1.68\times10^{-7}i & -4.20\times10^{-5} - 8.05\times10^{-5}i& -8.03\times10^{-5} + 4.38\times10^{-5}i\\
        2.11\times10^{-7} + 4.46\times10^{-7}i & -1.55\times10^{-5} + 2.90\times10^{-5}i& -2.90\times10^{-5} - 1.55\times10^{-5}i
    \end{pmatrix}
\end{equation}

\section{High-T bias term}\label{app:thT}
The thermal fermionic function:
\begin{equation}
    J_{B}\left(\frac{m^2}{T^2}\right)=
    \int_0^\infty dx\,x^2
    \ln\!\left[1 - e^{-\sqrt{x^2+ \frac{m^2}{T^2}}}\right],
\end{equation}
In the high-temperature limit, $M_{N_i}^2 \ll T^2$, the thermal fermionic function can be expanded as
\begin{eqnarray}
J_F\left(\frac{M_{N_i}^2}{T^2}\right) = \frac{7\pi^4}{360} -\frac{\pi^2}{24}\frac{M_{N_i}^2}{T^2} -\frac{1}{32}\frac{M_{N_i}^4}{T^4}
\ln\left(\frac{M_{N_i}^2}{a_F T^2}\right)+\mathcal{O}\left(\frac{M_{N_i}^6}{T^6}\right).
\end{eqnarray}
where 
$a_F=\pi^2\exp\!\left(\frac{3}{2}-2\gamma_E\right)$. The bias term for this case can be written as
\begin{equation}\label{eq:vbiasT}
    \Delta V^T_{\rm bias} = \sum_{ j>i=1}^{3}V^{\rm high}_{T}\left(v_{\phi},\theta_{j}\right) - V^{\rm high}_{T}(v_{\phi},\theta_{i}),
\end{equation}
with
\begin{eqnarray}
V^{\rm high}_{T}(v_{\phi},\theta)
=\tilde{V}^{\rm high}_{T}-
\frac{T^2 v_\phi |K^{\rm high}_T|}{12\sqrt{2}}
\cos(\theta+\delta^{\rm high}_T)\,,
\end{eqnarray}
where
\begin{eqnarray}
K_T^{\rm high} \equiv {\rm Tr}\!\left(M_R^\dagger Y\right), \qquad \delta^{\rm high}_T =\text{Arg}\left(K^{\rm high}_T\right).
\end{eqnarray}

\label{Bibliography}
\bibliographystyle{JHEP}
\bibliography{Refs.bib}

@article{Fukuda:1998mi,
    author = "Fukuda, Y. and others",
    collaboration = "Super-Kamiokande",
    title = "{Evidence for oscillation of atmospheric neutrinos}",
    eprint = "hep-ex/9807003",
    archivePrefix = "arXiv",
    reportNumber = "BU-98-17, ICRR-REPORT-422-98-18, UCI-98-8, KEK-PREPRINT-98-95, LSU-HEPA-5-98, UMD-98-003, SBHEP-98-5, TKU-PAP-98-06, TIT-HPE-98-09",
    doi = "10.1103/PhysRevLett.81.1562",
    journal = "Phys. Rev. Lett.",
    volume = "81",
    pages = "1562--1567",
    year = "1998"
}

@article{Kobayashi:2018wkl,
    author = "Kobayashi, Tatsuo and Shimizu, Yusuke and Takagi, Kenta and Tanimoto, Morimitsu and Tatsuishi, Takuya H. and Uchida, Hikaru",
    title = "{Finite modular subgroups for fermion mass matrices and baryon/lepton number violation}",
    eprint = "1812.11072",
    archivePrefix = "arXiv",
    primaryClass = "hep-ph",
    reportNumber = "EPHOU-18-017, HUPD1810",
    doi = "10.1016/j.physletb.2019.05.034",
    journal = "Phys. Lett. B",
    volume = "794",
    pages = "114--121",
    year = "2019"
}

@article{Okada:2019xqk,
    author = "Okada, Hiroshi and Orikasa, Yuta",
    title = "{Modular $S_3$ symmetric radiative seesaw model}",
    eprint = "1907.04716",
    archivePrefix = "arXiv",
    primaryClass = "hep-ph",
    reportNumber = "APCTP Pre2019-017",
    doi = "10.1103/PhysRevD.100.115037",
    journal = "Phys. Rev. D",
    volume = "100",
    number = "11",
    pages = "115037",
    year = "2019"
}

@article{Mishra:2020gxg,
    author = "Mishra, Subhasmita",
    title = "{Neutrino mixing and Leptogenesis with modular $S_3$ symmetry in the framework of type III seesaw}",
    eprint = "2008.02095",
    archivePrefix = "arXiv",
    primaryClass = "hep-ph",
    month = "8",
    year = "2020"
}

@article{Kang:2026osw,
    author = "Kang, Sin Kyu and Kumar, Ranjeet and Okada, Hiroshi",
    title = "{Radiative Dirac Neutrino Masses from Modular $S_3$ Symmetry in an Axion Model}",
    eprint = "2601.22740",
    archivePrefix = "arXiv",
    primaryClass = "hep-ph",
    month = "1",
    year = "2026"
}

@article{Meloni:2023aru,
    author = "Meloni, Davide and Parriciatu, Matteo",
    title = "{A simplest modular S$_{3}$ model for leptons}",
    eprint = "2306.09028",
    archivePrefix = "arXiv",
    primaryClass = "hep-ph",
    doi = "10.1007/JHEP09(2023)043",
    journal = "JHEP",
    volume = "09",
    pages = "043",
    year = "2023"
}

@article{Marciano:2024nwm,
    author = "Marciano, Simone and Meloni, Davide and Parriciatu, Matteo",
    title = "{Minimal seesaw and leptogenesis with the smallest modular finite group}",
    eprint = "2402.18547",
    archivePrefix = "arXiv",
    primaryClass = "hep-ph",
    doi = "10.1007/JHEP05(2024)020",
    journal = "JHEP",
    volume = "05",
    pages = "020",
    year = "2024"
}

@article{Belfkir:2024wdn,
    author = "Belfkir, Mohamed and Loualidi, Mohamed Amin and Nasri, Salah",
    title = "{Fermion Masses and Mixing in Pati{\textendash}Salam Unification with~$S_3$~Modular Symmetry}",
    eprint = "2501.00302",
    archivePrefix = "arXiv",
    primaryClass = "hep-ph",
    doi = "10.1093/ptep/ptaf032",
    journal = "PTEP",
    volume = "2025",
    number = "3",
    pages = "033B05",
    year = "2025"
}

@article{Nomura:2023usj,
    author = "Nomura, Takaaki and Okada, Hiroshi and Otsuka, Hajime",
    title = "{Texture zeros realization in a three-loop radiative neutrino mass model from modular A4 symmetry}",
    eprint = "2309.13921",
    archivePrefix = "arXiv",
    primaryClass = "hep-ph",
    reportNumber = "KYUSHU-HET-268",
    doi = "10.1016/j.nuclphysb.2024.116579",
    journal = "Nucl. Phys. B",
    volume = "1004",
    pages = "116579",
    year = "2024"
}

@article{RickyDevi:2024ijc,
    author = "Ricky Devi, Maibam",
    title = "{Neutrino Masses and Higher Degree Siegel Modular Forms}",
    eprint = "2401.16257",
    archivePrefix = "arXiv",
    primaryClass = "hep-ph",
    month = "1",
    year = "2024"
}

@article{Gogoi:2023jzl,
    author = "Gogoi, Jotin and Sarma, Lavina and Das, Mrinal Kumar",
    title = "{Leptogenesis and dark matter in minimal inverse seesaw using $A_4$ modular symmetry}",
    eprint = "2311.09883",
    archivePrefix = "arXiv",
    primaryClass = "hep-ph",
    doi = "10.1140/epjc/s10052-024-13029-5",
    journal = "Eur. Phys. J. C",
    volume = "84",
    number = "7",
    pages = "689",
    year = "2024"
}

@article{Pathak:2024sei,
    author = "Pathak, Gourab and Das, Pritam and Das, Mrinal Kumar",
    title = "{Neutrino mass genesis in scoto-inverse seesaw with modular $A_4$}",
    eprint = "2411.13895",
    archivePrefix = "arXiv",
    primaryClass = "hep-ph",
    doi = "10.1140/epjc/s10052-025-14263-1",
    journal = "Eur. Phys. J. C",
    volume = "85",
    number = "5",
    pages = "569",
    year = "2025"
}

@article{Nomura:2024vus,
    author = "Nomura, Takaaki and Okada, Hiroshi",
    title = "{A More Novel Approach of Radiative Linear Seesaw in a Modular A4 Symmetry}",
    eprint = "2410.21843",
    archivePrefix = "arXiv",
    primaryClass = "hep-ph",
    doi = "10.1093/ptep/ptaf044",
    journal = "PTEP",
    volume = "2025",
    number = "4",
    pages = "043B04",
    year = "2025"
}

@article{Kashav:2021zir,
    author = "Kashav, Monal and Verma, Surender",
    title = "{Broken scaling neutrino mass matrix and leptogenesis based on A$_{4}$ modular invariance}",
    eprint = "2103.07207",
    archivePrefix = "arXiv",
    primaryClass = "hep-ph",
    doi = "10.1007/JHEP09(2021)100",
    journal = "JHEP",
    volume = "09",
    pages = "100",
    year = "2021"
}

@article{Kashav:2022kpk,
    author = "Kashav, Monal and Verma, Surender",
    title = "{On minimal realization of topological Lorentz structures with one-loop seesaw extensions in A$_{4}$ modular symmetry}",
    eprint = "2205.06545",
    archivePrefix = "arXiv",
    primaryClass = "hep-ph",
    doi = "10.1088/1475-7516/2023/03/010",
    journal = "JCAP",
    volume = "03",
    pages = "010",
    year = "2023"
}

@article{Kobayashi:2019gtp,
    author = "Kobayashi, Tatsuo and Nomura, Takaaki and Shimomura, Takashi",
    title = "{Type II seesaw models with modular $A_4$ symmetry}",
    eprint = "1912.00637",
    archivePrefix = "arXiv",
    primaryClass = "hep-ph",
    reportNumber = "EPHOU-19-018, KIAS-P19068, UME-PP-011",
    doi = "10.1103/PhysRevD.102.035019",
    journal = "Phys. Rev. D",
    volume = "102",
    number = "3",
    pages = "035019",
    year = "2020"
}

@article{Nomura:2023kwz,
    author = "Nomura, Takaaki and Okada, Hiroshi",
    title = "{Quark and lepton model with flavor specific dark matter and muon \(g - 2\) in modular \(A_{4}\) and hidden \(U(1)\) symmetries}",
    eprint = "2304.13361",
    archivePrefix = "arXiv",
    primaryClass = "hep-ph",
    doi = "10.1016/j.dark.2025.101986",
    journal = "Phys. Dark Univ.",
    volume = "49",
    pages = "101986",
    year = "2025"
}

@article{Kim:2023jto,
    author = "Kim, Jongkuk and Okada, Hiroshi",
    title = "{Fermi-LAT GeV excess and muon $g-2$ in a modular $A_4$ symmetry}",
    eprint = "2302.09747",
    archivePrefix = "arXiv",
    primaryClass = "hep-ph",
    month = "2",
    year = "2023"
}

@article{Devi:2023vpe,
    author = "Devi, Maibam Ricky",
    title = "{Retrieving texture zeros in 3+1 active-sterile neutrino framework under the action of $A_4$ modular-invariants}",
    eprint = "2303.04900",
    archivePrefix = "arXiv",
    primaryClass = "hep-ph",
    month = "3",
    year = "2023"
}

@article{Behera:2024ark,
    author = "Behera, Mitesh Kumar and Ittisamai, Pawin and Pongkitivanichkul, Chakrit and Uttayarat, Patipan",
    title = "{Neutrino phenomenology in the modular S3 seesaw model}",
    eprint = "2403.00593",
    archivePrefix = "arXiv",
    primaryClass = "hep-ph",
    doi = "10.1103/PhysRevD.110.035004",
    journal = "Phys. Rev. D",
    volume = "110",
    number = "3",
    pages = "035004",
    year = "2024"
}

@article{Behera:2025tpj,
    author = "Behera, Mitesh Kumar and Ittisamai, Pawin and Pongkitivanichkul, Chakrit and Uttayarat, Patipan",
    title = "{Phenomenology of inverse seesaw using $S_3$ modular symmetry}",
    eprint = "2504.12954",
    archivePrefix = "arXiv",
    primaryClass = "hep-ph",
    doi = "10.1140/epjc/s10052-025-15017-9",
    journal = "Eur. Phys. J. C",
    volume = "85",
    number = "11",
    pages = "1316",
    year = "2025"
}

@article{Dasgupta:2021ggp,
    author = "Dasgupta, Arnab and Nomura, Takaaki and Okada, Hiroshi and Popov, Oleg and Tanimoto, Morimitsu",
    title = "{Dirac Radiative Neutrino Mass with Modular Symmetry and Leptogenesis}",
    eprint = "2111.06898",
    archivePrefix = "arXiv",
    primaryClass = "hep-ph",
    reportNumber = "APCTP Pre2021 - 029, CTP-SCU/2021033",
    month = "11",
    year = "2021"
}

@article{CentellesChulia:2023osj,
    author = "Centelles Chuli{\'a}, Salvador and Kumar, Ranjeet and Popov, Oleg and Srivastava, Rahul",
    title = "{Neutrino mass sum rules from modular A4 symmetry}",
    eprint = "2308.08981",
    archivePrefix = "arXiv",
    primaryClass = "hep-ph",
    doi = "10.1103/PhysRevD.109.035016",
    journal = "Phys. Rev. D",
    volume = "109",
    number = "3",
    pages = "035016",
    year = "2024"
}

@article{deMedeirosVarzielas:2023crv,
    author = "de Medeiros Varzielas, I. and Levy, M. and Penedo, J. T. and Petcov, S. T.",
    title = "{Quarks at the modular S$_{4}$ cusp}",
    eprint = "2307.14410",
    archivePrefix = "arXiv",
    primaryClass = "hep-ph",
    reportNumber = "SISSA 11/2023/FISI, CFTP/23-002",
    doi = "10.1007/JHEP09(2023)196",
    journal = "JHEP",
    volume = "09",
    pages = "196",
    year = "2023"
}

@article{Ding:2021zbg,
    author = "Ding, Gui-Jun and King, Stephen F. and Yao, Chang-Yuan",
    title = "{Modular $S_4\times SU(5)$ GUT}",
    eprint = "2103.16311",
    archivePrefix = "arXiv",
    primaryClass = "hep-ph",
    reportNumber = "USTC-ICTS/PCFT-21-13",
    doi = "10.1103/PhysRevD.104.055034",
    journal = "Phys. Rev. D",
    volume = "104",
    number = "5",
    pages = "055034",
    year = "2021"
}

@article{King:2019vhv,
    author = "King, Stephen F. and Zhou, Ye-Ling",
    title = "{Trimaximal TM$_1$ mixing with two modular $S_4$ groups}",
    eprint = "1908.02770",
    archivePrefix = "arXiv",
    primaryClass = "hep-ph",
    doi = "10.1103/PhysRevD.101.015001",
    journal = "Phys. Rev. D",
    volume = "101",
    number = "1",
    pages = "015001",
    year = "2020"
}

@article{deMedeirosVarzielas:2022ihu,
    author = "de Medeiros Varzielas, Ivo and Louren{\c{c}}o, Jo{\~a}o",
    title = "{Two A5 modular symmetries for Golden Ratio 2 mixing}",
    eprint = "2206.14869",
    archivePrefix = "arXiv",
    primaryClass = "hep-ph",
    doi = "10.1016/j.nuclphysb.2022.115974",
    journal = "Nucl. Phys. B",
    volume = "984",
    pages = "115974",
    year = "2022"
}

@article{Yao:2020zml,
    author = "Yao, Chang-Yuan and Liu, Xiang-Gan and Ding, Gui-Jun",
    title = "{Fermion masses and mixing from the double cover and metaplectic cover of the $A_5$ modular group}",
    eprint = "2011.03501",
    archivePrefix = "arXiv",
    primaryClass = "hep-ph",
    reportNumber = "USTC-ICTS/PCFT-20-36",
    doi = "10.1103/PhysRevD.103.095013",
    journal = "Phys. Rev. D",
    volume = "103",
    number = "9",
    pages = "095013",
    year = "2021"
}

@article{Fukuda:2001nj,
    author = "Fukuda, S. and others",
    collaboration = "Super-Kamiokande",
    title = "{Solar B-8 and hep neutrino measurements from 1258 days of Super-Kamiokande data}",
    eprint = "hep-ex/0103032",
    archivePrefix = "arXiv",
    doi = "10.1103/PhysRevLett.86.5651",
    journal = "Phys. Rev. Lett.",
    volume = "86",
    pages = "5651--5655",
    year = "2001"
}

@article{DiValentino:2019dzu,
    author = "Di Valentino, Eleonora and Melchiorri, Alessandro and Silk, Joseph",
    title = "{Cosmological constraints in extended parameter space from the Planck 2018 Legacy release}",
    eprint = "1908.01391",
    archivePrefix = "arXiv",
    primaryClass = "astro-ph.CO",
    doi = "10.1088/1475-7516/2020/01/013",
    journal = "JCAP",
    volume = "01",
    pages = "013",
    year = "2020"
}

@article{Ahmad:2001an,
    author = "Ahmad, Q. R. and others",
    collaboration = "SNO",
    title = "{Measurement of the rate of $\nu_e+d \to p+p+e^-$ interactions produced by $^8$B solar neutrinos at the Sudbury Neutrino Observatory}",
    eprint = "nucl-ex/0106015",
    archivePrefix = "arXiv",
    reportNumber = "UPR-0240E",
    doi = "10.1103/PhysRevLett.87.071301",
    journal = "Phys. Rev. Lett.",
    volume = "87",
    pages = "071301",
    year = "2001"
}

@article{Ahmad:2002jz,
    author = "Ahmad, Q. R. and others",
    collaboration = "SNO",
    title = "{Direct evidence for neutrino flavor transformation from neutral current interactions in the Sudbury Neutrino Observatory}",
    eprint = "nucl-ex/0204008",
    archivePrefix = "arXiv",
    doi = "10.1103/PhysRevLett.89.011301",
    journal = "Phys. Rev. Lett.",
    volume = "89",
    pages = "011301",
    year = "2002"
}

@article{Esteban:2020cvm,
    author = "Esteban, Ivan and Gonzalez-Garcia, M. C. and Maltoni, Michele and Schwetz, Thomas and Zhou, Albert",
    title = "{The fate of hints: updated global analysis of three-flavor neutrino oscillations}",
    eprint = "2007.14792",
    archivePrefix = "arXiv",
    primaryClass = "hep-ph",
    reportNumber = "IFT-UAM/CSIC-112, YITP-SB-2020-21",
    doi = "10.1007/JHEP09(2020)178",
    journal = "JHEP",
    volume = "09",
    pages = "178",
    year = "2020"
}

@article{Ma:2001dn,
    author = "Ma, Ernest and Rajasekaran, G.",
    title = "{Softly broken A(4) symmetry for nearly degenerate neutrino masses}",
    eprint = "hep-ph/0106291",
    archivePrefix = "arXiv",
    reportNumber = "UCRHEP-T308",
    doi = "10.1103/PhysRevD.64.113012",
    journal = "Phys. Rev. D",
    volume = "64",
    pages = "113012",
    year = "2001"
}

@article{Cremades:2004wa,
    author = "Cremades, D. and Ibanez, L. E. and Marchesano, F.",
    title = "{Computing Yukawa couplings from magnetized extra dimensions}",
    eprint = "hep-th/0404229",
    archivePrefix = "arXiv",
    reportNumber = "FTUAM-04-7, IFT-UAM-CSIC-04-15, MAD-TH-04-4",
    doi = "10.1088/1126-6708/2004/05/079",
    journal = "JHEP",
    volume = "05",
    pages = "079",
    year = "2004"
}

@article{An:2012eh,
    author = "An, F. P. and others",
    collaboration = "Daya Bay",
    title = "{Observation of electron-antineutrino disappearance at Daya Bay}",
    eprint = "1203.1669",
    archivePrefix = "arXiv",
    primaryClass = "hep-ex",
    doi = "10.1103/PhysRevLett.108.171803",
    journal = "Phys. Rev. Lett.",
    volume = "108",
    pages = "171803",
    year = "2012"
}

@article{Ahn:2012nd,
    author = "Ahn, J. K. and others",
    collaboration = "RENO",
    title = "{Observation of Reactor Electron Antineutrino Disappearance in the RENO Experiment}",
    eprint = "1204.0626",
    archivePrefix = "arXiv",
    primaryClass = "hep-ex",
    doi = "10.1103/PhysRevLett.108.191802",
    journal = "Phys. Rev. Lett.",
    volume = "108",
    pages = "191802",
    year = "2012"
}

@inbook{Feruglio:2017spp,
    author = "Feruglio, Ferruccio",
    editor = "Levy, Aharon and Forte, Stefano and Ridolfi, Giovanni",
    title = "{Are neutrino masses modular forms?}",
    booktitle = "{From My Vast Repertoire ...}: {Guido Altarelli's Legacy}",
    eprint = "1706.08749",
    archivePrefix = "arXiv",
    primaryClass = "hep-ph",
    reportNumber = "DFPD-2017-TH-09",
    doi = "10.1142/9789813238053_0012",
    pages = "227--266",
    year = "2019"
}

@article{Hiramatsu:2013qaa,
    author = "Hiramatsu, Takashi and Kawasaki, Masahiro and Saikawa, Ken'ichi",
    title = "{On the estimation of gravitational wave spectrum from cosmic domain walls}",
    eprint = "1309.5001",
    archivePrefix = "arXiv",
    primaryClass = "astro-ph.CO",
    reportNumber = "ICRR-REPORT-659-2013-8, IPMU13-0182, YITP-13-87",
    doi = "10.1088/1475-7516/2014/02/031",
    journal = "JCAP",
    volume = "02",
    pages = "031",
    year = "2014"
}

@article{Singh:2024imk,
    author = "Singh, Labh and Kashav, Monal and Verma, Surender",
    title = "{Minimal type-I Dirac seesaw and leptogenesis under A4 modular invariance}",
    eprint = "2405.07165",
    archivePrefix = "arXiv",
    primaryClass = "hep-ph",
    doi = "10.1016/j.nuclphysb.2024.116666",
    journal = "Nucl. Phys. B",
    volume = "1007",
    pages = "116666",
    year = "2024"
}

@article{Qu:2024rns,
    author = "Qu, Bu-Yao and Ding, Gui-Jun",
    title = "{Non-holomorphic modular flavor symmetry}",
    eprint = "2406.02527",
    archivePrefix = "arXiv",
    primaryClass = "hep-ph",
    doi = "10.1007/JHEP08(2024)136",
    journal = "JHEP",
    volume = "08",
    pages = "136",
    year = "2024"
}

@article{deSalas:2020pgw,
    author = "de Salas, P. F. and Forero, D. V. and Gariazzo, S. and Mart{\'\i}nez-Mirav{\'e}, P. and Mena, O. and Ternes, C. A. and T{\'o}rtola, M. and Valle, J. W. F.",
    title = "{2020 global reassessment of the neutrino oscillation picture}",
    eprint = "2006.11237",
    archivePrefix = "arXiv",
    primaryClass = "hep-ph",
    doi = "10.1007/JHEP02(2021)071",
    journal = "JHEP",
    volume = "02",
    pages = "071",
    year = "2021"
}

@article{Capozzi:2021fjo,
    author = "Capozzi, Francesco and Di Valentino, Eleonora and Lisi, Eligio and Marrone, Antonio and Melchiorri, Alessandro and Palazzo, Antonio",
    title = "{Unfinished fabric of the three neutrino paradigm}",
    eprint = "2107.00532",
    archivePrefix = "arXiv",
    primaryClass = "hep-ph",
    doi = "10.1103/PhysRevD.104.083031",
    journal = "Phys. Rev. D",
    volume = "104",
    number = "8",
    pages = "083031",
    year = "2021"
}

@article{Esteban:2024eli,
    author = "Esteban, Ivan and Gonzalez-Garcia, M. C. and Maltoni, Michele and Martinez-Soler, Ivan and Pinheiro, Jo{\~a}o Paulo and Schwetz, Thomas",
    title = "{NuFit-6.0: updated global analysis of three-flavor neutrino oscillations}",
    eprint = "2410.05380",
    archivePrefix = "arXiv",
    primaryClass = "hep-ph",
    reportNumber = "IFT-UAM/CSIC-24-140, YITP-SB-2024-24, IPPP/24/64, IPPP/24/64, IFT-UAM/CSIC-24-140, YITP-SB-2024-24",
    doi = "10.1007/JHEP12(2024)216",
    journal = "JHEP",
    volume = "12",
    pages = "216",
    year = "2024"
}

@article{Planck:2018vyg,
    author = "Aghanim, N. and others",
    collaboration = "Planck",
    title = "{Planck 2018 results. VI. Cosmological parameters}",
    eprint = "1807.06209",
    archivePrefix = "arXiv",
    primaryClass = "astro-ph.CO",
    doi = "10.1051/0004-6361/201833910",
    journal = "Astron. Astrophys.",
    volume = "641",
    pages = "A6",
    year = "2020",
    note = "[Erratum: Astron.Astrophys. 652, C4 (2021)]"
}

@article{Hahn:2006hr,
    author = "Hahn, T.",
    title = "{Routines for the diagonalization of complex matrices}",
    eprint = "physics/0607103",
    archivePrefix = "arXiv",
    reportNumber = "MPP-2006-85",
    month = "7",
    year = "2006"
}

@article{Hochmuth:2007wq,
    author = "Hochmuth, K. A. and Petcov, S. T. and Rodejohann, W.",
    title = "{U(PMNS) = U**dagger (l) U(nu)}",
    eprint = "0706.2975",
    archivePrefix = "arXiv",
    primaryClass = "hep-ph",
    reportNumber = "MPP-2007-27, SISSA-44-2007-EP",
    doi = "10.1016/j.physletb.2007.08.072",
    journal = "Phys. Lett. B",
    volume = "654",
    pages = "177--188",
    year = "2007"
}

@online{iminuit,
	author = {Piti Ongmongkolkul (@piti118) and Christoph Deil (@cdeil) and Hans Dembinski (@HDembinski) and @Dapid and Chris Burr (@chrisburr) and Andrew (@energynumbers) and Fabian Rost (@fabianrost84) and Alex Pearce (@alexpearce) and Lukas Geiger (@lgeiger) and Omar Zapata (@omazapa)},
	title = {iminuit - MINUIT from Python},
	year = {2012--},
	url = {https://github.com/iminuit/iminuit},
	note = {[Online; accessed 2018.03.05]}
}

@article{Nomura:2025ovm,
    author = "Nomura, Takaaki and Okada, Hiroshi and Wang, Xing-Yu",
    title = "{A radiative neutrino mass model with leptoquarks under non-holomorphic modular A$_{4}$ symmetry}",
    eprint = "2504.21404",
    archivePrefix = "arXiv",
    primaryClass = "hep-ph",
    doi = "10.1007/JHEP09(2025)163",
    journal = "JHEP",
    volume = "09",
    pages = "163",
    year = "2025"
}

@article{Nomura:2025raf,
    author = "Nomura, Takaaki and Okada, Hiroshi",
    title = "{Neutrino mass model at a three-loop level from a non-holomorphic modular A $_{4}$ symmetry}",
    eprint = "2506.02639",
    archivePrefix = "arXiv",
    primaryClass = "hep-ph",
    doi = "10.1088/1674-1137/ae15ee",
    journal = "Chin. Phys. C",
    volume = "50",
    number = "2",
    pages = "023108",
    year = "2026"
}

@article{Nomura:2024nwh,
    author = "Nomura, Takaaki and Okada, Hiroshi",
    title = "{Zee model in a non-holomorphic modular A4 symmetry}",
    eprint = "2412.18095",
    archivePrefix = "arXiv",
    primaryClass = "hep-ph",
    doi = "10.1016/j.physletb.2025.139618",
    journal = "Phys. Lett. B",
    volume = "867",
    pages = "139618",
    year = "2025"
}

@article{ParticleDataGroup:2024cfk,
    author = "Navas, S. and others",
    collaboration = "Particle Data Group",
    title = "{Review of particle physics}",
    doi = "10.1103/PhysRevD.110.030001",
    journal = "Phys. Rev. D",
    volume = "110",
    number = "3",
    pages = "030001",
    year = "2024"
}

@article{Cohen:1990it,
    author = "Cohen, Andrew G. and Kaplan, David B. and Nelson, Ann E.",
    title = "{Baryogenesis at the weak phase transition}",
    reportNumber = "NSF-ITP-90-85, UCSD-PTH-90-09, BUHEP-90-15",
    doi = "10.1016/0550-3213(91)90395-E",
    journal = "Nucl. Phys. B",
    volume = "349",
    pages = "727--742",
    year = "1991"
}

@article{Turok:1990in,
    author = "Turok, Neil and Zadrozny, John",
    title = "{Dynamical generation of baryons at the electroweak transition}",
    reportNumber = "PUPT-90-1183",
    doi = "10.1103/PhysRevLett.65.2331",
    journal = "Phys. Rev. Lett.",
    volume = "65",
    pages = "2331--2334",
    year = "1990"
}

@article{Dine:1990fj,
    author = "Dine, Michael and Huet, Patrick and Singleton, Jr, Robert L. and Susskind, Leonard",
    title = "{Creating the baryon asymmetry at the electroweak phase transition}",
    reportNumber = "SCIPP-90-31, SLAC-PUB-5386",
    doi = "10.1016/0370-2693(91)91905-B",
    journal = "Phys. Lett. B",
    volume = "257",
    pages = "351--356",
    year = "1991"
}

@article{Cohen:1991iu,
    author = "Cohen, Andrew G. and Kaplan, D. B. and Nelson, A. E.",
    title = "{Spontaneous baryogenesis at the weak phase transition}",
    reportNumber = "UCSD-PTH-91-11, BUHEP-91-5",
    doi = "10.1016/0370-2693(91)91711-4",
    journal = "Phys. Lett. B",
    volume = "263",
    pages = "86--92",
    year = "1991"
}

@article{Trodden:1998ym,
    author = "Trodden, Mark",
    title = "{Electroweak baryogenesis}",
    eprint = "hep-ph/9803479",
    archivePrefix = "arXiv",
    reportNumber = "CWRU-P6-98",
    doi = "10.1103/RevModPhys.71.1463",
    journal = "Rev. Mod. Phys.",
    volume = "71",
    pages = "1463--1500",
    year = "1999"
}

@article{Affleck:1984fy,
    author = "Affleck, Ian and Dine, Michael",
    title = "{A New Mechanism for Baryogenesis}",
    reportNumber = "Print-84-0574 (PRINCETON)",
    doi = "10.1016/0550-3213(85)90021-5",
    journal = "Nucl. Phys. B",
    volume = "249",
    pages = "361--380",
    year = "1985"
}

@article{Sakharov:1967dj,
      author         = "Sakharov, A. D.",
      title          = "{Violation of CP Invariance, C asymmetry, and baryon
                        asymmetry of the universe}",
      journal        = "Pisma Zh. Eksp. Teor. Fiz.",
      volume         = "5",
      year           = "1967",
      pages          = "32-35",
      doi            = "10.1070/PU1991v034n05ABEH002497",
      note           = "[Usp. Fiz. Nauk161,no.5,61(1991)]",
      SLACcitation   = "%%CITATION = ZFPRA,5,32;%%"
}

@article{Fukugita:1986hr,
      author         = "Fukugita, M. and Yanagida, T.",
      title          = "{Baryogenesis Without Grand Unification}",
      journal        = "Phys. Lett.",
      volume         = "B174",
      year           = "1986",
      pages          = "45-47",
      doi            = "10.1016/0370-2693(86)91126-3",
      reportNumber   = "RIFP-641",
      SLACcitation   = "%%CITATION = PHLTA,B174,45;%%"
}

@article{Kuzmin:1985mm,
      author         = "Kuzmin, V. A. and Rubakov, V. A. and Shaposhnikov, M. E.",
      title          = "{On the Anomalous Electroweak Baryon Number
                        Nonconservation in the Early Universe}",
      journal        = "Phys. Lett.",
      volume         = "155B",
      year           = "1985",
      pages          = "36",
      doi            = "10.1016/0370-2693(85)91028-7",
      reportNumber   = "IC/85/8",
      SLACcitation   = "%%CITATION = PHLTA,155B,36;%%"
}

@article{Zhang:2025dsa,
    author = "Zhang, Xianshuo and Reyimuaji, Yakefu",
    title = "{Inverse seesaw model in nonholomorphic modular A4 flavor symmetry}",
    eprint = "2507.06945",
    archivePrefix = "arXiv",
    primaryClass = "hep-ph",
    doi = "10.1103/17p3-bw5r",
    journal = "Phys. Rev. D",
    volume = "112",
    number = "7",
    pages = "075050",
    year = "2025"
}

@article{Thomas:2022hyj,
    author = "Thomas, Arun Mathew and Choudhury, Debajyoti",
    title = "{Neutron oscillation and baryogenesis from six dimensions}",
    eprint = "2205.03846",
    archivePrefix = "arXiv",
    primaryClass = "hep-ph",
    doi = "10.1103/PhysRevD.106.L031701",
    journal = "Phys. Rev. D",
    volume = "106",
    number = "3",
    pages = "L031701",
    year = "2022"
}

@article{Pilaftsis:1997jf,
    author = "Pilaftsis, Apostolos",
    title = "{CP violation and baryogenesis due to heavy Majorana neutrinos}",
    eprint = "hep-ph/9707235",
    archivePrefix = "arXiv",
    reportNumber = "MPI-PHT-97-30",
    doi = "10.1103/PhysRevD.56.5431",
    journal = "Phys. Rev. D",
    volume = "56",
    pages = "5431--5451",
    year = "1997"
}

@article{Pilaftsis:2003gt,
    author = "Pilaftsis, Apostolos and Underwood, Thomas E. J.",
    title = "{Resonant leptogenesis}",
    eprint = "hep-ph/0309342",
    archivePrefix = "arXiv",
    reportNumber = "MC-TH-2003-09",
    doi = "10.1016/j.nuclphysb.2004.05.029",
    journal = "Nucl. Phys. B",
    volume = "692",
    pages = "303--345",
    year = "2004"
}

@article{Dev:2017wwc,
    author = "Dev, Bhupal and Garny, Mathias and Klaric, Juraj and Millington, Peter and Teresi, Daniele",
    title = "{Resonant enhancement in leptogenesis}",
    eprint = "1711.02863",
    archivePrefix = "arXiv",
    primaryClass = "hep-ph",
    reportNumber = "TUM-HEP-1110-17",
    doi = "10.1142/S0217751X18420034",
    journal = "Int. J. Mod. Phys. A",
    volume = "33",
    pages = "1842003",
    year = "2018"
}

@article{Hugle:2018qbw,
    author = "Hugle, Thomas and Platscher, Moritz and Schmitz, Kai",
    title = "{Low-Scale Leptogenesis in the Scotogenic Neutrino Mass Model}",
    eprint = "1804.09660",
    archivePrefix = "arXiv",
    primaryClass = "hep-ph",
    doi = "10.1103/PhysRevD.98.023020",
    journal = "Phys. Rev. D",
    volume = "98",
    number = "2",
    pages = "023020",
    year = "2018"
}

@article{Alanne:2018brf,
    author = "Alanne, Tommi and Hugle, Thomas and Platscher, Moritz and Schmitz, Kai",
    title = "{Low-scale leptogenesis assisted by a real scalar singlet}",
    eprint = "1812.04421",
    archivePrefix = "arXiv",
    primaryClass = "hep-ph",
    doi = "10.1088/1475-7516/2019/03/037",
    journal = "JCAP",
    volume = "03",
    pages = "037",
    year = "2019"
}

@article{Kusenko:2014uta,
    author = "Kusenko, Alexander and Schmitz, Kai and Yanagida, Tsutomu T.",
    title = "{Leptogenesis via Axion Oscillations after Inflation}",
    eprint = "1412.2043",
    archivePrefix = "arXiv",
    primaryClass = "hep-ph",
    reportNumber = "IPMU-14-0352",
    doi = "10.1103/PhysRevLett.115.011302",
    journal = "Phys. Rev. Lett.",
    volume = "115",
    number = "1",
    pages = "011302",
    year = "2015"
}

@article{Hambye:2016sby,
    author = "Hambye, Thomas and Teresi, Daniele",
    title = "{Higgs doublet decay as the origin of the baryon asymmetry}",
    eprint = "1606.00017",
    archivePrefix = "arXiv",
    primaryClass = "hep-ph",
    reportNumber = "ULB-TH-16-08",
    doi = "10.1103/PhysRevLett.117.091801",
    journal = "Phys. Rev. Lett.",
    volume = "117",
    number = "9",
    pages = "091801",
    year = "2016"
}

@article{Datta:2022jic,
    author = "Datta, Arghyajit and Roshan, Rishav and Sil, Arunansu",
    title = "{Effects of Reheating on Charged Lepton Yukawa Equilibration and Leptogenesis}",
    eprint = "2206.10650",
    archivePrefix = "arXiv",
    primaryClass = "hep-ph",
    doi = "10.1103/PhysRevLett.132.061802",
    journal = "Phys. Rev. Lett.",
    volume = "132",
    number = "6",
    pages = "061802",
    year = "2024"
}

@article{Bhandari:2023wit,
    author = "Bhandari, Dipendu and Datta, Arghyajit and Sil, Arunansu",
    title = "{Leptogenesis from a phase transition in a dynamical vacuum}",
    eprint = "2312.13157",
    archivePrefix = "arXiv",
    primaryClass = "hep-ph",
    doi = "10.1103/PhysRevD.110.115008",
    journal = "Phys. Rev. D",
    volume = "110",
    number = "11",
    pages = "115008",
    year = "2024"
}

@article{King:2024idj,
    author = "King, Stephen F. and Manna, Soumen Kumar and Roshan, Rishav and Sil, Arunansu",
    title = "{Leptogenesis with Majoron dark matter}",
    eprint = "2412.14121",
    archivePrefix = "arXiv",
    primaryClass = "hep-ph",
    doi = "10.1103/PhysRevD.111.095008",
    journal = "Phys. Rev. D",
    volume = "111",
    number = "9",
    pages = "095008",
    year = "2025"
}

@article{Bhattacharya:2023kws,
    author = "Bhattacharya, Subhaditya and Mondal, Niloy and Roshan, Rishav and Vatsyayan, Drona",
    title = "{Leptogenesis, dark matter and gravitational waves from discrete symmetry breaking}",
    eprint = "2312.15053",
    archivePrefix = "arXiv",
    primaryClass = "hep-ph",
    doi = "10.1088/1475-7516/2024/06/029",
    journal = "JCAP",
    volume = "06",
    pages = "029",
    year = "2024"
}

@article{Bhattacharya:2024ohh,
    author = "Bhattacharya, Subhaditya and Mahanta, Devabrat and Mondal, Niloy and Pradhan, Dipankar",
    title = "{Two-component dark matter and low scale thermal Leptogenesis}",
    eprint = "2412.21202",
    archivePrefix = "arXiv",
    primaryClass = "hep-ph",
    doi = "10.1088/1475-7516/2025/09/032",
    journal = "JCAP",
    volume = "09",
    pages = "032",
    year = "2025"
}

@article{Bhattacharya:2025uaa,
    author = "Bhattacharya, Subhaditya and Mondal, Niloy and Sil, Arunansu",
    title = "{Exploring Leptogenesis, WIMP Dark Matter, and Gravitational Waves in an extended Scalar Framework}",
    eprint = "2512.02672",
    archivePrefix = "arXiv",
    primaryClass = "hep-ph",
    month = "12",
    year = "2025"
}

@article{Barman:2026yhc,
    author = "Barman, Basabendu and Basu, Arindam and Borah, Debasish and Das, Nayan",
    title = "{Leptogenesis with sub-electroweak-scale reheating temperature}",
    eprint = "2607.09282",
    archivePrefix = "arXiv",
    primaryClass = "hep-ph",
    month = "7",
    year = "2026"
}

@article{Kang:2022psa,
    author = "Kang, Dong Woo and Kim, Jongkuk and Nomura, Takaaki and Okada, Hiroshi",
    title = "{Natural mass hierarchy among three heavy Majorana neutrinos for resonant leptogenesis under modular A$_{4}$ symmetry}",
    eprint = "2205.08269",
    archivePrefix = "arXiv",
    primaryClass = "hep-ph",
    reportNumber = "KIAS-P22026, APCTP Pre2022 - 005",
    doi = "10.1007/JHEP07(2022)050",
    journal = "JHEP",
    volume = "07",
    pages = "050",
    year = "2022"
}

@article{Borah:2017qdu,
    author = "Borah, Debasish and Das, Mrinal Kumar and Mukherjee, Ananya",
    title = "{Common origin of nonzero $\theta_{13}$ and baryon asymmetry of the Universe in a TeV scale seesaw model with $A_4$ flavor symmetry}",
    eprint = "1711.02445",
    archivePrefix = "arXiv",
    primaryClass = "hep-ph",
    doi = "10.1103/PhysRevD.97.115009",
    journal = "Phys. Rev. D",
    volume = "97",
    number = "11",
    pages = "115009",
    year = "2018"
}

@article{Branco:2009by,
    author = "Branco, G. C. and Gonzalez Felipe, R. and Rebelo, M. N. and Serodio, H.",
    title = "{Resonant leptogenesis and tribimaximal leptonic mixing with A(4) symmetry}",
    eprint = "0904.3076",
    archivePrefix = "arXiv",
    primaryClass = "hep-ph",
    doi = "10.1103/PhysRevD.79.093008",
    journal = "Phys. Rev. D",
    volume = "79",
    pages = "093008",
    year = "2009"
}

@article{Adhikary:2014qba,
    author = "Adhikary, Biswajit and Chakraborty, Mainak and Ghosal, Ambar",
    title = "{Flavored leptogenesis with quasidegenerate neutrinos in a broken cyclic symmetric model}",
    eprint = "1407.6173",
    archivePrefix = "arXiv",
    primaryClass = "hep-ph",
    doi = "10.1103/PhysRevD.93.113001",
    journal = "Phys. Rev. D",
    volume = "93",
    number = "11",
    pages = "113001",
    year = "2016"
}

@article{Karmakar:2015jza,
    author = "Karmakar, Biswajit and Sil, Arunansu",
    title = "{Spontaneous CP violation in lepton-sector: A common origin for $\theta_{13}$, the Dirac CP phase, and leptogenesis}",
    eprint = "1509.07090",
    archivePrefix = "arXiv",
    primaryClass = "hep-ph",
    doi = "10.1103/PhysRevD.93.013006",
    journal = "Phys. Rev. D",
    volume = "93",
    number = "1",
    pages = "013006",
    year = "2016"
}

@article{Datta:2021zzf,
    author = "Datta, Arghyajit and Karmakar, Biswajit and Sil, Arunansu",
    title = "{Flavored leptogenesis and neutrino mass with A$_{4}$ symmetry}",
    eprint = "2106.06773",
    archivePrefix = "arXiv",
    primaryClass = "hep-ph",
    doi = "10.1007/JHEP12(2021)051",
    journal = "JHEP",
    volume = "12",
    pages = "051",
    year = "2021"
}

@article{Parriciatu:2024dhb,
    author = "Parriciatu, Matteo",
    title = "{A minimalistic perspective on neutrino CP-violation and Leptogenesis: Modular Invariance}",
    doi = "10.22323/1.476.0244",
    journal = "PoS",
    volume = "ICHEP2024",
    pages = "244",
    year = "2025"
}

@article{Pathak:2025zdp,
    author = "Pathak, Gourab and Das, Mrinal Kumar",
    title = "{Matter-antimatter asymmetry in minimal inverse seesaw framework with A$_{4}$ modular symmetry}",
    eprint = "2505.03000",
    archivePrefix = "arXiv",
    primaryClass = "hep-ph",
    doi = "10.1088/1361-6471/ae42d7",
    journal = "J. Phys. G",
    volume = "53",
    number = "2",
    pages = "025004",
    year = "2026"
}

@article{Nanda:2025lem,
    author = "Nanda, Swaraj Kumar and Ricky Devi, Maibam and Patra, Sudhanwa",
    title = "{Non-Holomorphic $A_4$ Modular Symmetry in Type-I Seesaw: Implications for Neutrino Masses and Leptogenesis}",
    eprint = "2509.22108",
    archivePrefix = "arXiv",
    primaryClass = "hep-ph",
    month = "9",
    year = "2025"
}

@article{Tavartkiladze:2025oiq,
    author = "Tavartkiladze, Zurab",
    title = "{Minimal modular flavor symmetry and lepton textures near fixed points}",
    eprint = "2512.24804",
    archivePrefix = "arXiv",
    primaryClass = "hep-ph",
    doi = "10.1103/rhm2-g9rq",
    journal = "Phys. Rev. D",
    volume = "113",
    number = "9",
    pages = "095023",
    year = "2026"
}

@article{Tapender:2026ets,
    author = "Tapender and Verma, Surender",
    title = "{Tri-Resonant Leptogenesis in a Non-Holomorphic Modular A$_4$ Scotogenic Model}",
    eprint = "2602.17243",
    archivePrefix = "arXiv",
    primaryClass = "hep-ph",
    month = "2",
    year = "2026"
}

@article{Priya:2026ehe,
    author = "Priya and Chauhan, B. C. and Kumar, Deepak and Nomura, Takaaki",
    title = "{Predictions of Modular Symmetry Fixed Points on Neutrino Masses, Mixing, and Leptogenesis}",
    eprint = "2604.04585",
    archivePrefix = "arXiv",
    primaryClass = "hep-ph",
    month = "4",
    year = "2026"
}

@article{Priya:2026vpo,
    author = "Priya and Singh, Labh and Chauhan, B. C. and Verma, Surender",
    title = "{Radiative Lifting of $\mathbb{Z}_3$ Domain-Wall Degeneracy in a Type-III Seesaw Model: Implications for Leptogenesis and Gravitational Waves}",
    eprint = "2606.22642",
    archivePrefix = "arXiv",
    primaryClass = "hep-ph",
    month = "6",
    year = "2026"
}

@article{Batra:2026pef,
    author = "Batra, Cheshta and Mandal, Rusa and Rawat, Kunal and Tong, Tom",
    title = "{Connecting Flavor and Baryon Asymmetry via Leptogenesis in Effective Froggatt-Nielsen Theory}",
    eprint = "2603.05372",
    archivePrefix = "arXiv",
    primaryClass = "hep-ph",
    reportNumber = "SI-HEP-2026-04",
    month = "3",
    year = "2026"
}

@article{Wei:2022poh,
    author = "Wei, Dongdong and Jiang, Yun",
    title = "{Domain wall networks from first-order phase transitions and gravitational waves}",
    eprint = "2208.07186",
    archivePrefix = "arXiv",
    primaryClass = "hep-ph",
    doi = "10.1103/PhysRevD.110.123505",
    journal = "Phys. Rev. D",
    volume = "110",
    number = "12",
    pages = "123505",
    year = "2024"
}

@article{Borboruah:2022eex,
    author = "Borboruah, Z. A. and Yajnik, U. A.",
    title = "{Left-right symmetry breaking and gravitational waves: A tale of two phase transitions}",
    eprint = "2212.05829",
    archivePrefix = "arXiv",
    primaryClass = "astro-ph.CO",
    doi = "10.1103/PhysRevD.110.043016",
    journal = "Phys. Rev. D",
    volume = "110",
    number = "4",
    pages = "043016",
    year = "2024"
}

@article{Fornal:2024avx,
    author = "Fornal, Bartosz and Polynice, Dyori and Thompson, Luka",
    title = "{Probing the neutrino seesaw scale with gravitational waves}",
    eprint = "2406.16463",
    archivePrefix = "arXiv",
    primaryClass = "hep-ph",
    doi = "10.1103/PhysRevD.110.095013",
    journal = "Phys. Rev. D",
    volume = "110",
    number = "9",
    pages = "095013",
    year = "2024"
}

@article{Roshan:2026xpf,
    author = "Roshan, Rishav and Saha, Indrajit",
    title = "{Twin-peaked gravitational wave signal from a dark sector phase transition}",
    eprint = "2603.15829",
    archivePrefix = "arXiv",
    primaryClass = "hep-ph",
    month = "3",
    year = "2026"
}

@article{Buchmuller:2003gz,
    author = "Buchmuller, W. and Di Bari, P. and Plumacher, M.",
    title = "{The Neutrino mass window for baryogenesis}",
    eprint = "hep-ph/0302092",
    archivePrefix = "arXiv",
    reportNumber = "DESY-03-001, UAB-FT-540, CERN-TH-2003-016",
    doi = "10.1016/S0550-3213(03)00449-8",
    journal = "Nucl. Phys. B",
    volume = "665",
    pages = "445--468",
    year = "2003"
}

@article{Buchmuller:2002rq,
    author = "Buchmuller, W. and Di Bari, P. and Plumacher, M.",
    title = "{Cosmic microwave background, matter - antimatter asymmetry and neutrino masses}",
    eprint = "hep-ph/0205349",
    archivePrefix = "arXiv",
    reportNumber = "DESY-02-058, OUTP-02-23-P",
    doi = "10.1016/S0550-3213(02)00737-X",
    journal = "Nucl. Phys. B",
    volume = "643",
    pages = "367--390",
    year = "2002",
    note = "[Erratum: Nucl.Phys.B 793, 362 (2008)]"
}

@article{Davidson:2008bu,
    author = "Davidson, Sacha and Nardi, Enrico and Nir, Yosef",
    title = "{Leptogenesis}",
    eprint = "0802.2962",
    archivePrefix = "arXiv",
    primaryClass = "hep-ph",
    doi = "10.1016/j.physrep.2008.06.002",
    journal = "Phys. Rept.",
    volume = "466",
    pages = "105--177",
    year = "2008"
}

@article{Priya:2025wdm,
    author = "Priya and Singh, Labh and Chauhan, B. C. and Verma, Surender",
    title = "{Type-III seesaw in non-holomorphic modular symmetry and leptogenesis}",
    eprint = "2508.05047",
    archivePrefix = "arXiv",
    primaryClass = "hep-ph",
    doi = "10.1007/JHEP01(2026)036",
    journal = "JHEP",
    volume = "01",
    pages = "036",
    year = "2026"
}

@article{Griest:1990kh,
    author = "Griest, Kim and Seckel, David",
    title = "{Three exceptions in the calculation of relic abundances}",
    reportNumber = "CFPA-TH-90-001A, BA-90-79",
    doi = "10.1103/PhysRevD.43.3191",
    journal = "Phys. Rev. D",
    volume = "43",
    pages = "3191--3203",
    year = "1991"
}

@article{Baker:2015qna,
    author = "Baker, Michael J. and others",
    title = "{The Coannihilation Codex}",
    eprint = "1510.03434",
    archivePrefix = "arXiv",
    primaryClass = "hep-ph",
    reportNumber = "MITP-15-078",
    doi = "10.1007/JHEP12(2015)120",
    journal = "JHEP",
    volume = "12",
    pages = "120",
    year = "2015"
}

@article{Nomura:2024vzw,
    author = "Nomura, Takaaki and Okada, Hiroshi and Popov, Oleg",
    title = "{Non-holomorphic modular A4 symmetric scotogenic model}",
    eprint = "2409.12547",
    archivePrefix = "arXiv",
    primaryClass = "hep-ph",
    doi = "10.1016/j.physletb.2024.139171",
    journal = "Phys. Lett. B",
    volume = "860",
    pages = "139171",
    year = "2025"
}

@article{Nomura:2024atp,
    author = "Nomura, Takaaki and Okada, Hiroshi",
    title = "{Type-II seesaw of a non-holomorphic modular A4 symmetry}",
    eprint = "2408.01143",
    archivePrefix = "arXiv",
    primaryClass = "hep-ph",
    doi = "10.1016/j.physletb.2025.139763",
    journal = "Phys. Lett. B",
    volume = "868",
    pages = "139763",
    year = "2025"
}

@article{Manohar:2020nzp,
    author = "Manohar, Aneesh V. and Nardoni, Emily",
    title = "{Renormalization Group Improvement of the Effective Potential: an EFT Approach}",
    eprint = "2010.15806",
    archivePrefix = "arXiv",
    primaryClass = "hep-ph",
    doi = "10.1007/JHEP04(2021)093",
    journal = "JHEP",
    volume = "04",
    pages = "093",
    year = "2021"
}

@article{Ellis:2020awk,
    author = "Ellis, John and Lewicki, Marek and No, Jos{\'e} Miguel",
    title = "{Gravitational waves from first-order cosmological phase transitions: lifetime of the sound wave source}",
    eprint = "2003.07360",
    archivePrefix = "arXiv",
    primaryClass = "hep-ph",
    reportNumber = "KCL-PH-TH/2020-04, CERN-TH-2020-016, IFT-UAM/CSIC-20-35",
    doi = "10.1088/1475-7516/2020/07/050",
    journal = "JCAP",
    volume = "07",
    pages = "050",
    year = "2020"
}

@article{Wainwright:2011kj,
    author = "Wainwright, Carroll L.",
    title = "{CosmoTransitions: Computing Cosmological Phase Transition Temperatures and Bubble Profiles with Multiple Fields}",
    eprint = "1109.4189",
    archivePrefix = "arXiv",
    primaryClass = "hep-ph",
    doi = "10.1016/j.cpc.2012.04.004",
    journal = "Comput. Phys. Commun.",
    volume = "183",
    pages = "2006--2013",
    year = "2012"
}

@article{Caprini:2015zlo,
    author = "Caprini, Chiara and others",
    title = "{Science with the space-based interferometer eLISA. II: Gravitational waves from cosmological phase transitions}",
    eprint = "1512.06239",
    archivePrefix = "arXiv",
    primaryClass = "astro-ph.CO",
    reportNumber = "DESY-15-246",
    doi = "10.1088/1475-7516/2016/04/001",
    journal = "JCAP",
    volume = "04",
    pages = "001",
    year = "2016"
}

@article{Das:2026zuo,
    author = "Das, Jaydeb and Niyogi, Saurabh and Srivastava, Tripurari",
    title = "{Revisiting singlet fermion dark matter with a scalar portal: connecting Higgs phenomenology and strong electroweak phase transition}",
    eprint = "2601.13147",
    archivePrefix = "arXiv",
    primaryClass = "hep-ph",
    doi = "10.1088/1475-7516/2026/06/018",
    journal = "JCAP",
    volume = "06",
    pages = "018",
    year = "2026"
}

@article{Davidson:2002qv,
    author = "Davidson, Sacha and Ibarra, Alejandro",
    title = "{A Lower bound on the right-handed neutrino mass from leptogenesis}",
    eprint = "hep-ph/0202239",
    archivePrefix = "arXiv",
    reportNumber = "OUTP-02-10P, IPPP-02-16, DCPT-02-32",
    doi = "10.1016/S0370-2693(02)01735-5",
    journal = "Phys. Lett. B",
    volume = "535",
    pages = "25--32",
    year = "2002"
}

@article{Flanz:1996fb,
    author = "Flanz, Marion and Paschos, Emmanuel A. and Sarkar, Utpal and Weiss, Jan",
    title = "{Baryogenesis through mixing of heavy Majorana neutrinos}",
    eprint = "hep-ph/9607310",
    archivePrefix = "arXiv",
    doi = "10.1016/S0370-2693(96)01337-8",
    journal = "Phys. Lett. B",
    volume = "389",
    pages = "693--699",
    year = "1996"
}

@article{Iso:2010mv,
    author = "Iso, Satoshi and Okada, Nobuchika and Orikasa, Yuta",
    title = "{Resonant Leptogenesis in the Minimal B-L Extended Standard Model at TeV}",
    eprint = "1011.4769",
    archivePrefix = "arXiv",
    primaryClass = "hep-ph",
    reportNumber = "KEK-TH-1422",
    doi = "10.1103/PhysRevD.83.093011",
    journal = "Phys. Rev. D",
    volume = "83",
    pages = "093011",
    year = "2011"
}

@article{Qi:2022fzs,
    author = "Qi, XinXin and Sun, Hao",
    title = "{Interplay between dark matter and leptogenesis in a common framework}",
    eprint = "2208.13345",
    archivePrefix = "arXiv",
    primaryClass = "hep-ph",
    doi = "10.1007/JHEP09(2023)118",
    journal = "JHEP",
    volume = "09",
    pages = "118",
    year = "2023"
}

@article{Das:2024gua,
    author = "Das, Arindam and Orikasa, Yuta",
    title = "{Resonant leptogenesis in minimal U(1)X extensions of the Standard Model}",
    eprint = "2407.05644",
    archivePrefix = "arXiv",
    primaryClass = "hep-ph",
    doi = "10.1016/j.physletb.2025.139395",
    journal = "Phys. Lett. B",
    volume = "864",
    pages = "139395",
    year = "2025"
}

@article{LISA:2017pwj,
    author = "Amaro-Seoane, Pau and others",
    collaboration = "LISA",
    title = "{Laser Interferometer Space Antenna}",
    eprint = "1702.00786",
    archivePrefix = "arXiv",
    primaryClass = "astro-ph.IM",
    month = "2",
    year = "2017"
}

@article{Yagi:2011wg,
    author = "Yagi, Kent and Seto, Naoki",
    title = "{Detector configuration of DECIGO/BBO and identification of cosmological neutron-star binaries}",
    eprint = "1101.3940",
    archivePrefix = "arXiv",
    primaryClass = "astro-ph.CO",
    doi = "10.1103/PhysRevD.83.044011",
    journal = "Phys. Rev. D",
    volume = "83",
    pages = "044011",
    year = "2011",
    note = "[Erratum: Phys.Rev.D 95, 109901 (2017)]"
}

@article{Nakayama:2009ce,
    author = "Nakayama, Kazunori and Yokoyama, Jun'ichi",
    title = "{Gravitational Wave Background and Non-Gaussianity as a Probe of the Curvaton Scenario}",
    eprint = "0910.0715",
    archivePrefix = "arXiv",
    primaryClass = "astro-ph.CO",
    reportNumber = "ICRR-REPORT-551, RESCEU-26-09",
    doi = "10.1088/1475-7516/2010/01/010",
    journal = "JCAP",
    volume = "01",
    pages = "010",
    year = "2010"
}

@article{Punturo:2010zz,
    author = "Punturo, M. and others",
    editor = "Ricci, Fulvio",
    title = "{The Einstein Telescope: A third-generation gravitational wave observatory}",
    doi = "10.1088/0264-9381/27/19/194002",
    journal = "Class. Quant. Grav.",
    volume = "27",
    pages = "194002",
    year = "2010"
}

@article{Reitze:2019iox,
    author = "Reitze, David and others",
    title = "{Cosmic Explorer: The U.S. Contribution to Gravitational-Wave Astronomy beyond LIGO}",
    eprint = "1907.04833",
    archivePrefix = "arXiv",
    primaryClass = "astro-ph.IM",
    reportNumber = "LIGO-P1900316",
    journal = "Bull. Am. Astron. Soc.",
    volume = "51",
    number = "7",
    pages = "035",
    year = "2019"
}

@article{LIGOScientific:2014pky,
    author = "Aasi, J. and others",
    collaboration = "LIGO Scientific",
    title = "{Advanced LIGO}",
    eprint = "1411.4547",
    archivePrefix = "arXiv",
    primaryClass = "gr-qc",
    doi = "10.1088/0264-9381/32/7/074001",
    journal = "Class. Quant. Grav.",
    volume = "32",
    pages = "074001",
    year = "2015"
}

@article{VIRGO:2014yos,
    author = "Acernese, F. and others",
    collaboration = "VIRGO",
    title = "{Advanced Virgo: a second-generation interferometric gravitational wave detector}",
    eprint = "1408.3978",
    archivePrefix = "arXiv",
    primaryClass = "gr-qc",
    doi = "10.1088/0264-9381/32/2/024001",
    journal = "Class. Quant. Grav.",
    volume = "32",
    number = "2",
    pages = "024001",
    year = "2015"
}

@article{KAGRA:2018plz,
    author = "Akutsu, T. and others",
    collaboration = "KAGRA",
    title = "{KAGRA: 2.5 Generation Interferometric Gravitational Wave Detector}",
    eprint = "1811.08079",
    archivePrefix = "arXiv",
    primaryClass = "gr-qc",
    reportNumber = "JGW-P1809243",
    doi = "10.1038/s41550-018-0658-y",
    journal = "Nature Astron.",
    volume = "3",
    number = "1",
    pages = "35--40",
    year = "2019"
}

@article{NANOGrav:2023gor,
    author = "Agazie, Gabriella and others",
    collaboration = "NANOGrav",
    title = "{The NANOGrav 15 yr Data Set: Evidence for a Gravitational-wave Background}",
    eprint = "2306.16213",
    archivePrefix = "arXiv",
    primaryClass = "astro-ph.HE",
    doi = "10.3847/2041-8213/acdac6",
    journal = "Astrophys. J. Lett.",
    volume = "951",
    number = "1",
    pages = "L8",
    year = "2023"
}

@article{Schmitz:2020syl,
    author = "Schmitz, Kai",
    title = "{New Sensitivity Curves for Gravitational-Wave Signals from Cosmological Phase Transitions}",
    eprint = "2002.04615",
    archivePrefix = "arXiv",
    primaryClass = "hep-ph",
    reportNumber = "CERN-TH-2020-018",
    doi = "10.1007/JHEP01(2021)097",
    journal = "JHEP",
    volume = "01",
    pages = "097",
    year = "2021"
}

@article{Kang:2017mkl,
    author = "Kang, Zhaofeng and Ko, P. and Matsui, Toshinori",
    title = "{Strong first order EWPT $\&$ strong gravitational waves in Z$_{3}$-symmetric singlet scalar extension}",
    eprint = "1706.09721",
    archivePrefix = "arXiv",
    primaryClass = "hep-ph",
    reportNumber = "KIAS-P17047",
    doi = "10.1007/JHEP02(2018)115",
    journal = "JHEP",
    volume = "02",
    pages = "115",
    year = "2018"
}

@article{Kannike:2019mzk,
    author = "Kannike, Kristjan and Loos, Kaius and Raidal, Martti",
    title = "{Gravitational wave signals of pseudo-Goldstone dark matter in the $\mathbb{Z}_{3}$ complex singlet model}",
    eprint = "1907.13136",
    archivePrefix = "arXiv",
    primaryClass = "hep-ph",
    doi = "10.1103/PhysRevD.101.035001",
    journal = "Phys. Rev. D",
    volume = "101",
    number = "3",
    pages = "035001",
    year = "2020"
}

@article{Ghosh:2024ing,
    author = "Ghosh, Dilip Kumar and Mukherjee, Koustav and Mukherjee, Shourya",
    title = "{Electroweak phase transition in two scalar singlet model with pNGB dark matter}",
    eprint = "2409.00192",
    archivePrefix = "arXiv",
    primaryClass = "hep-ph",
    doi = "10.1007/JHEP01(2025)078",
    journal = "JHEP",
    volume = "01",
    pages = "078",
    year = "2025"
}

@inproceedings{Quiros:1999jp,
    author = "Quiros, Mariano",
    title = "{Finite temperature field theory and phase transitions}",
    booktitle = "{ICTP Summer School in High-Energy Physics and Cosmology}",
    eprint = "hep-ph/9901312",
    archivePrefix = "arXiv",
    reportNumber = "IEM-FT-187-99",
    pages = "187--259",
    month = "1",
    year = "1999"
}

@article{Gelmini:2020bqg,
    author = "Gelmini, Graciela B. and Pascoli, Silvia and Vitagliano, Edoardo and Zhou, Ye-Ling",
    title = "{Gravitational wave signatures from discrete flavor symmetries}",
    eprint = "2009.01903",
    archivePrefix = "arXiv",
    primaryClass = "hep-ph",
    doi = "10.1088/1475-7516/2021/02/032",
    journal = "JCAP",
    volume = "02",
    pages = "032",
    year = "2021"
}

@article{Press:1989yh,
    author = "Press, William H. and Ryden, Barbara S. and Spergel, David N.",
    title = "{Dynamical Evolution of Domain Walls in an Expanding Universe}",
    reportNumber = "NSF-ITP-89-51, CFA-1870",
    doi = "10.1086/168151",
    journal = "Astrophys. J.",
    volume = "347",
    pages = "590--604",
    year = "1989"
}

@article{Everett:1982nm,
    author = "Everett, Allen E. and Vilenkin, Alexander",
    title = "{Left-right Symmetric Theories and Vacuum Domain Walls and Strings}",
    reportNumber = "TUTP-82-4",
    doi = "10.1016/0550-3213(82)90135-3",
    journal = "Nucl. Phys. B",
    volume = "207",
    pages = "43--53",
    year = "1982"
}

@article{Garagounis:2002kt,
    author = "Garagounis, Theodore and Hindmarsh, Mark",
    title = "{Scaling in numerical simulations of domain walls}",
    eprint = "hep-ph/0212359",
    archivePrefix = "arXiv",
    reportNumber = "SUSX-TH-02-029",
    doi = "10.1103/PhysRevD.68.103506",
    journal = "Phys. Rev. D",
    volume = "68",
    pages = "103506",
    year = "2003"
}

@article{Borah:2025bfa,
    author = "Borah, Debasish and Saha, Indrajit",
    title = "{Gravitational waves from seesaw assisted collapsing domain walls}",
    eprint = "2512.22339",
    archivePrefix = "arXiv",
    primaryClass = "hep-ph",
    doi = "10.1016/j.physletb.2026.140618",
    journal = "Phys. Lett. B",
    volume = "879",
    pages = "140618",
    year = "2026"
}

@article{Kawasaki:2014sqa,
    author = "Kawasaki, Masahiro and Saikawa, Ken'ichi and Sekiguchi, Toyokazu",
    title = "{Axion dark matter from topological defects}",
    eprint = "1412.0789",
    archivePrefix = "arXiv",
    primaryClass = "hep-ph",
    reportNumber = "ICRR-REPORT-696-2014-22, IPMU14-0348",
    doi = "10.1103/PhysRevD.91.065014",
    journal = "Phys. Rev. D",
    volume = "91",
    number = "6",
    pages = "065014",
    year = "2015"
}

@article{Saikawa:2017hiv,
    author = "Saikawa, Ken'ichi",
    title = "{A review of gravitational waves from cosmic domain walls}",
    eprint = "1703.02576",
    archivePrefix = "arXiv",
    primaryClass = "hep-ph",
    reportNumber = "DESY-17-036",
    doi = "10.3390/universe3020040",
    journal = "Universe",
    volume = "3",
    number = "2",
    pages = "40",
    year = "2017"
}

@article{Roshan:2024qnv,
    author = "Roshan, Rishav and White, Graham",
    title = "{Using gravitational waves to see the first second of the Universe}",
    eprint = "2401.04388",
    archivePrefix = "arXiv",
    primaryClass = "hep-ph",
    doi = "10.1103/RevModPhys.97.015001",
    journal = "Rev. Mod. Phys.",
    volume = "97",
    number = "1",
    pages = "015001",
    year = "2025"
}

@article{Stauffer:1978kr,
    author = "Stauffer, D.",
    title = "{Scaling theory of percolation clusters}",
    doi = "10.1016/0370-1573(79)90060-7",
    journal = "Phys. Rept.",
    volume = "54",
    pages = "1--74",
    year = "1979"
}

@article{Wu:2022stu,
    author = "Wu, Yongcheng and Xie, Ke-Pan and Zhou, Ye-Ling",
    title = "{Collapsing domain walls beyond Z2}",
    eprint = "2204.04374",
    archivePrefix = "arXiv",
    primaryClass = "hep-ph",
    doi = "10.1103/PhysRevD.105.095013",
    journal = "Phys. Rev. D",
    volume = "105",
    number = "9",
    pages = "095013",
    year = "2022"
}

@article{Cyburt:2015mya,
    author = "Cyburt, Richard H. and Fields, Brian D. and Olive, Keith A. and Yeh, Tsung-Han",
    title = "{Big Bang Nucleosynthesis: 2015}",
    eprint = "1505.01076",
    archivePrefix = "arXiv",
    primaryClass = "astro-ph.CO",
    reportNumber = "UMN-TH-3432-15, FTPI-MINN-15-19",
    doi = "10.1103/RevModPhys.88.015004",
    journal = "Rev. Mod. Phys.",
    volume = "88",
    pages = "015004",
    year = "2016"
}

@article{Abazajian:2019eic,
    author = "Abazajian, Kevork and others",
    title = "{CMB-S4 Science Case, Reference Design, and Project Plan}",
    eprint = "1907.04473",
    archivePrefix = "arXiv",
    primaryClass = "astro-ph.IM",
    reportNumber = "FERMILAB-PUB-19-431-AE-SCD",
    month = "7",
    year = "2019"
}

@article{Deng:2020dnf,
    author = "Deng, Xin and Liu, Xuewen and Yang, Jing and Zhou, Ruiyu and Bian, Ligong",
    title = "{Heavy dark matter and Gravitational waves}",
    eprint = "2012.15174",
    archivePrefix = "arXiv",
    primaryClass = "hep-ph",
    doi = "10.1103/PhysRevD.103.055013",
    journal = "Phys. Rev. D",
    volume = "103",
    number = "5",
    pages = "055013",
    year = "2021"
}

@article{Hattori:2015xla,
    author = "Hattori, Hironori and Kobayashi, Tatsuo and Omoto, Naoya and Seto, Osamu",
    title = "{Entropy production by domain wall decay in the NMSSM}",
    eprint = "1510.03595",
    archivePrefix = "arXiv",
    primaryClass = "hep-ph",
    reportNumber = "HGU-CAP-039, EPHOU-15-014",
    doi = "10.1103/PhysRevD.92.103518",
    journal = "Phys. Rev. D",
    volume = "92",
    number = "10",
    pages = "103518",
    year = "2015"
}

@misc{amaroseoane2017laserinterferometerspaceantenna,
      title={Laser Interferometer Space Antenna}, 
      author={Pau Amaro-Seoane and Heather Audley and Stanislav Babak and John Baker and Enrico Barausse and Peter Bender and Emanuele Berti and Pierre Binetruy and Michael Born and Daniele Bortoluzzi and Jordan Camp and Chiara Caprini and Vitor Cardoso and Monica Colpi and John Conklin and Neil Cornish and Curt Cutler and Karsten Danzmann and Rita Dolesi and Luigi Ferraioli and Valerio Ferroni and Ewan Fitzsimons and Jonathan Gair and Lluis Gesa Bote and Domenico Giardini and Ferran Gibert and Catia Grimani and Hubert Halloin and Gerhard Heinzel and Thomas Hertog and Martin Hewitson and Kelly Holley-Bockelmann and Daniel Hollington and Mauro Hueller and Henri Inchauspe and Philippe Jetzer and Nikos Karnesis and Christian Killow and Antoine Klein and Bill Klipstein and Natalia Korsakova and Shane L Larson and Jeffrey Livas and Ivan Lloro and Nary Man and Davor Mance and Joseph Martino and Ignacio Mateos and Kirk McKenzie and Sean T McWilliams and Cole Miller and Guido Mueller and Germano Nardini and Gijs Nelemans and Miquel Nofrarias and Antoine Petiteau and Paolo Pivato and Eric Plagnol and Ed Porter and Jens Reiche and David Robertson and Norna Robertson and Elena Rossi and Giuliana Russano and Bernard Schutz and Alberto Sesana and David Shoemaker and Jacob Slutsky and Carlos F. Sopuerta and Tim Sumner and Nicola Tamanini and Ira Thorpe and Michael Troebs and Michele Vallisneri and Alberto Vecchio and Daniele Vetrugno and Stefano Vitale and Marta Volonteri and Gudrun Wanner and Harry Ward and Peter Wass and William Weber and John Ziemer and Peter Zweifel},
      year={2017},
      eprint={1702.00786},
      archivePrefix={arXiv},
      primaryClass={astro-ph.IM},
      url={https://arxiv.org/abs/1702.00786}, 
}

@article{Kawamura:2020pcg,
    author = "Kawamura, Seiji and others",
    title = "{Current status of space gravitational wave antenna DECIGO and B-DECIGO}",
    eprint = "2006.13545",
    archivePrefix = "arXiv",
    primaryClass = "gr-qc",
    doi = "10.1093/ptep/ptab019",
    journal = "PTEP",
    volume = "2021",
    number = "5",
    pages = "05A105",
    year = "2021"
}

@article{Sesana:2019vho,
    author = "Sesana, Alberto and others",
    title = "{Unveiling the gravitational universe at $\mu$-Hz frequencies}",
    eprint = "1908.11391",
    archivePrefix = "arXiv",
    primaryClass = "astro-ph.IM",
    doi = "10.1007/s10686-021-09709-9",
    journal = "Exper. Astron.",
    volume = "51",
    number = "3",
    pages = "1333--1383",
    year = "2021"
}

@article{Garcia-Bellido:2021zgu,
    author = "Garcia-Bellido, Juan and Murayama, Hitoshi and White, Graham",
    title = "{Exploring the early Universe with Gaia and Theia}",
    eprint = "2104.04778",
    archivePrefix = "arXiv",
    primaryClass = "hep-ph",
    reportNumber = "IFT-UAM/CSIC-2021-038",
    doi = "10.1088/1475-7516/2021/12/023",
    journal = "JCAP",
    volume = "12",
    number = "12",
    pages = "023",
    year = "2021"
}

@article{Weltman:2018zrl,
    author = "Weltman, A. and others",
    title = "{Fundamental physics with the Square Kilometre Array}",
    eprint = "1810.02680",
    archivePrefix = "arXiv",
    primaryClass = "astro-ph.CO",
    doi = "10.1017/pasa.2019.42",
    journal = "Publ. Astron. Soc. Austral.",
    volume = "37",
    pages = "e002",
    year = "2020"
}

@article{Crowder:2005nr,
    author = "Crowder, Jeff and Cornish, Neil J.",
    title = "{Beyond LISA: Exploring future gravitational wave missions}",
    eprint = "gr-qc/0506015",
    archivePrefix = "arXiv",
    doi = "10.1103/PhysRevD.72.083005",
    journal = "Phys. Rev. D",
    volume = "72",
    pages = "083005",
    year = "2005"
}

@article{Corbin:2005ny,
    author = "Corbin, Vincent and Cornish, Neil J.",
    title = "{Detecting the cosmic gravitational wave background with the big bang observer}",
    eprint = "gr-qc/0512039",
    archivePrefix = "arXiv",
    doi = "10.1088/0264-9381/23/7/014",
    journal = "Class. Quant. Grav.",
    volume = "23",
    pages = "2435--2446",
    year = "2006"
}

@article{Harry:2006fi,
    author = "Harry, G. M. and Fritschel, P. and Shaddock, D. A. and Folkner, W. and Phinney, E. S.",
    title = "{Laser interferometry for the big bang observer}",
    doi = "10.1088/0264-9381/23/15/008",
    journal = "Class. Quant. Grav.",
    volume = "23",
    pages = "4887--4894",
    year = "2006",
    note = "[Erratum: Class.Quant.Grav. 23, 7361 (2006)]"
}

@article{Seto:2001qf,
    author = "Seto, Naoki and Kawamura, Seiji and Nakamura, Takashi",
    title = "{Possibility of direct measurement of the acceleration of the universe using 0.1-Hz band laser interferometer gravitational wave antenna in space}",
    eprint = "astro-ph/0108011",
    archivePrefix = "arXiv",
    doi = "10.1103/PhysRevLett.87.221103",
    journal = "Phys. Rev. Lett.",
    volume = "87",
    pages = "221103",
    year = "2001"
}

@article{Kawamura:2006up,
    author = "Kawamura, S. and others",
    editor = "Mio, N.",
    title = "{The Japanese space gravitational wave antenna DECIGO}",
    doi = "10.1088/0264-9381/23/8/S17",
    journal = "Class. Quant. Grav.",
    volume = "23",
    pages = "S125--S132",
    year = "2006"
}

@article{LIGOScientific:2016wof,
    author = "Abbott, Benjamin P and others",
    collaboration = "LIGO Scientific",
    title = "{Exploring the Sensitivity of Next Generation Gravitational Wave Detectors}",
    eprint = "1607.08697",
    archivePrefix = "arXiv",
    primaryClass = "astro-ph.IM",
    reportNumber = "LIGO-P1600143",
    doi = "10.1088/1361-6382/aa51f4",
    journal = "Class. Quant. Grav.",
    volume = "34",
    number = "4",
    pages = "044001",
    year = "2017"
}

@article{Hild:2010id,
    author = "Hild, S. and others",
    title = "{Sensitivity Studies for Third-Generation Gravitational Wave Observatories}",
    eprint = "1012.0908",
    archivePrefix = "arXiv",
    primaryClass = "gr-qc",
    doi = "10.1088/0264-9381/28/9/094013",
    journal = "Class. Quant. Grav.",
    volume = "28",
    pages = "094013",
    year = "2011"
}

@article{Sathyaprakash:2012jk,
    author = "Sathyaprakash, B. and others",
    editor = "Hannam, Mark and Sutton, Patrick and Hild, Stefan and van den Broeck, Chris",
    title = "{Scientific Objectives of Einstein Telescope}",
    eprint = "1206.0331",
    archivePrefix = "arXiv",
    primaryClass = "gr-qc",
    doi = "10.1088/0264-9381/29/12/124013",
    journal = "Class. Quant. Grav.",
    volume = "29",
    pages = "124013",
    year = "2012",
    note = "[Erratum: Class.Quant.Grav. 30, 079501 (2013)]"
}

@article{Maggiore:2019uih,
    author = "Maggiore, Michele and others",
    title = "{Science Case for the Einstein Telescope}",
    eprint = "1912.02622",
    archivePrefix = "arXiv",
    primaryClass = "astro-ph.CO",
    doi = "10.1088/1475-7516/2020/03/050",
    journal = "JCAP",
    volume = "03",
    pages = "050",
    year = "2020"
}

@article{LIGOScientific:2014qfs,
    author = "Aasi, J. and others",
    collaboration = "LIGO Scientific, VIRGO",
    title = "{Characterization of the LIGO detectors during their sixth science run}",
    eprint = "1410.7764",
    archivePrefix = "arXiv",
    primaryClass = "gr-qc",
    doi = "10.1088/0264-9381/32/11/115012",
    journal = "Class. Quant. Grav.",
    volume = "32",
    number = "11",
    pages = "115012",
    year = "2015"
}

@article{LIGOScientific:2016jlg,
    author = "Abbott, Benjamin P. and others",
    collaboration = "LIGO Scientific, Virgo",
    title = "{Upper Limits on the Stochastic Gravitational-Wave Background from Advanced LIGO\textquoteright{}s First Observing Run}",
    eprint = "1612.02029",
    archivePrefix = "arXiv",
    primaryClass = "gr-qc",
    doi = "10.1103/PhysRevLett.118.121101",
    journal = "Phys. Rev. Lett.",
    volume = "118",
    number = "12",
    pages = "121101",
    year = "2017",
    note = "[Erratum: Phys.Rev.Lett. 119, 029901 (2017)]"
}

@article{Moore:2014eua,
    author = "Moore, Christopher J. and Taylor, Stephen R. and Gair, Jonathan R.",
    title = "{Estimating the sensitivity of pulsar timing arrays}",
    eprint = "1406.5199",
    archivePrefix = "arXiv",
    primaryClass = "astro-ph.IM",
    doi = "10.1088/0264-9381/32/5/055004",
    journal = "Class. Quant. Grav.",
    volume = "32",
    number = "5",
    pages = "055004",
    year = "2015"
}

@article{Maggiore:1999vm,
    author = "Maggiore, Michele",
    title = "{Gravitational wave experiments and early universe cosmology}",
    eprint = "gr-qc/9909001",
    archivePrefix = "arXiv",
    reportNumber = "IFUP-TH-20-99",
    doi = "10.1016/S0370-1573(99)00102-7",
    journal = "Phys. Rept.",
    volume = "331",
    pages = "283--367",
    year = "2000"
}

@article{Allen:1997ad,
    author = "Allen, Bruce and Romano, Joseph D.",
    title = "{Detecting a stochastic background of gravitational radiation: Signal processing strategies and sensitivities}",
    eprint = "gr-qc/9710117",
    archivePrefix = "arXiv",
    reportNumber = "WISC-MILW-97-TH-14",
    doi = "10.1103/PhysRevD.59.102001",
    journal = "Phys. Rev. D",
    volume = "59",
    pages = "102001",
    year = "1999"
}

@article{Roshan:2026yon,
    author = "Roshan, Rishav",
    title = "{Imprint of domain wall annihilation on induced gravitational waves}",
    eprint = "2604.25726",
    archivePrefix = "arXiv",
    primaryClass = "hep-ph",
    month = "4",
    year = "2026"
}

@article{Chakrabarty:2022yzp,
    author = "Chakrabarty, Nabarun and Roy, Himadri and Srivastava, Tripurari",
    title = "{Single-step first order phase transition and gravitational waves in a SIMP dark matter scenario}",
    eprint = "2212.09659",
    archivePrefix = "arXiv",
    primaryClass = "hep-ph",
    doi = "10.1016/j.nuclphysb.2023.116392",
    journal = "Nucl. Phys. B",
    volume = "998",
    pages = "116392",
    year = "2024"
}

@article{Ellis:2022lft,
    author = "Ellis, John and Lewicki, Marek and Merchand, Marco and No, Jos{\'e} Miguel and Zych, Mateusz",
    title = "{The scalar singlet extension of the Standard Model: gravitational waves versus baryogenesis}",
    eprint = "2210.16305",
    archivePrefix = "arXiv",
    primaryClass = "hep-ph",
    reportNumber = "CERN-TH-2022-150, KCL-PH-TH/2022-48, IFT{\textendash}UAM/CSIC{\textendash}22-133",
    doi = "10.1007/JHEP01(2023)093",
    journal = "JHEP",
    volume = "01",
    pages = "093",
    year = "2023"
}

@article{Ellis:2018mja,
    author = "Ellis, John and Lewicki, Marek and No, Jos{\'e} Miguel",
    title = "{On the Maximal Strength of a First-Order Electroweak Phase Transition and its Gravitational Wave Signal}",
    eprint = "1809.08242",
    archivePrefix = "arXiv",
    primaryClass = "hep-ph",
    reportNumber = "KCL-PH-TH/2018-46, CERN-TH/2018-197, IFT-UAM/CSIC-18-94, CERN-TH-2018-197",
    doi = "10.1088/1475-7516/2019/04/003",
    journal = "JCAP",
    volume = "04",
    pages = "003",
    year = "2019"
}

@article{Croon:2018erz,
    author = "Croon, Djuna and Sanz, Ver{\'o}nica and White, Graham",
    title = "{Model Discrimination in Gravitational Wave spectra from Dark Phase Transitions}",
    eprint = "1806.02332",
    archivePrefix = "arXiv",
    primaryClass = "hep-ph",
    doi = "10.1007/JHEP08(2018)203",
    journal = "JHEP",
    volume = "08",
    pages = "203",
    year = "2018"
}

@article{Beniwal:2018hyi,
    author = "Beniwal, Ankit and Lewicki, Marek and White, Martin and Williams, Anthony G.",
    title = "{Gravitational waves and electroweak baryogenesis in a global study of the extended scalar singlet model}",
    eprint = "1810.02380",
    archivePrefix = "arXiv",
    primaryClass = "hep-ph",
    reportNumber = "ADP-18-27-T1075, KCL-PH-TH/2018-54",
    doi = "10.1007/JHEP02(2019)183",
    journal = "JHEP",
    volume = "02",
    pages = "183",
    year = "2019"
}

@article{Kobakhidze:2016mch,
    author = "Kobakhidze, Archil and Manning, Adrian and Yue, Jason",
    title = "{Gravitational waves from the phase transition of a nonlinearly realized electroweak gauge symmetry}",
    eprint = "1607.00883",
    archivePrefix = "arXiv",
    primaryClass = "hep-ph",
    doi = "10.1142/S0218271817501140",
    journal = "Int. J. Mod. Phys. D",
    volume = "26",
    number = "10",
    pages = "1750114",
    year = "2017"
}

@article{Mazumdar:2018dfl,
    author = "Mazumdar, Anupam and White, Graham",
    title = "{Review of cosmic phase transitions: their significance and experimental signatures}",
    eprint = "1811.01948",
    archivePrefix = "arXiv",
    primaryClass = "hep-ph",
    doi = "10.1088/1361-6633/ab1f55",
    journal = "Rept. Prog. Phys.",
    volume = "82",
    number = "7",
    pages = "076901",
    year = "2019"
}

@article{Kamionkowski:1993fg,
    author = "Kamionkowski, Marc and Kosowsky, Arthur and Turner, Michael S.",
    title = "{Gravitational radiation from first order phase transitions}",
    eprint = "astro-ph/9310044",
    archivePrefix = "arXiv",
    reportNumber = "IASSNS-HEP-93-44, FERMILAB-PUB-93-235-A",
    doi = "10.1103/PhysRevD.49.2837",
    journal = "Phys. Rev. D",
    volume = "49",
    pages = "2837--2851",
    year = "1994"
}

@article{Srivastava:2025oer,
    author = "Srivastava, Tripurari and Das, Jaydeb and Ghosh, Anupam and Chaudhuri, Arnab",
    title = "{Electroweak phase transition, gravitational waves and collider probes in multi-scalar dark matter scenarios}",
    eprint = "2507.05917",
    archivePrefix = "arXiv",
    primaryClass = "hep-ph",
    doi = "10.1088/1475-7516/2026/02/032",
    journal = "JCAP",
    volume = "02",
    pages = "032",
    year = "2026"
}

@article{Choudhury:2026laq,
    author = "Choudhury, Debajyoti and Das, Jaydeb and Srivastava, Tripurari",
    title = "{Solving Cosmological Puzzles using Finite Temperature $ν$SMEFT}",
    eprint = "2604.21492",
    archivePrefix = "arXiv",
    primaryClass = "hep-ph",
    month = "4",
    year = "2026"
}

@article{Bertone:2004pz,
    author = "Bertone, Gianfranco and Hooper, Dan and Silk, Joseph",
    title = "{Particle dark matter: Evidence, candidates and constraints}",
    eprint = "hep-ph/0404175",
    archivePrefix = "arXiv",
    reportNumber = "FERMILAB-PUB-04-047-A",
    doi = "10.1016/j.physrep.2004.08.031",
    journal = "Phys. Rept.",
    volume = "405",
    pages = "279--390",
    year = "2005"
}

@article{Rubakov:1996vz,
    author = "Rubakov, V. A. and Shaposhnikov, M. E.",
    title = "{Electroweak baryon number nonconservation in the early universe and in high-energy collisions}",
    eprint = "hep-ph/9603208",
    archivePrefix = "arXiv",
    reportNumber = "CERN-TH-96-13, INR-0913-96",
    doi = "10.1070/PU1996v039n05ABEH000145",
    journal = "Usp. Fiz. Nauk",
    volume = "166",
    pages = "493--537",
    year = "1996"
}

@article{Klinkhamer:1984di,
    author = "Klinkhamer, Frans R. and Manton, N. S.",
    title = "{A Saddle Point Solution in the Weinberg-Salam Theory}",
    reportNumber = "NSF-ITP-84-57",
    doi = "10.1103/PhysRevD.30.2212",
    journal = "Phys. Rev. D",
    volume = "30",
    pages = "2212",
    year = "1984"
}

@article{Manton:1983nd,
    author = "Manton, N. S.",
    title = "{Topology in the Weinberg-Salam Theory}",
    reportNumber = "NSF-ITP-83-64",
    doi = "10.1103/PhysRevD.28.2019",
    journal = "Phys. Rev. D",
    volume = "28",
    pages = "2019",
    year = "1983"
}

@article{Chaudhuri:2022sis,
    author = "Chaudhuri, Arnab and Das, Jaydeb",
    title = "{Study of entropy production due to electroweak phase transition in Z2 symmetric extension of the Standard Model}",
    eprint = "2206.08699",
    archivePrefix = "arXiv",
    primaryClass = "hep-ph",
    doi = "10.1103/PhysRevD.106.095016",
    journal = "Phys. Rev. D",
    volume = "106",
    number = "9",
    pages = "095016",
    year = "2022"
}

@article{Chaudhuri:2025cjp,
    author = "Chaudhuri, Arnab and Mishra, Priya and Mohanta, Rukmani",
    title = "{Gravitational Wave Signatures of U(1)$_{X}$ Breaking and Right-Handed Neutrino Dynamics}",
    eprint = "2508.09835",
    archivePrefix = "arXiv",
    primaryClass = "hep-ph",
    doi = "10.1088/1475-7516/2026/07/061",
    journal = "JCAP",
    volume = "07",
    pages = "061",
    year = "2026"
}

@article{Borah:2023zsb,
    author = "Borah, Pankaj and Ghosh, Pradipta and Roy, Sourov and Saha, Abhijit Kumar",
    title = "{Electroweak phase transition in a right-handed neutrino superfield extended NMSSM}",
    eprint = "2301.05061",
    archivePrefix = "arXiv",
    primaryClass = "hep-ph",
    doi = "10.1007/JHEP08(2023)029",
    journal = "JHEP",
    volume = "08",
    pages = "029",
    year = "2023"
}

@article{Borah:2024emz,
    author = "Borah, Pankaj and Ghosh, Pradipta and Saha, Abhijit Kumar",
    title = "{Prospecting bipartite dark matter through gravitational waves}",
    eprint = "2412.17141",
    archivePrefix = "arXiv",
    primaryClass = "hep-ph",
    doi = "10.1088/1475-7516/2025/05/035",
    journal = "JCAP",
    volume = "05",
    pages = "035",
    year = "2025"
}

@article{Borah:2024lml,
    author = "Borah, Debasish and Jyoti Das, Suruj and Saha, Indrajit",
    title = "{Dark matter from phase transition generated PBH evaporation with gravitational waves signatures}",
    eprint = "2401.12282",
    archivePrefix = "arXiv",
    primaryClass = "hep-ph",
    doi = "10.1103/PhysRevD.110.035014",
    journal = "Phys. Rev. D",
    volume = "110",
    number = "3",
    pages = "035014",
    year = "2024"
}

@article{Nicolis:2003tg,
    author = "Nicolis, Alberto",
    title = "{Relic gravitational waves from colliding bubbles and cosmic turbulence}",
    eprint = "gr-qc/0303084",
    archivePrefix = "arXiv",
    reportNumber = "IEM-FT-230-03",
    doi = "10.1088/0264-9381/21/4/L05",
    journal = "Class. Quant. Grav.",
    volume = "21",
    pages = "L27",
    year = "2004"
}

@article{Linde:1981zj,
    author = "Linde, Andrei D.",
    title = "{Decay of the False Vacuum at Finite Temperature}",
    reportNumber = "LEBEDEV-81-265",
    doi = "10.1016/0550-3213(83)90072-X",
    journal = "Nucl. Phys. B",
    volume = "216",
    pages = "421",
    year = "1983",
    note = "[Erratum: Nucl.Phys.B 223, 544 (1983)]"
}

@article{Kehayias:2009tn,
    author = "Kehayias, John and Profumo, Stefano",
    title = "{Semi-Analytic Calculation of the Gravitational Wave Signal From the Electroweak Phase Transition for General Quartic Scalar Effective Potentials}",
    eprint = "0911.0687",
    archivePrefix = "arXiv",
    primaryClass = "hep-ph",
    doi = "10.1088/1475-7516/2010/03/003",
    journal = "JCAP",
    volume = "03",
    pages = "003",
    year = "2010"
}

@article{Bhattacharyya:2026esv,
    author = "Bhattacharyya, Arka and Biswas, Sanjoy and Niyogi, Saurabh",
    title = "{Electroweak phase transitions in a $U(1)_D$ extension of the standard model with dimension-six operators: Gravitational waves and LHC signatures}",
    eprint = "2603.18583",
    archivePrefix = "arXiv",
    primaryClass = "hep-ph",
    month = "3",
    year = "2026"
}

\end{document}